\documentclass[11pt]{article}

\usepackage{jheppub} 
\usepackage{graphicx,color}
\usepackage{amsmath}
\usepackage{slashed}
\usepackage{multirow}
\usepackage{autobreak}
\usepackage[normalem]{ulem}
\usepackage{hyperref}
\usepackage{cleveref}
\usepackage{slashed}
\usepackage{amssymb}
\usepackage{epsfig}
\usepackage{array}
\usepackage{mathtools}
\usepackage{caption}
\usepackage[utf8]{inputenc}
\usepackage{mathtools, nccmath}
\usepackage{float}
\usepackage[section]{placeins}
\usepackage[utf8]{inputenc}
\usepackage[T1]{fontenc}
\usepackage{lmodern}
\usepackage[x11names]{xcolor}
\usepackage{framed}
\usepackage{float}

\colorlet{shadecolor}{blue!10}

\newcommand{\minitab}[2][l]{\begin{tabular}{#1}#2\end{tabular}}

\SetSymbolFont{largesymbols}{bold}{OMX}{txex}{b}{n}

\DeclareMathAlphabet{\mathpzc}{OT1}{pzc}{m}{it}

\newcommand{\be}{\begin{eqnarray*}}
	\newcommand{\ee}{\end{eqnarray*}}

\newcommand{\gev}{{\text{GeV}}}
\newcommand{\tev}{{\text{TeV}}}

\usepackage{parskip}

\newcommand{\ba}{\begin{array}}
	\newcommand{\ea}{\end{array}}
\newcommand{\bd}{\begin{displaymath}}
	\newcommand{\ed}{\end{displaymath}}
\newcommand{\besub}{\begin{subequations}}
	\newcommand{\eesub}{\end{subequations}}

\def\q2 {q^2}

\def\vvevof#1{\left\lgroup\!#1\!\right\rgroup}
\def\op{\mathpzc{o}}
\def\Op{\mathpzc{O}}

\def\Ff{\!{\textbf{\textit{f}}}}
\def\ft{f_{\tt theo}}

\def\fft{{\mathfrak{f}}_{\tt theo}}

\def\Nt{\Sigma_{\tt theo}}

\def\Gg{\pmb{$g$}}

\def\er{{\mathfrak N}} 
\def\lum{{\mathfrak{L}}_{\tt int} } 
\def\ert{{\mathfrak N}_{\tt theo}} 

\def\lum{{\mathfrak{L}}_{\tt int} } 
\def\Gg{\!{\textbf{\textit{g}}}}

\def\ft{f_{\tt theo}}

\def\fft{{\mathfrak{f}}_{\tt theo}}
\def\FFt{{\mathfrak{P}}}

\def\Nt{\Sigma_{\tt theo}}
\def\op{\mathpzc{o}}

\def\ert{{\mathfrak N}_{\tt theo}} 

\def\vvevof#1{\left\lgroup\!#1\!\right\rgroup}
\def\hc{\text{H.c.}}
\def\gz{g_{\tt z}}
\def\sw{s_{\tt w}}
\def\stw{s_{\tt 2w}}
\def\mz{m_{\tt z}}

\def\vp{\varphi}

\def\ct{{\tt C}}

\definecolor{lightblue}{rgb}{.5,.5,1}

\def\bt{\begin{table}}
	\def\et{\end{table}}

\usepackage{bbm}  
\usepackage{amsmath}                                                

\newcommand{\nc}{\newcommand}
\nc{\beq}{\begin{equation}}  \nc{\eeq}{\end{equation}}
\nc{\bea}{\begin{eqnarray}}  \nc{\eea}{\end{eqnarray}}
\nc{\baa}{\begin{array}}     \nc{\eaa}{\end{array}}
\nc{\bit}{\begin{itemize}}   \nc{\eit}{\end{itemize}}
\nc{\ben}{\begin{enumerate}} \nc{\een}{\end{enumerate}}
\nc{\bce}{\begin{center}}    \nc{\ece}{\end{center}}
\nc{\bpm}{\begin{pmatrix}}   \nc{\epm}{\end{pmatrix}}
\nc{\bvt}{\begin{verbatim}}  \nc{\evt}{\end{verbatim}}
\nc{\bal}{\begin{align}}
\def\mcr{\nonumber\\[6pt]}

\def\half{\frac12}	

\def\to{\rightarrow}

\def\boldoverdot{\,{\raise6pt\hbox{\bf.}\!\!\!\!\>}}
\def\re{{\bf Re}}
\def\im{{\bf Im}}

\def\then{{\quad\Rightarrow\quad}}

\def\ccal{{\cal C}}

\def\lcal{{\cal L}}

\def\ocal{{\cal O}}

\def\ucal{{\cal U}}

\def\ee{{\bf e}}

\def\zBB{{\mathbbm Z}}
\usepackage{bm}
\def\xibf{{\bm\xi}}		
\def\vev{vacuum expectation value}

\def\diag{\hbox{\diag}}

\def\gev{\hbox{GeV}}
\def\tev{\hbox{TeV}}

\def\vevof#1{\left\langle #1 \right\rangle}

\def\doubleundertext#1{
{\undertext{\vphantom{y}#1}}\par\nobreak\vskip-\the\baselineskip\vskip4pt%
\undertext{\hbox to 2in{}}}
\def\inbox#1{\vbox{\hrule\hbox{\vrule\kern5pt
     \vbox{\kern5pt#1\kern5pt}\kern5pt\vrule}\hrule}}
\def\sqr#1#2{{\vcenter{\hrule height.#2pt
      \hbox{\vrule width.#2pt height#1pt \kern#1pt
         \vrule width.#2pt}
      \hrule height.#2pt}}}

\def\today{\ifcase\month\or
  January\or February\or March\or April\or May\or June\or
  July\or August\or September\or October\or November\or December\fi
  \space\number\day, \number\year}
\def\pmb#1{\setbox0=\hbox{#1}%
  \kern-.025em\copy0\kern-\wd0
  \kern.05em\copy0\kern-\wd0
  \kern-.025em\raise.0433em\box0 }

\def\pmbb#1{\setbox0=\hbox{#1}%
  \kern-.02em\copy0\kern-\wd0
  \kern.04em\copy0\kern-\wd0
  \kern-.02em\raise.03464em\box0 }
\def\up#1{^{\left( #1 \right) }}
\def\inv#1{\frac1{#1}}
\def\sumprime_#1{\setbox0=\hbox{$\scriptstyle{#1}$}
  \setbox2=\hbox{$\displaystyle{\sum}$}
  \setbox4=\hbox{${}'\mathsurround=0pt$}
  \dimen0=.5\wd0 \advance\dimen0 by-.5\wd2
  \ifdim\dimen0>0pt
  \ifdim\dimen0>\wd4 \kern\wd4 \else\kern\dimen0\fi\fi
\mathop{{\sum}'}_{\kern-\wd4 #1}}

\begin{document}


\title{Optimal determination of New Physics couplings: A comparative study}

%
\author[a]{Subhaditya Bhattacharya,}
\author[a]{Sahabub Jahedi,}
\author[b]{Jose Wudka}

\affiliation[a]{Department of Physics, Indian Institute of Technology Guwahati, Assam 781039, India}
\affiliation[b]{ Department of Physics and Astrophysics, University of California, Riverside, California 92521, USA }

\emailAdd{subhab@iitg.ac.in}
\emailAdd{sahabub@iitg.ac.in}
\emailAdd{jose.wudka@ucr.edu}

\abstract
{
We study the determination of new physics (NP) parameters using the optimal observable technique (OOT) in situations where the standard model (SM) dominates over the NP effects, and when the NP dominates over the SM contribution, using the 2-Higgs doublet model as an illustrative example; for the case of SM domination we extend our results using an effective theory parameterization of NP effects. For the case of SM dominance we concentrate on $ t\bar t$ production in an $e^+e^-$ collider, while for the case of NP dominance we consider both $t\bar t$ production and pair production of charged scalars, also in an $e^+e^-$ collider. We discuss the effects of the efficiency of background reduction, luminosity and beam polarization, and provide a comparison of the optimal uncertainties with those obtained using a standard $ \chi^2$ analysis of (Monte Carlo generated) collider data.
} 


\maketitle
\flushbottom

\section{Introduction}
\label{sec:intro}
Although the Standard Model (SM) of particle physics is now complete after the discovery of Higgs Boson \cite{ATLAS:2012yve, CMS:2012qbp} at the Large Hadron Collider (LHC), 
it leaves several questions unanswered, keeping the quest for new physics (NP) alive. However, LHC hasn't been able to pin down on any such NP yet, where searches are complicated by the huge hadronic activity and QCD backgrounds for many signals of NP. On the other hand, $e^+e^-$ colliders are ideal for  probing such cases. Accordingly, there are many proposals for such machines: the ILC \cite{Behnke:2013xla,Baer:2013cma,ILC:2019gyn}, CLIC \cite{CLICdp:2018cto,Roloff:2018dqu,Aicheler:2018arh}, CEPC \cite{CEPC-SPPCStudyGroup:2015csa,CEPC-SPPCStudyGroup:2015esa} and FCC \cite{TLEPDesignStudyWorkingGroup:2013myl,FCC:2018evy}. Precision measurements in particular, can be performed neatly at $e^+e^-$ machines due to the complete knowledge of the centre-of-mass energy, availability of the beam polarization, and lack of hadronic activitie and parton distribution function uncertainties. The top quark, being the heaviest SM particle, provides one of the most effective windows to probe NP; current measurements are still consistent with important deviations from SM prediction.

The Optimal Observable Technique (OOT) \cite{Atwood:1991ka,Davier:1992nw,Diehl:1993br,Gunion:1996vv} is a powerful tool for determining the smallest statistical uncertainty to which any NP couplings can be determined in a given experimental environment; the degree to which any two models can be differentiated can also be predicted. The OOT has been applied to a variety of cases such as Higgs physics \cite{Hagiwara:2000tk,Dutta:2008bh,Hioki:2007jc}, top-quark physics at  $e^+\,e^-$ colliders \cite{Grzadkowski:1996pc,Grzadkowski:1997cj,Grzadkowski:1998bh,Grzadkowski:1999kx,Grzadkowski:2000nx}, 
$\gamma \gamma$ colliders \cite{Grzadkowski:2004iw,Grzadkowski:2005ye}, and $e \gamma$ colliders \cite{Cao:2006pu}. 
A few works have appeared using this approach in top-quark physics at the LHC \cite{Gunion:1998hm,Hioki:2012vn,Hioki:2014eca} and various aspects of flavour physics \cite{Bhattacharya:2015ida,Calcuttawala:2017usw,Calcuttawala:2018wgo}. Recently, the OOT has been used in studying heavy charged fermion \cite{Bhattacharya:2021ltd} 
production and and di-boson production \cite{Jahedi:2022duc,Jahedi:2023myu} at the $e^+\,e^-$ collider. We also note that there exists some other statistical 
techniques, which uses optimisation for NP extraction, such as multivariate analysis \cite{Holmstrom:1995bt,TMVA:2007ngy,Voss:2007jxm,Bhat:2010zz}, matrix method \cite{D0:2004rvt,D0:2004lvh,Fiedler:2010sg,Artoisenet:2010cn}, etc.

Optimal uncertainties of NP couplings depend on the relative contributions of the NP and SM for the process under observation. In this paper we will consider two complementary situations: in the first the NP generates a subdominant correction to the SM, while in the second the NP dominates. These two cases can be realized, for example, in a 2-Higgs doublet model (see, {\it e.g.} \cite{Gunion:1989we,Branco:2011iw}), where, in the first, the scale of the non-SM scalars is large compared to the electroweak scale and with the collider energy; while in the second, that same scale is sufficiently low for non-SM particles to be directly produced. The specific illustrations we consider are, for the first case, $ t\bar t$ production, and for the second, production of charged-scalar pairs.

The study of processes where the NP is not observed directly ({\it cf.} the first case above) is most conveniently done using the so-called SM effective field theory (SMEFT) approach. In our discussion below we will use this language when applicable, with the connection to a specific two-doublet model briefly spelled-out; the results thus obtained, even if motivated by a specific extension of the SM, will be more general. Within the SMEFT the Lagrangian takes the form
\beq
\mathcal{L}_{\tt eff}=\mathcal{L}_{\tt SM} + \frac{1}{\Lambda^{d-4}}\sum_{i}c_i O^d_{i}\,,
\label{eq:eft}
\eeq
where $c_i$'s are the dimensionless ``Wilson coefficients''  and the $O^d_i$ are gauge-invariant, dimension-$d$ effective operators constructed using SM fields and their (covariant) derivatives; $\Lambda$ denotes the scale of NP which, by consistency of this approach, must lie above the available energy. The lists of operators of dimension five \cite{Weinberg:1979sa}, six  \cite{Buchmuller:1985jz,Grzadkowski:2010es}, seven  \cite{Lehman:2014jma,Bhattacharya:2015vja}, eight  \cite{Li:2020gnx,Murphy:2020rsh,Murphy:2020cly}  and  nine  \cite{Li:2020xlh,Liao:2020jmn,Li:2020tsi} are already available. SMEFT approach has been used exhaustively to examine different properties of top-quark physics at the LHC \cite{Aguilar-Saavedra:2010uur,Durieux:2014xla,Buckley:2015nca,Buckley:2015lku,Degrande:2018fog,Chala:2018agk,DHondt:2018cww,Aguilar-Saavedra:2018ksv,Hartland:2019bjb,Neumann:2019kvk,Maltoni:2019aot,Brivio:2019ius}, and also at the lepton colliders \cite{Kane:1991bg,Grzadkowski:1997cj,Brzezinski:1997av,Boos:1999ca,Rontsch:2015una,Janot:2015yza,Englert:2017dev,Durieux:2018tev,Durieux:2018ggn,Jafari:2019seq,Bissmann:2020mfi}.
%

The goals of the paper are, first, to provide an estimate of how well the coefficients can be measured for the two above cases; 
and second, to compare the optimal uncertainties with standard collider estimates in order to provide a 
measure of how far the experimental analysis can be improved (for this comparison we use Monte-Carlo generated `data'). The collider analysis we use 
is basic, aimed at comparing and contrasting the results obtained by both methods; it can certainly be improved, a point on which we comment in \cref{sec:summary}.

Our paper is organized as follows: in the next section we summarize the results of the OOT; we then study the case of $t\bar{t}$ production (\cref{sec:ttb}) as an example of SM dominance over the NP, and that of charged scalars where the NP dominates over the SM (\cref{sec:model}); a summary is presented in \cref{sec:summary}.

\section{Optimal uncertainties}
\label{sec:OOTSMdom}
In this section we provide a brief summary of the statistical uncertainties of the NP physics in the OOT approach. The results here are a straightforward extension of the OOT expression obtained in \cite{Diehl:1993br,Gunion:1996vv,Bhattacharya:2021ltd}; the derivation is summarized in the appendix.

The theoretical cross-section for a process  involving both SM and NP contributions takes the general form
\beq
\ocal(\phi) = \frac{d\sigma_{\tt theo}}{d\phi}=\frac{d\sigma_{\tt SM}}{d\phi}+\sum_{i}g_i f_i(\phi)\,,
\label{eq:st}
\eeq   
where $g_i$ are model-dependent coefficients (that, in general, are non-linear functions of the NP couplings),  
the $f_i$ are  functions of the phase-space variables $\phi $; for example, for a  $2\to2$ scattering process such $\phi$ can be chosen 
as the scattering angle in the center-of-mass frame. The choice of the $ f_i $'s are not unique although necessarily linearly independent, but all observable results are unambiguous. 

We now consider a collider experiment with integrated luminosity $\lum$ and where the events of interest follow a Poisson distribution. In this case the optimal covariance matrix for the coefficients $ g_i $ is given by ,
\beq
V=\inv\lum M^{-1}\,, \qquad M_{ij}=\int \frac{f_i(\phi)f_j(\phi)}{\ocal(\phi)}d\phi\,.
\label{eq:V.M}
\eeq  
If the SM contribution dominates the expression \cref{eq:st}, one can replace $ \ocal \to \ocal_{\tt SM} = d\sigma_{\tt SM}/d\phi $ and the above expression reduces to the well-known result \cite{Diehl:1993br}.

In a model where the coefficients take some specific values $ g_i\up0 $, the corresponding statistical uncertainty for is determined by the $\chi^2$ function given by
\beq
\chi^2=\epsilon \sum_{i}\delta g_i \delta g_j (V_0^{-1})_{ij}, \quad \delta g_i=g_i-g_i\up0, \quad V_0=V|_{g=g\up0};
\label{eq:chi2}
\eeq  
where $ \epsilon $ is the efficiency for the process being considered; this factor takes into account the branching ratios of the produced particles to the signal states, as well as the effect of 
kinematic cuts used to segregate the signal from the SM background contamination. Specifically, the region $ \chi^2\le n$ determines the region in parameter space around $ g_i\up0 $ where the optimal statistical uncertainty of the $ g_i $ is below $\sqrt{n} \sigma $. For cases where the NP contribution is small, i.e. $ \ocal \simeq\ocal_{\tt SM} $,  $V_0 $ will be independent of $ g_1\up0$, and so will the $ \chi^2\le n$ region; however, in many cases of interest the SM dominates but the NP contribution to $ \ocal $ is not negligible, and the dependence of $V_0$ on $ g_1\up0$ 
may be important. It is worth mntioning that there are several studies applying the OOT to cases where NP effect is  small, so that NP parameters can be neglected in the covariance matrix ($V_0$) 
\cite{Diehl:1993br,Grzadkowski:1997cj,Grzadkowski:1998bh,Grzadkowski:1999kx,Grzadkowski:2000nx,Cao:2006pu,Gunion:1998hm,Hioki:2007jc}.
\section{Example of SM dominance: $t\, \bar t$ production at  $e^+e^-$ colliders}
\label{sec:ttb}
As an application of the OOT we consider $t\bar t $ production at an $ e^+ e^-$ future collider like the ILC assuming any NP contributions are subdominant. These conditions are realized, for example, in the so-called ``flipped'' two-Higgs doublet model \cite{Barger:1989fj,Grossman:1994jb,Akeroyd:1994ga,Akeroyd:1996he,Akeroyd:1998ui,Aoki:2009ha} (for a review see \cite{Wang:2022yhm}) where one scalar doublet couples to down-type quarks, while the second doublet couples to leptons and up-type quarks. If this second doublet is assumed to be heavy~\footnote{More specifically, we assume the second doublet has a small expectation value, large mass, and small mixing with the other doublet.} it will generate an effective interaction of the form $(\bar l e)(\bar q u) $, where $l,\,q$ are the lepton and quark left-handed doublets, and $e,\,u$ the right-handed lepton and up-quark singlets, respectively. As noted in the Introduction, this situation can be studied in an effective theory approach, with the advantage that the results are not tied to a specific model. Thus, we consider first the possible low-energy manifestation of such UV complete model ({\it i.e.} below the scale of the heavy scalar) using an effective theory parameterization; within this context we will use the OOT to derive the minimum statistical uncertainty to which the Wilson couplings in the effective theory can be measured.

\begin{figure}[htb!]
	$$
	\includegraphics[scale=0.33]{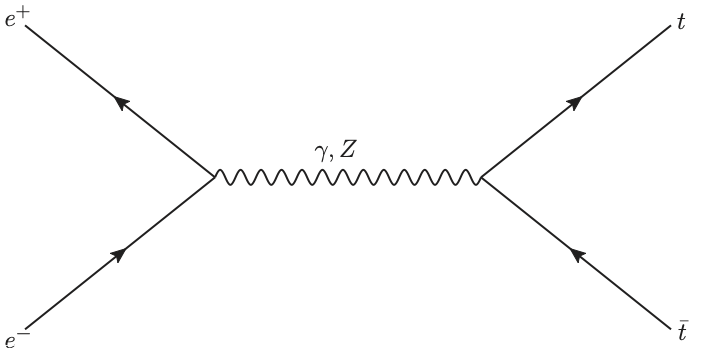} \quad
	\includegraphics[scale=0.38]{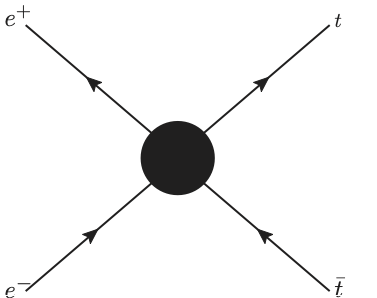}
	$$
	\caption{Top-quark pair production at $e^+\,e^-$ collider. Left: SM contribution; Right: SMEFT contribution.}
	\label{fig:ttdiag}
\end{figure}

In the SM, pair-production of top-quark at the  $e^+\,e^-$ colliders is generated at tree-level by  $\gamma$ and $Z$ mediation, as shown in the left side of 
\cref{fig:ttdiag}. The SMEFT contributions can be separated into two classes: those that modify the $eeZ$ and $ttZ$  vertices, and those that generate 4 fermion $eett$ vertices (right side of \cref{fig:ttdiag}). 

The relevant effective operators (in the so-called Warsaw-basis parameterization \cite{Grzadkowski:2010es}) are:
\beq
\begin{array}{c|lll}
\text{group} & \multicolumn{2}{c}{\text{Operator}} & \text{vertex} \cr
\hline\hline
\multirow{3}{*}{I}
& Q_{\vp l}\up1 &= (i \vp^\dagger D_\mu\vp + \hc )(\overline l_p \gamma^\mu l_r) & \phantom{-}\half {\tt v}^2 \gz \bar e \slashed Z P_L e\cr
& Q_{\vp l}\up3 &= (i \vp^\dagger \tau^I D_\mu\vp + \hc)(\overline l_p \tau^I \gamma^\mu l_r) & - \half {\tt v}^2 \gz \bar e \slashed Z P_L e\cr
& Q_{\vp e}     &= (i \vp^\dagger D_\mu\vp + \hc )(\overline e_p \gamma^\mu e_r) & \phantom{-}\half {\tt v}^2 \gz \bar e \slashed Z P_R e\cr
\hline
\multirow{3}{*}{II}
& Q_{\vp q}\up1 &= (i \vp^\dagger D_\mu\vp + \hc )(\overline q_p \gamma^\mu q_r) & \phantom{-}\half {\tt v}^2 \gz  \bar t \slashed Z P_L t\cr
& Q_{\vp q}\up3 &= (i \vp^\dagger \tau^I D_\mu\vp + \hc)(\overline q_p \tau^I \gamma^\mu q_r) & -  \half {\tt v}^2 \gz \bar t \slashed Z P_L t\cr
& Q_{\vp u}     &= (i \vp^\dagger D_\mu\vp + \hc )(\overline u_p \gamma^\mu u_r) & \phantom{-}\half {\tt v}^2 \gz \bar t \slashed Z P_R t\cr
\hline
\multirow{3}{*}{III}
& Q_{lq}\up1 &= (\overline l_p \gamma_\mu l_r)(\overline q_s \gamma^\mu q_u) &  \phantom{-}(\overline e \gamma_\mu P_L e )(\overline t \gamma^\mu P_L t) \cr
& Q_{lq}\up3 &= (\overline l_p \gamma_\mu \tau^I l_r)(\overline q_s \gamma^\mu \tau^I q_u) & -(\overline e \gamma_\mu P_L e )(\overline t \gamma^\mu P_L t) \cr
& Q_{lu} &= (\overline l_p \gamma_\mu l_r)(\overline u_s \gamma^\mu u_u)  & \phantom{-}(\overline e \gamma_\mu P_L e )(\overline t \gamma^\mu P_R t) \cr
& Q_{qe} &= (\overline q_p \gamma_\mu q_r)(\overline e_s \gamma^\mu e_u)  &\phantom{-}(\overline e \gamma_\mu P_R e )(\overline t \gamma^\mu P_L t) \cr
& Q_{eu} &= (\overline e_p \gamma_\mu e_r)(\overline u_s \gamma^\mu u_u)  & \phantom{-}(\overline e \gamma_\mu P_R e )(\overline t \gamma^\mu P_R t) \cr
\hline
\multirow{3}{*}{IV}
& Q_{lequ}\up1  &= (\overline l_p^j e_r) \epsilon_{jk} (\overline q_s^k u_u) & - (\overline e P_R e)  (\overline t P_R t)\cr
& Q_{lequ}\up3 &= (\overline l_p^j \sigma_{\mu\nu} e_r) \epsilon_{jk} (\overline q_s^k \sigma^{\mu\nu} u_u) &- (\overline e \sigma_{\mu\nu} P_R e)  (\overline t \sigma^{\mu\nu} P_R t) 
\end{array}
\label{eq:op.table}
\eeq
where $\gz = \sqrt{g^2 + g'{}^2} $ and $p,\,r,\,s,\,u$ are family indices.

In general, the operators in \cref{eq:op.table} are generated by different types of NP so that they need not have a common scale $ \Lambda$. The experimental constraints on the $Zee$ 
coupling allow an $ \sim 0.1\%$ deviation from the the SM prediction  that corresponds to $ \Lambda \sim 7~\tev$ (for a unit Wilson coefficient) for group I operators \cite{ParticleDataGroup:2022pth}. 
The constraints on the $ttZ$ coupling (group II) are significantly weaker $ \Lambda \sim 1 \, \tev$ \cite{CMS:2021aly}; the constraints on the 4-fermion operators (groups III and IV) will be of the same order. 

Operators in group IV are unique in that, for the process at hand, they do not interfere with the SM, or with the operators in the other groups, and this will allow a future $e^+e^-$ collider 
to differentiate their contributions from those generated by other types of new physics. These operators offer a convenient method to investigate a class NP effects by suppressing SM contributions in $t\bar{t}$ production for a judicious choice of beam polarization~\footnote{As noted above, operators in group IV are generated by interesting types of new physics; for the current reaction their contributions do not interfere with the SM, which allows a clean evaluation of the effects of polarization -- they produce $O(1/\Lambda^4)$ corrections. Operators in the  I, II and III are generated by {\em different} types of new physics ({\it e.g.} additional $Z$ gauge bosons) and do interfere with the SM, giving rise to $O(1/\Lambda^2)$ contributions. There are also contributions from dimension 8 operators which can generate $O(1/\Lambda^4)$ corrections as well, but these occur only in EFT-SM interference terms and do not correspond to the types of new physics being considered here.}. It is one of the goals of this paper to illustrate this feature using the OOT as a tool; to simplify the 
discussion we will then consider the effects of these operators, ignoring those that may be generated by those in groups I, II and III. 
Our effective Lagrangian then takes the form
\beq
\lcal_{\tt eff} =  \frac{c_1}{\Lambda^2}Q_{lequ}\up1 +  \frac{c_2}{\Lambda^2}Q_{lequ}\up3  + \text{H.c.} \,;
\label{eq:leff}
\eeq
where $c_{1,2}$ are  dimensionless (Wilson) coefficients, and $\Lambda$ is the scale of new physics\footnote{Within the context of the flipped two-Higgs doublet model, $ \Lambda$ denotes the scale of the heavy scalar and $c_1$ the product of its Yukawa couplings to the leptons and up-type quarks.}. 

The helicity  amplitude~\footnote{For the helicity amplitude calculation see \cite{Vega:1995cc}.} $M(\lambda_{e^-}, \lambda_{e^+};\, \lambda_t, \lambda_{\bar t} ) $ for this process is given by  
\bal
M(\lambda,\,- \lambda; \lambda',\, -\lambda') &= e_0^2 \left( \frac{\lambda  \lambda' + \cos\theta}{\beta_z^2} \right) \left[ \frac23 \beta_z^2-  \left( \frac{4 \sw^2-1  +   \lambda }{4\stw^2} \right) \left( 1- \frac83 \sw^2  - \beta_t \lambda'\right) \right], \mcr
M(\lambda,\, \lambda ; \lambda', \, -\lambda') &=\frac{4 c_2 m_t \sqrt{s}}{\Lambda^2}  \, \lambda \sin \theta \, , \mcr
M(\lambda,\,- \lambda; \lambda', \, \lambda') &= \frac{2 e_0^2  m_t \lambda' \sin\theta}{\sqrt{s}} \left[\frac23 -\left( \frac{4 \sw ^2-1  +   \lambda}{4\stw^2 \,  \beta_z^2 } \right) \left(1- \frac83 \sw ^2 \right)  \right], \mcr
M(\lambda, \, \lambda; \lambda',\, \lambda') &=\frac{ s}{2\Lambda^2}(\lambda \lambda' \beta_t-1) \left( c_1 + 4c_2\lambda \lambda' \, \cos \theta\right);
\label{eq:amp}
\end{align}
where $\lambda_i=\pm1$ indicates the helicity of particle $i$, $e_0$  the  $U(1)_{\tt em}$ coupling constant, $\sqrt{s}$  the CM energy, $\sw = \sin(\theta_{\tt w} ),\, \stw= \sin(2\theta_{\tt w} )$ ($ \theta_{\tt w} $ denotes the weak mixing angle), $m_t$  the top-quark mass, $m_z$  the $Z$-boson mass and
\beq
\beta_z = \sqrt{1 - \frac{m_z^2}{s}} ; \qquad \beta_t = \sqrt{1 - \frac{4m_t^2}{s}}.
\eeq
In the above expressions, electron mass has taken to be zero; we also assumed that $ c_{1,2}$ are real, if this is not the case then one must replace $ c_a \to \re c_a +i \lambda_e \im c_a $. We note here that, for zero electron mass, the SM  contributes only to the opposite helicity amplitudes $ \lambda_{e^-} = - \lambda_{e^+}$, whereas scalar and tensor mediated effective operators  contribute only to same helicity amplitudes, so there is no EFT-SM interference; this is not the case for  possible contributions from operators in groups I-III in \cref{eq:op.table}.


The corresponding differential cross-section when the $ e^\pm$ beams have partial polarizations $P_{e^\pm} $ (with $ -1\le P_{e^\pm}\le1$) is given by
\beq
\frac{d\sigma(P_{e^+},\,P_{e^-})}{d\Omega} = \sum_{\lambda_e^+= \pm1 } \sum_{ \lambda_e^-= \pm1} \frac{(1+ \lambda_{e^-} P_{e^-})(1+ \lambda_{e^+} P_{e^+})}4 \left( \frac{d\sigma}{d\Omega}\right)_{\lambda_{e^-}, \lambda_{e^+} } \,;
\label{eq:difcs}
\eeq
where $\left(  d\sigma/d\Omega \right)_{\lambda_{e^-}, \lambda_{e^+} }$ is the cross section obtained from \cref{eq:amp} by summing over $ \lambda_t,\,\lambda_{\bar t} $.
 For $ s\gg m_t^2 $ and $ \sw^2 \simeq0.25 $ it is easy to see that the SM contribution to the total cross section has the form
\beq
\sigma(P_{e^+},\,P_{e^-})_{\tt SM} \propto 1-\mathfrak{a}^2 +\left( \mathfrak{a}-P_{e^-} \right)\left( \mathfrak{a}+P_{e^+} \right) \,, \quad \mathfrak{a} \sim 0.0652\,;
\label{eq:sm.pol}
\eeq
which vanishes when $ P_{e^+}=P_{e^-}=\pm1 $ and has a maximum when $ P_{e^\pm}=\pm1 $; the SM polarized cross section lies above the unpolarized one when $P_{e^-} \!\lesssim \mathfrak{a}$ and $P_{e^+} \!\gtrsim - \mathfrak{a}$, or $P_{e^-} \! \gtrsim \mathfrak{a} $ and $P_{e^+} \!\lesssim -\mathfrak{a}$.

Explicitly, the SM contribution is given by
\bal
\frac{d\sigma_{\tt SM}}{d\Omega}&=\frac{\alpha^2_0(1-P_{e^-}P_{e^+})}{3s}\bigg\{  1+\ct( \xi_1-P_{\tt eff} \xi_2 ) +\inv4(\xi_1^2-2P_{\tt eff} \xi_1 \xi_2+\xi_2^2) \bigg(\ct ^2+\frac{\beta_t^2}{2-\beta_t^2}  \bigg)   \mcr
&- \left[ \xi_2 (1+\ct \xi_1) - \inv4 P_{\tt eff} \left( 4\xi_1 - (2\xi_1^2 - \xi_2^2)\ct   \right) \right]\beta_t \cos \theta + \mcr
& +   \left[ 1+\ct \left( \xi_1 -P_{\tt eff}\xi_2\right) +\frac{2\ct+1}4 (\xi_1^2 - 2 P_{\tt eff} \xi_1 \xi_2  +\xi_2^2)  + \frac{\ct(2-\ct)}2 P_{\tt eff}\xi_1 \xi_2  \right] \beta_t^2 \cos^2 \theta \biggr\}\,;
\end{align}
where 
\beq
P_{\tt eff} = \frac{P_{e^-}-P_{e^+}}{1-P_{e^-}P_{e^+}}\, , \qquad \xi_1=\inv{2s^2_{\tt 2w}\beta_z^2}, \qquad \xi_2=\frac{4 \sw^2-1}{2s^2_{\tt 2w}\beta_z^2}, \qquad \ct=3-12 s_{\tt w}^2\,;
\eeq
and, using the notation of \cref{eq:st}, effective operator contribution is determined by the following choice of $f_i$ and $g_i$:
\bal
\{f_1,\,f_2,\,f_3\} &= \frac{3 \beta_t s}{256 \pi^2 \Lambda^4} \left\{  (1 + \beta_t^2),\, 16 \beta_t   \cos \theta,\, 32 \beta_t^2  \cos^2\theta \right\}\,, \mcr
\{g_1,\,g_2,\,g_3\} &= (1+P_{e^-}P_{e^+}) \left\{  \bigg(c_1^2+16 \frac{1-\beta_t^2}{1+\beta_t^2} c_2^2\bigg),\, -c_1 c_2,\,c_2^2 \right\}\,.
\label{fi}
\end{align}

\begin{figure}[H]
	\centering
	$$
	\includegraphics[width=6cm,height=4.5cm]{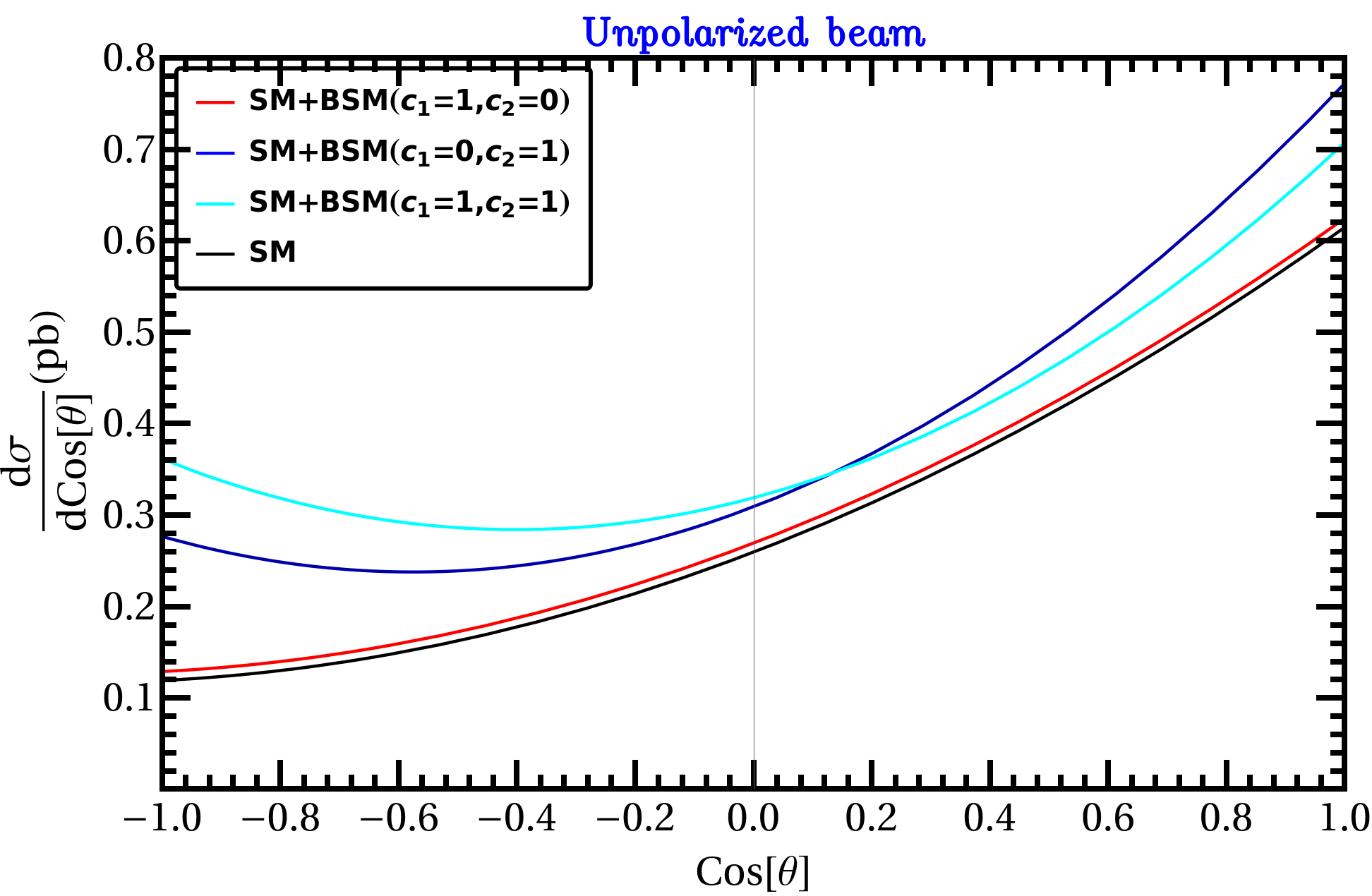}
	\includegraphics[width=6cm,height=4.5cm]{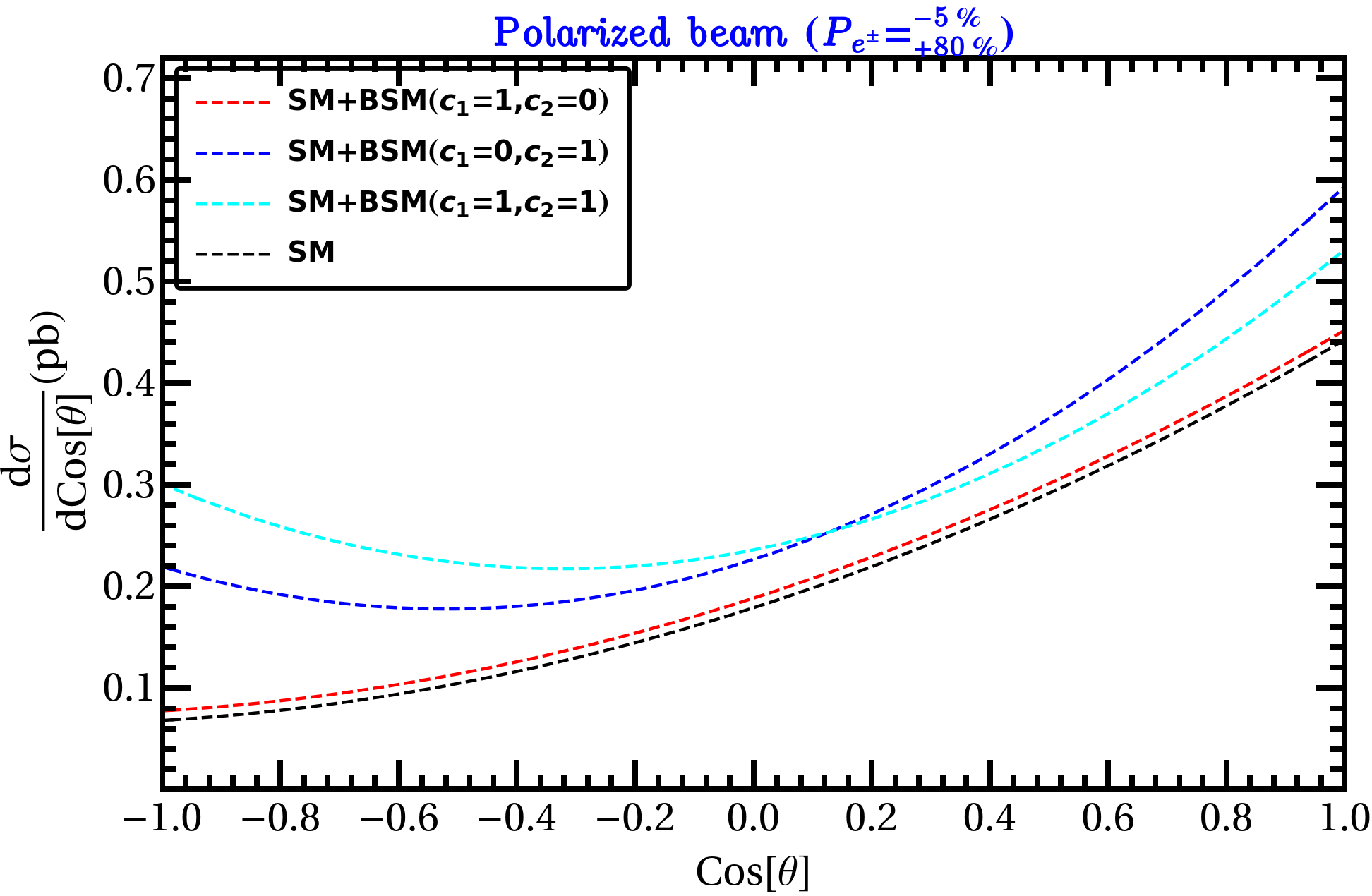}
	$$
	$$
	\includegraphics[width=6cm,height=4.15cm]{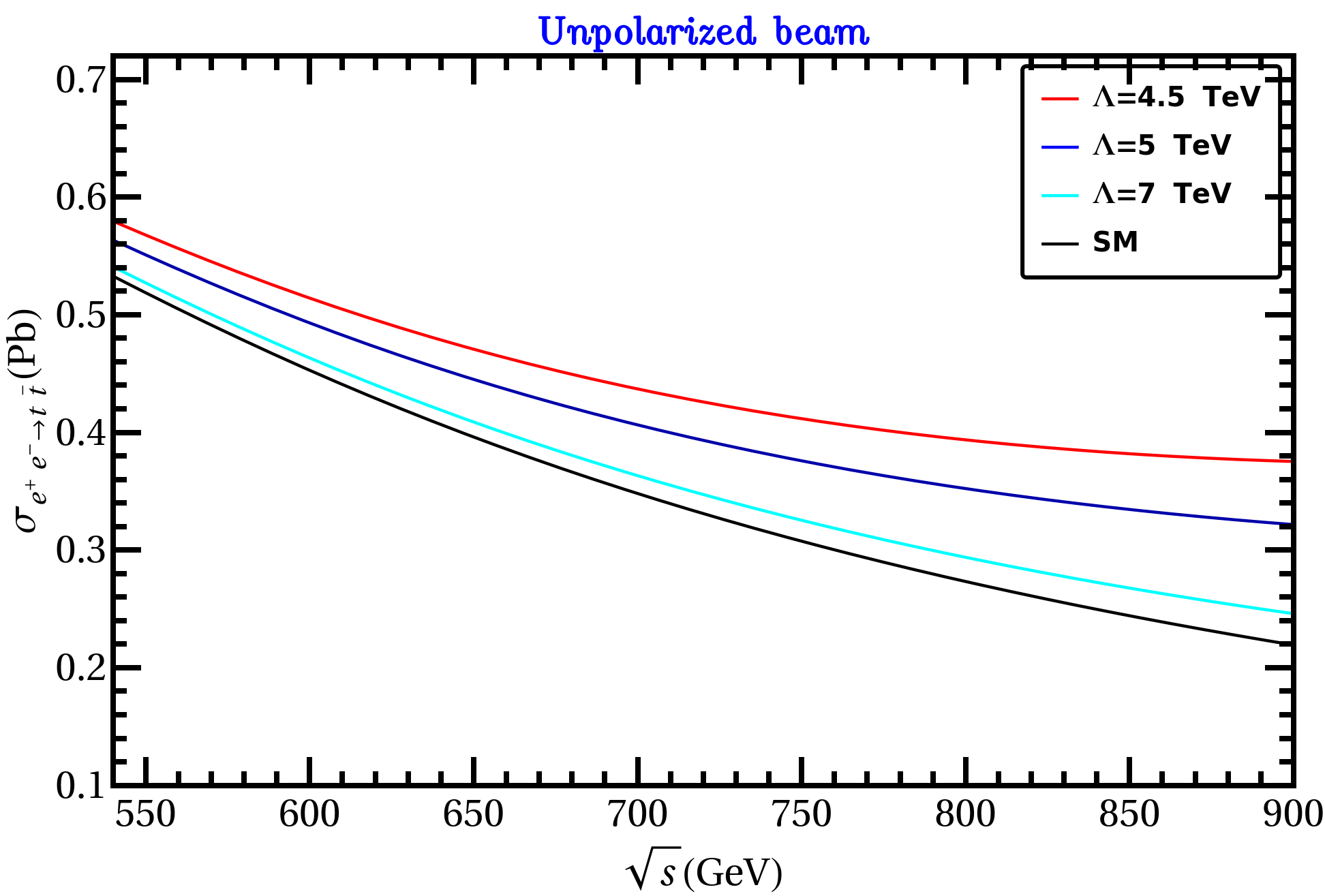}
	\includegraphics[width=6cm,height=4.15cm]{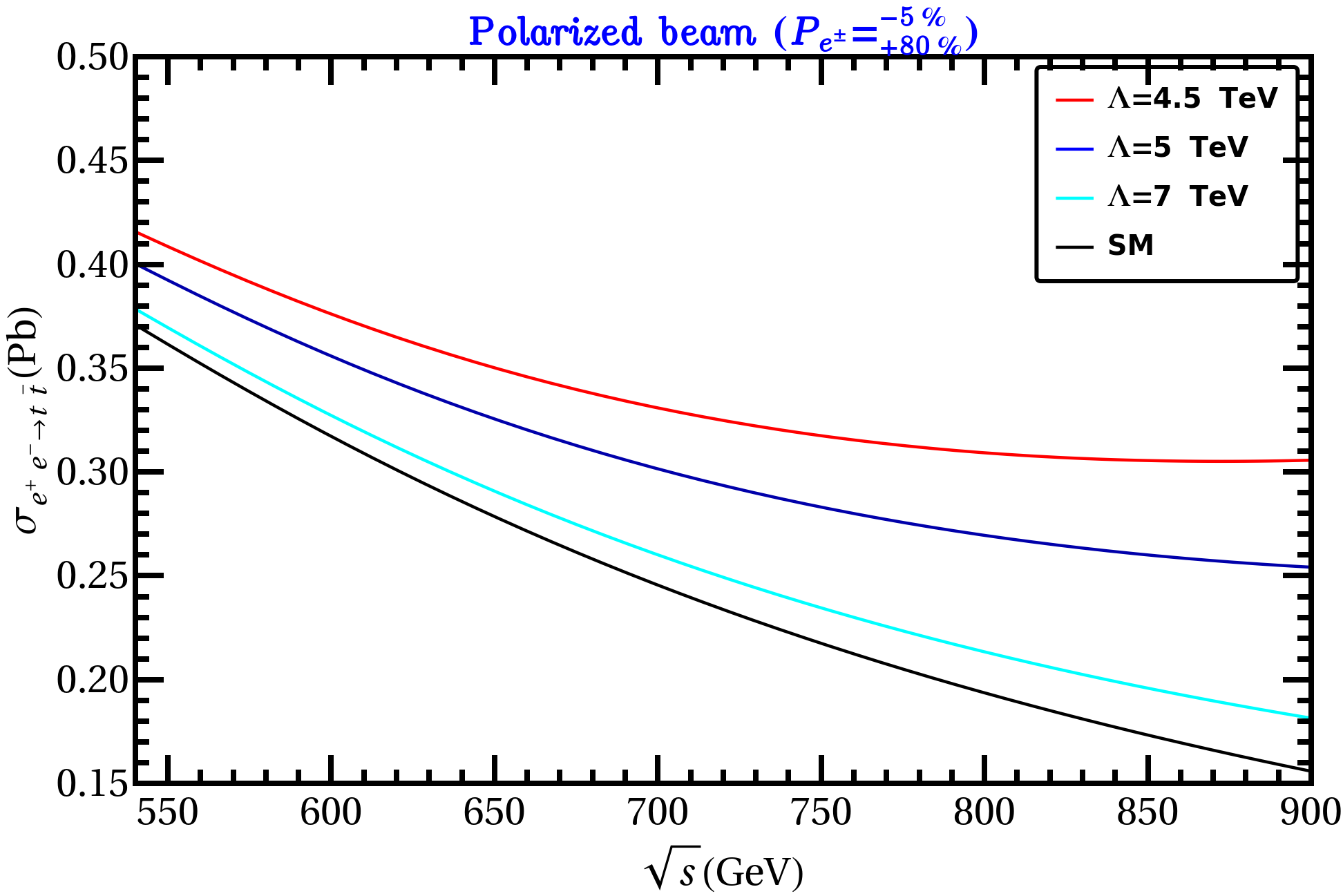}
	$$
	$$
	\includegraphics[width=6cm,height=4.15cm]{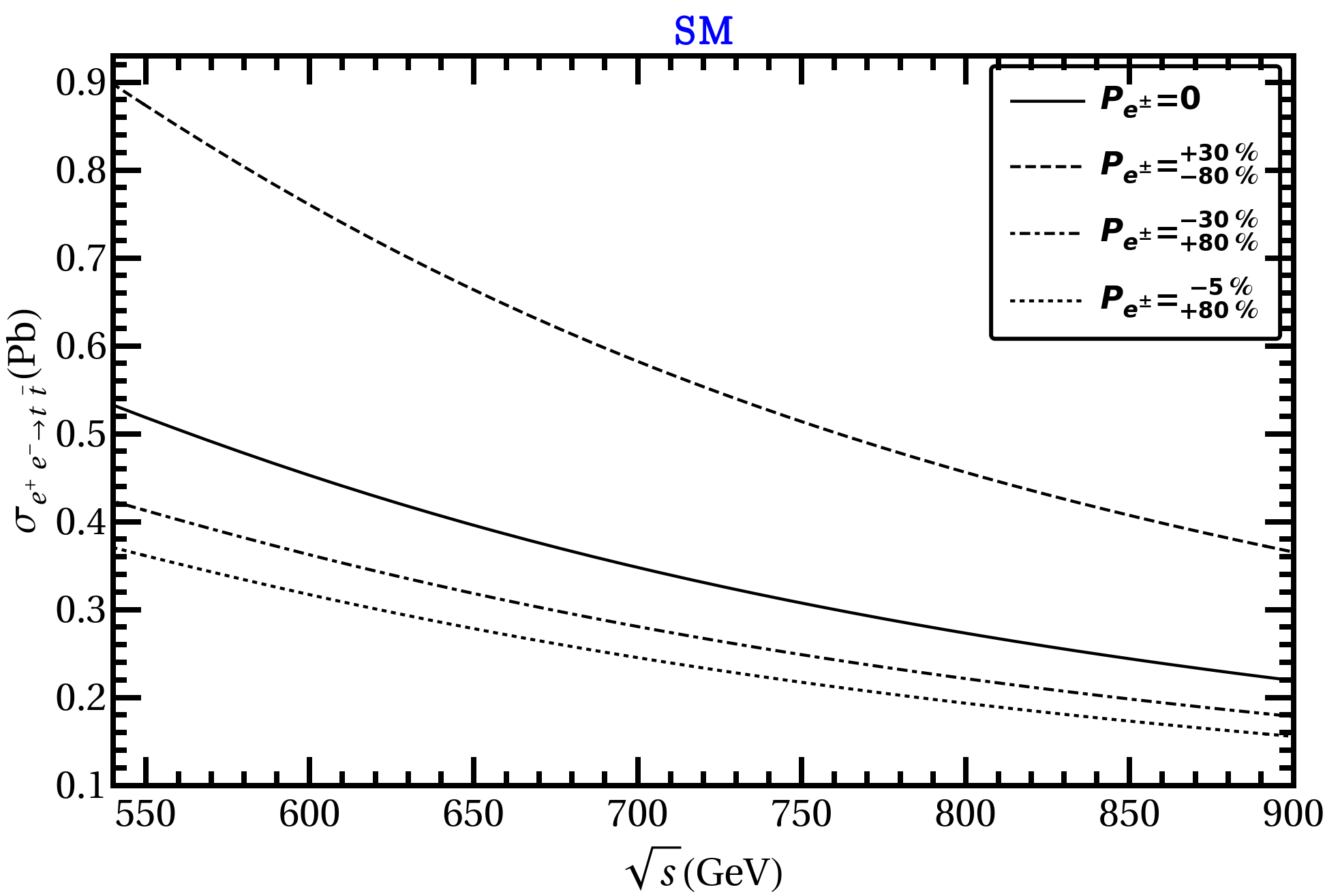}
	\includegraphics[width=6cm,height=4.15cm]{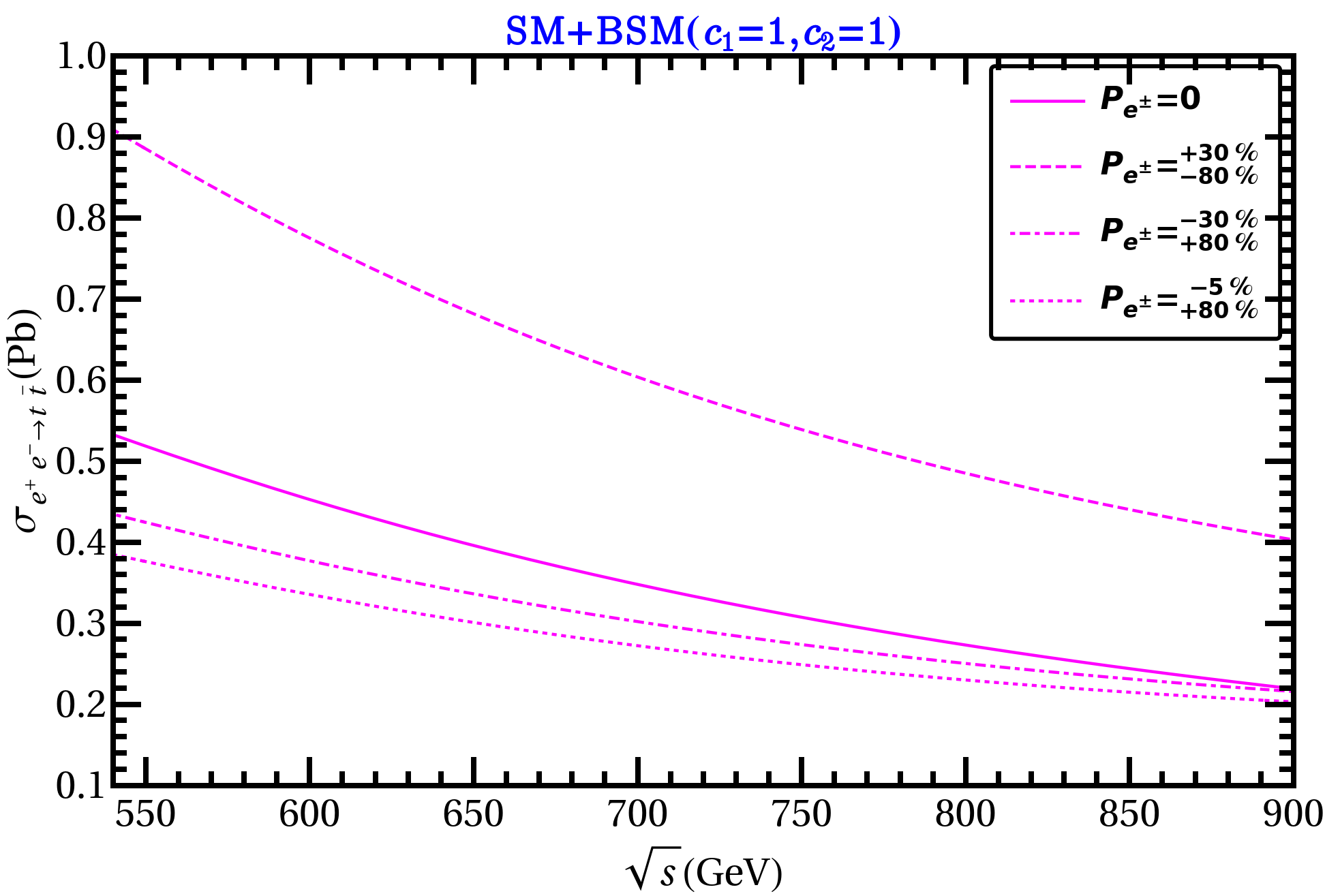}
	$$
	$$
	\includegraphics[width=6cm,height=4.15cm]{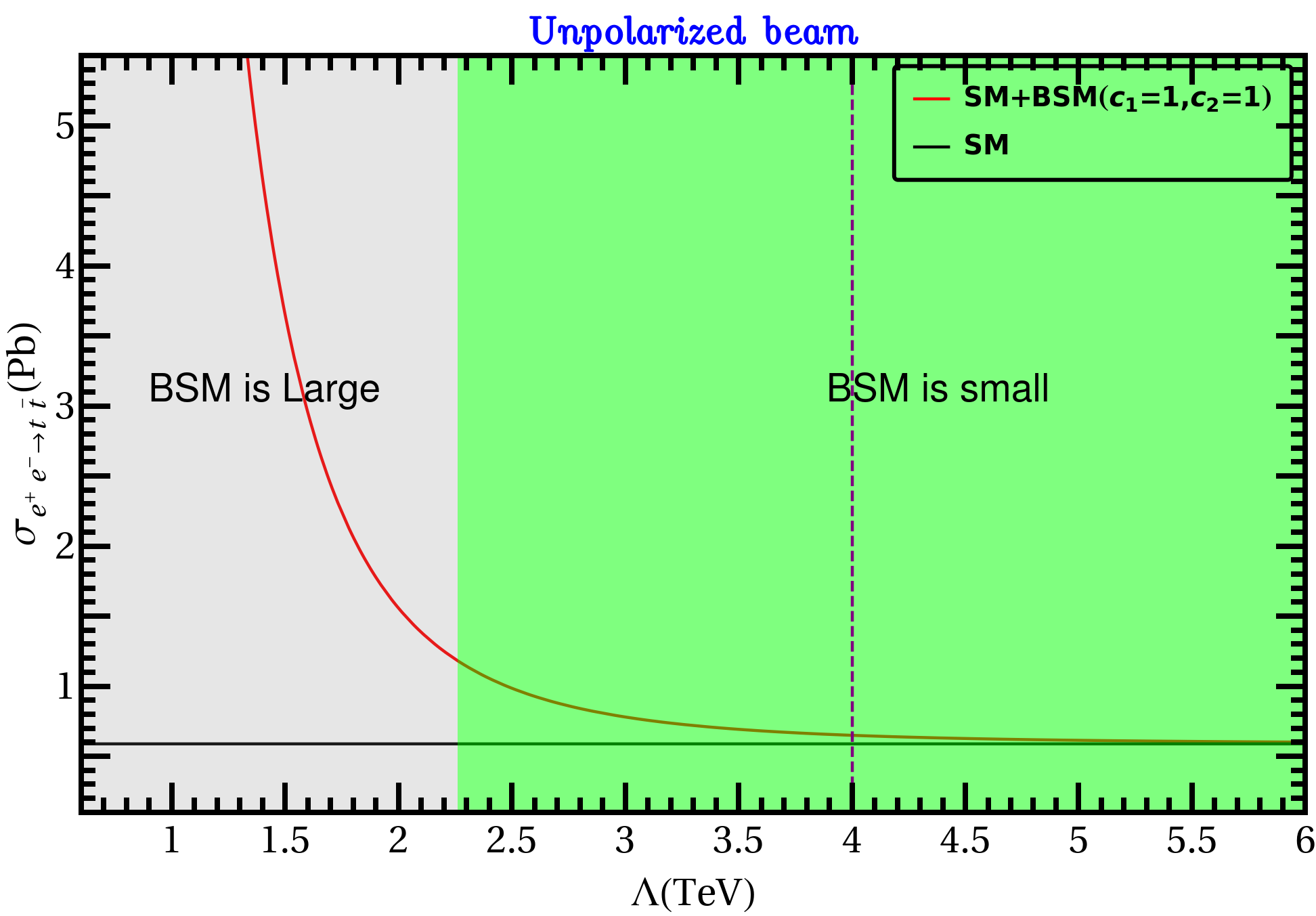}
	\includegraphics[width=6cm,height=4.15cm]{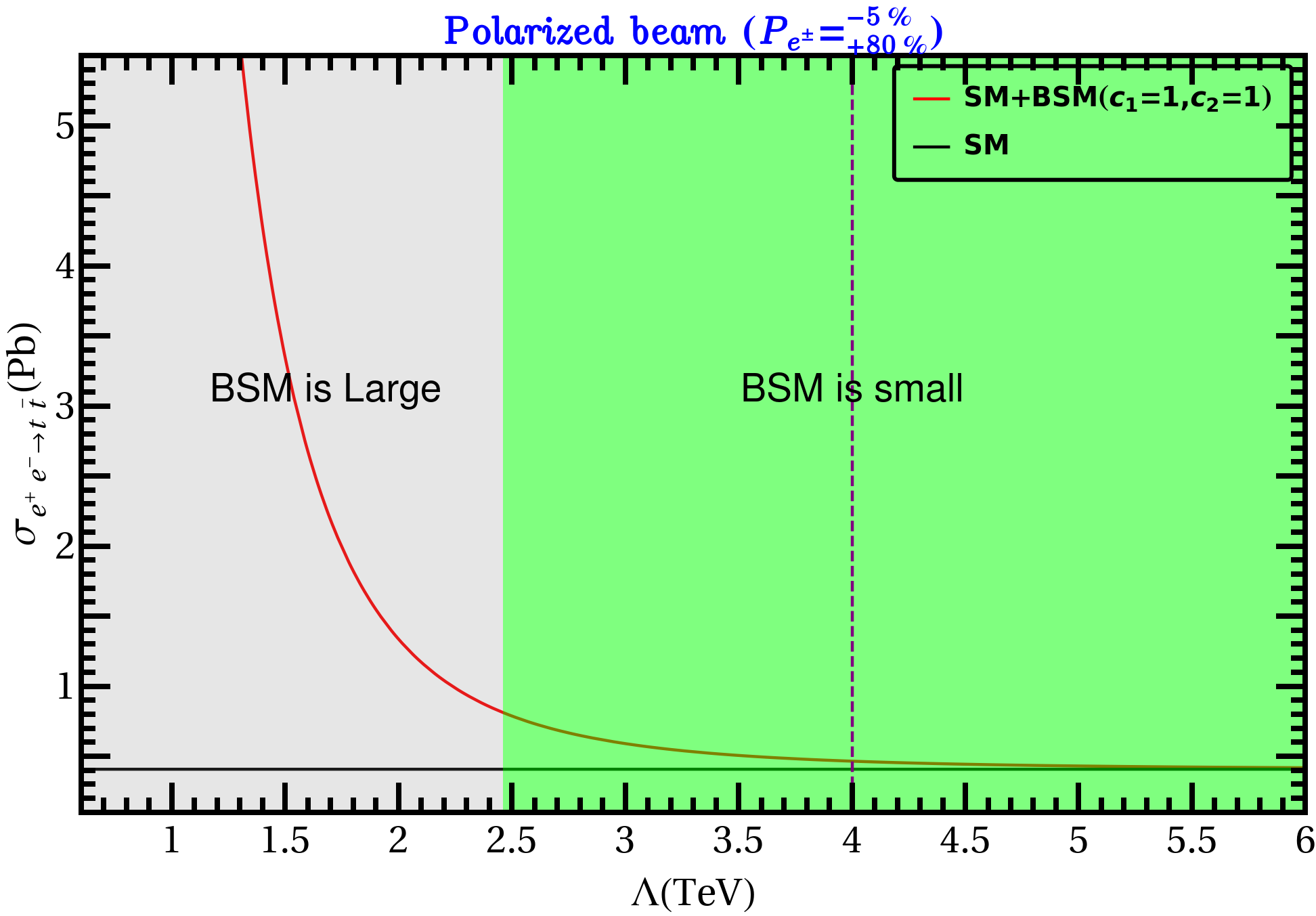}
	$$
	\caption{\small Plots of the $ e^+ e^-\to\bar t t$ cross-section. Top row: angular dependence for CM energy $\sqrt{s} = 500$ GeV, and different values of $\Lambda$ for 
		unpolarized (left) and polarized with $ P_{e^\pm}=_{+80\%}^{-5\%}$ (right) beams. Second row: total cross section as a function of $ \sqrt{s}$ for $c_1= c_2=1$ and several 
		values of $\Lambda$, for unpolarized (left) and unpolarized (right) beams. Third row: dependence on the beam polarization for the SM (left) and the SM + EFT with 
		$ c_1=c_2=1$ and $\Lambda =4$ TeV (right). Bottom row: comparison of the SM total cross section (black horizontal line) with the SM + EFT with $ c_1=c_2=1$ as a function of 
		$\Lambda$ for unpolarized (left) and polarized (right) beams; the region labeled ``BSM small'' corresponds to $ \sigma_{\tt SM + EFT}  > 2 \sigma_{\tt SM}  $.}
	\label{fig:ttprod}
\end{figure}

We plot in \cref{fig:ttprod} this differential cross section and the associated cross section for various choices of the model parameters and beam polarizations. As expected, the SM cross section decreases as the CM energy $\sqrt{s} $ increases, and the polarized SM cross section for $P_{e^\pm}=_{+80\%}^{-5\%}$ is smaller than the unpolarized one, in accordance with \cref{eq:sm.pol}. We therefore will use this choice of beam polarizations for the optimal estimation of the Wilson coefficients, and compare it to that with unpolarized beams.

It also evident that as $\Lambda$ increases the total cross-section approaches the SM value. We identify a value $ \Lambda = \Lambda_{\tt boundary}$ corresponding to $ \sigma_{\tt tot} = 2 \sigma_{\tt SM} $, so that the process is SM-dominated when $ \Lambda > \Lambda_{\tt boundary}$; using \cref{fig:ttprod} we find $ \Lambda_{\tt boundary} \sim2\, \tev $ and, being interested in situations where the EFT represents a correction to the SM, we will consider NP scales above this value. This also shows why $t\bar{t}$ production is such an effective process to probe NP above TeV scale. For our explicit calculations we use  the following collider parameters: 
\beq
\Lambda = 4 \, \tev \mcr \,; \qquad \sqrt{s} = 500 \, \gev\,; \qquad \lum =1000 \, \text{fb}^{-1}\,.
\label{eq:params}
\eeq

\subsection{Collider analysis}
\label{sec:eft.col}

In this section, we provide an event level signal and background analysis of top quark pair production at $e^-e^+$ machine. 
Our main goal here is to determine the efficiency $ \epsilon$ with which we approximate the optimal observable.

\begin{figure}[htb!]
	$$
	\includegraphics[width=7cm,height=5cm]{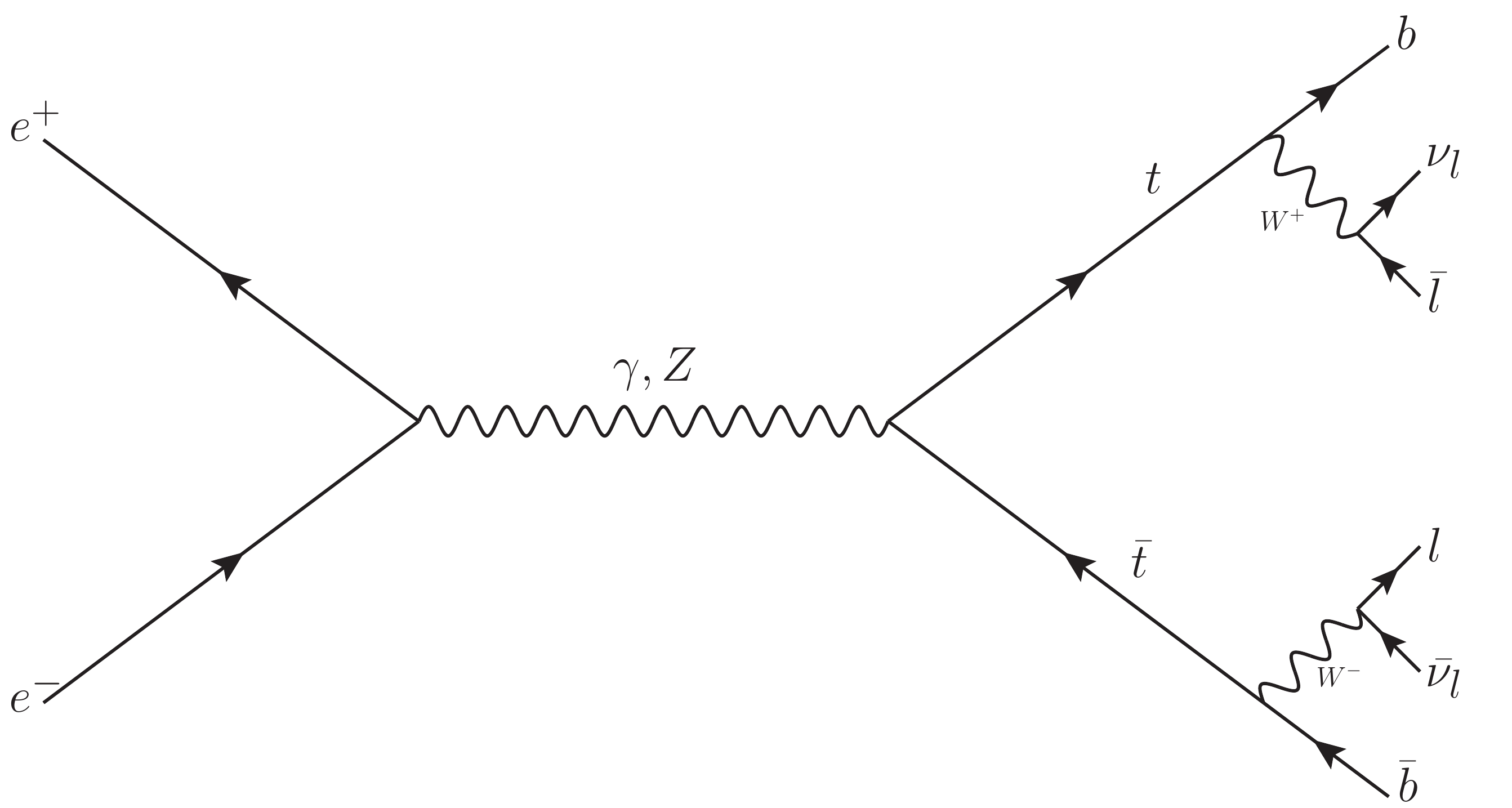} \quad 
	\includegraphics[width=7cm,height=5cm]{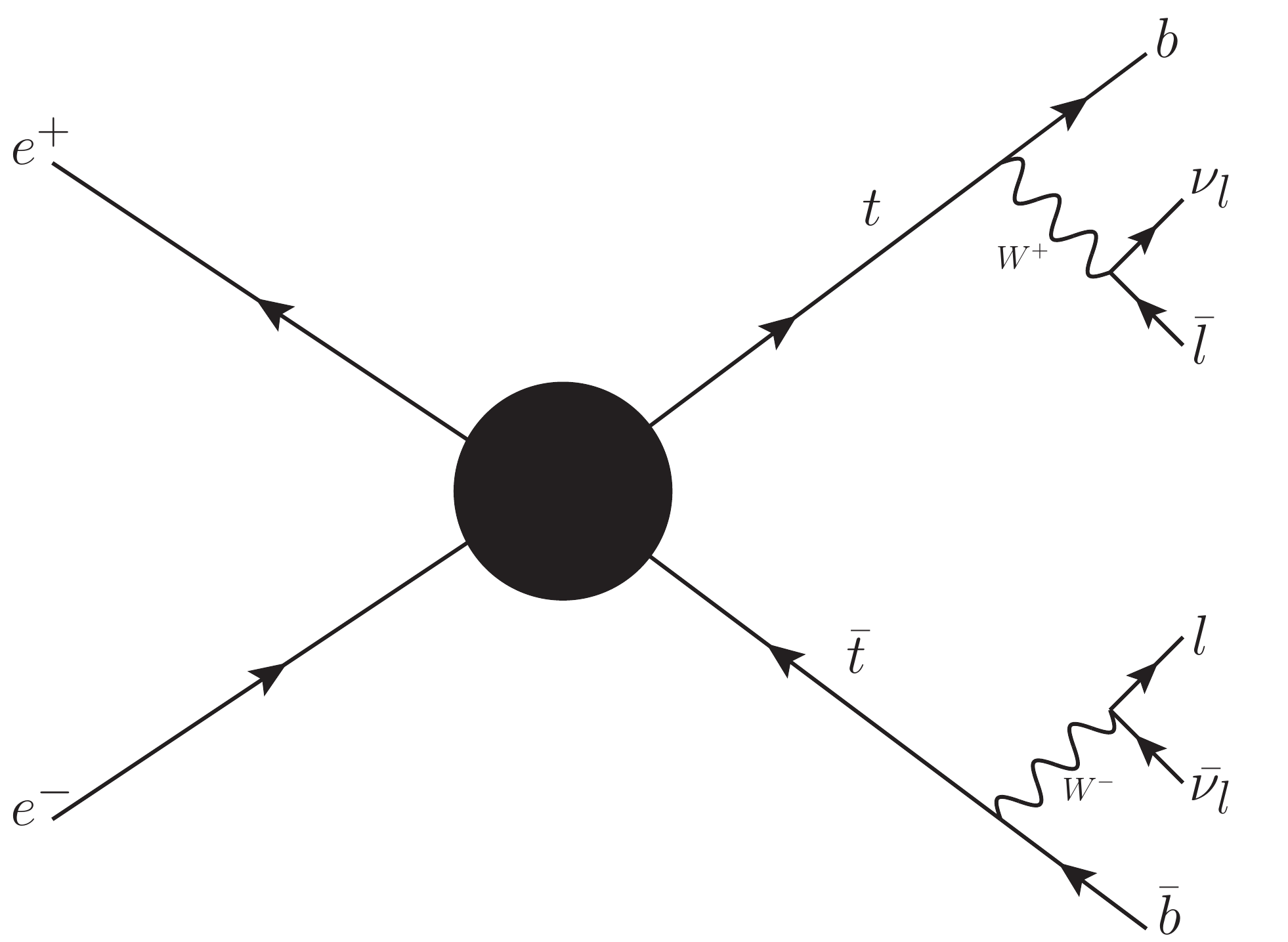}
	$$
	\caption{\small Production and decay of top-quark at $e^+ \, e^-$ colliders for $2l~+2b$ + missing energy signal.}
	\label{fig:signal}
\end{figure}

As noted earlier, the SM process is mediated by $Z$ or photon in s-channel exchange, while the NP effects we consider are generated 
by a contact 4-fermion interaction; as in the previous sections we will restrict ourselves to the NP effects in \cref{eq:leff}. 
We will consider only the leptonic decays of the $W$ bosons that follow from the decay of the $ \bar t\,t$ pair, namely,
\beq
e^+e^- \longrightarrow\, t \, \bar t \to \left(b W^+ \right)\, \left(\bar b W^- \right) \to \left(b l'{}^+ \nu_{l'} \right)\, \left(\bar b l^- \bar\nu_l \right)\,; \quad l,\,l' = e,\,\mu\,,
\eeq
(see \cref{fig:signal}). Thus the signature will be two opposite-sign leptons of same/different flavors + two b jets+ missing energy ($\slashed E$). 
The leading (non-interfering) SM background contributions are generated by  $Zh$, $ZZ$ and $W^+  W^-  Z$ production.

\begin{table}[htb!]
	\centering
{\renewcommand{\arraystretch}{1.4}%
	\begin{tabular}{|c|c|c|c|}
		\hline
		\multirow{2}*{BPs} &
		\multirow{2}*{Input model} &
		\multicolumn{2}{c|}{Cross-section (fb)} \\
		\cline{3-4}
		&& $P_{e^\pm}=0$ & $P_{e^\pm}=_{+80\%}^{-5\%}$ \\
		\hline
		BP1 & $c_1=1, \, c_2=0$ & 597.8 & 415.4    \\
		BP2 & $c_1=0, \, c_2=1$  & 645.7 & 461.4   \\
		BP3 & $c_1=1, \, c_2=1$  & 651.9 &  467.4  \\
		\hline
	\end{tabular}}
	\caption{Total cross-section at $e^+e^-$ collider for different benchmark points for unpolarized and polarized $P_{e^\pm}=_{+80\%}^{-5\%}$ beams with $\sqrt{s}=500$ GeV. All the benchmark points have $\Lambda=4$ TeV. }
	\label{tab:BPs}
\end{table}

\begin{figure}[htb!]
	$$
	\includegraphics[width=6cm,height=4.5cm]{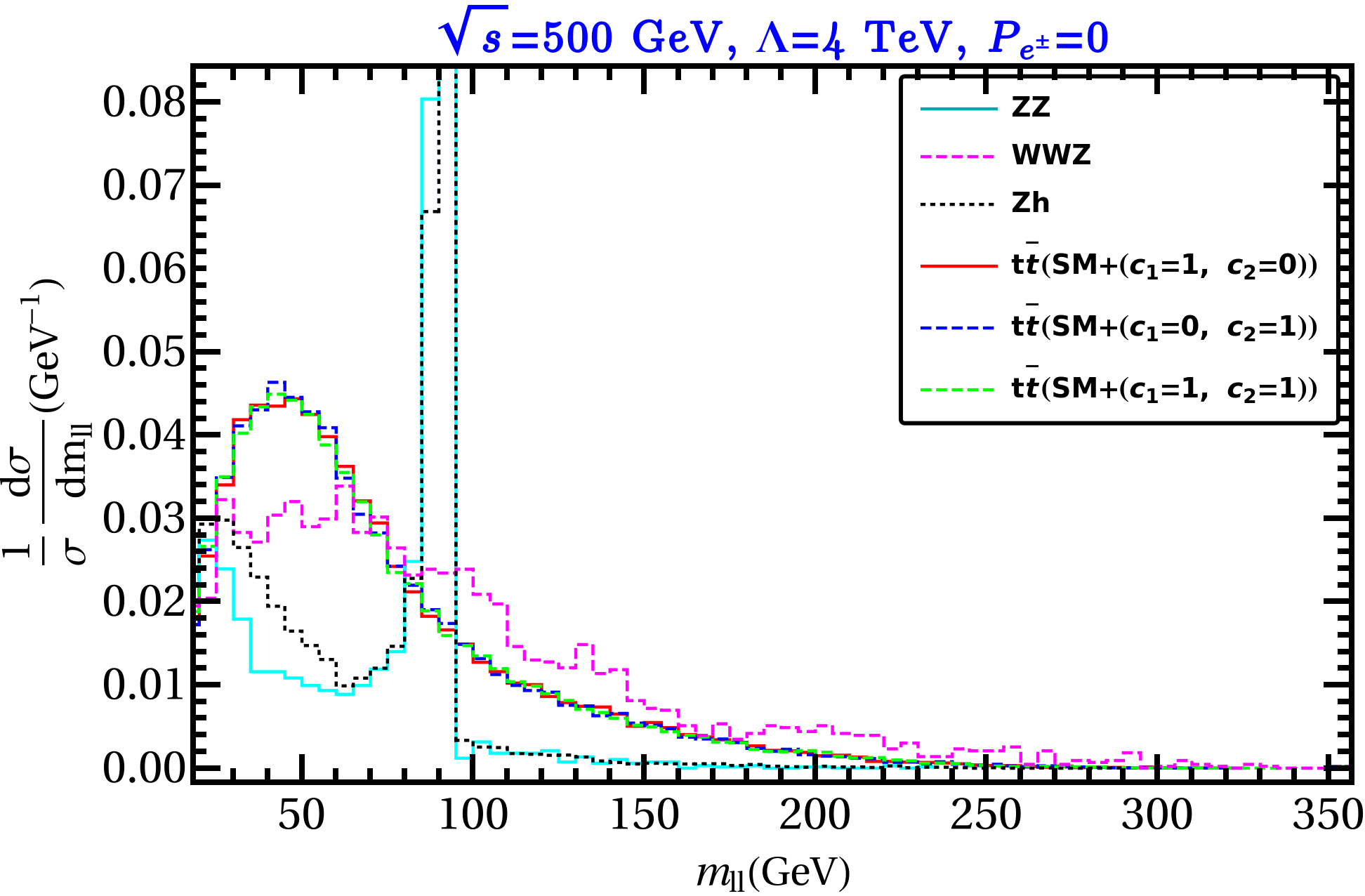}
	\includegraphics[width=6cm,height=4.5cm]{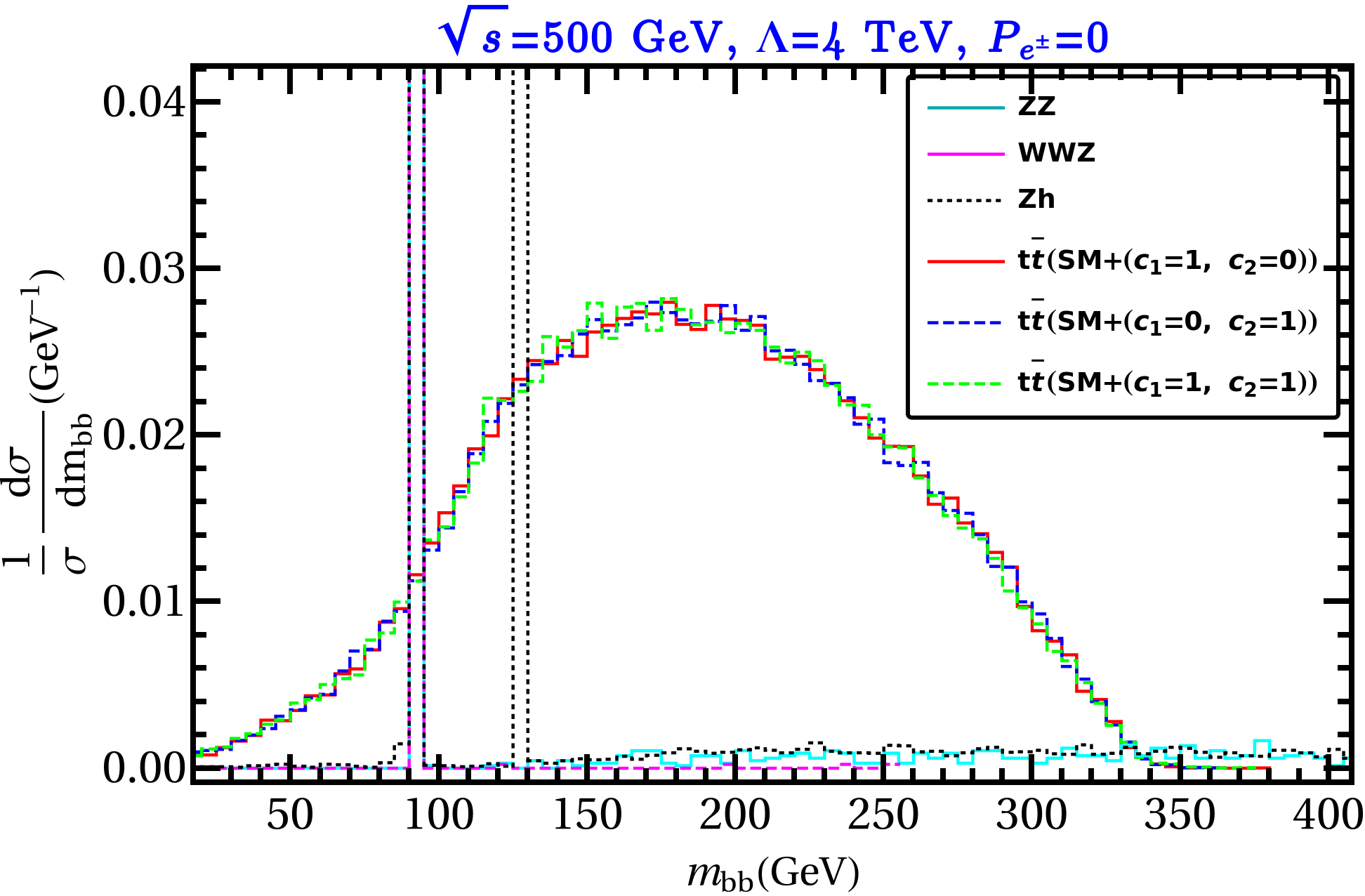}
	$$
	$$
	\includegraphics[width=6cm,height=4.5cm]{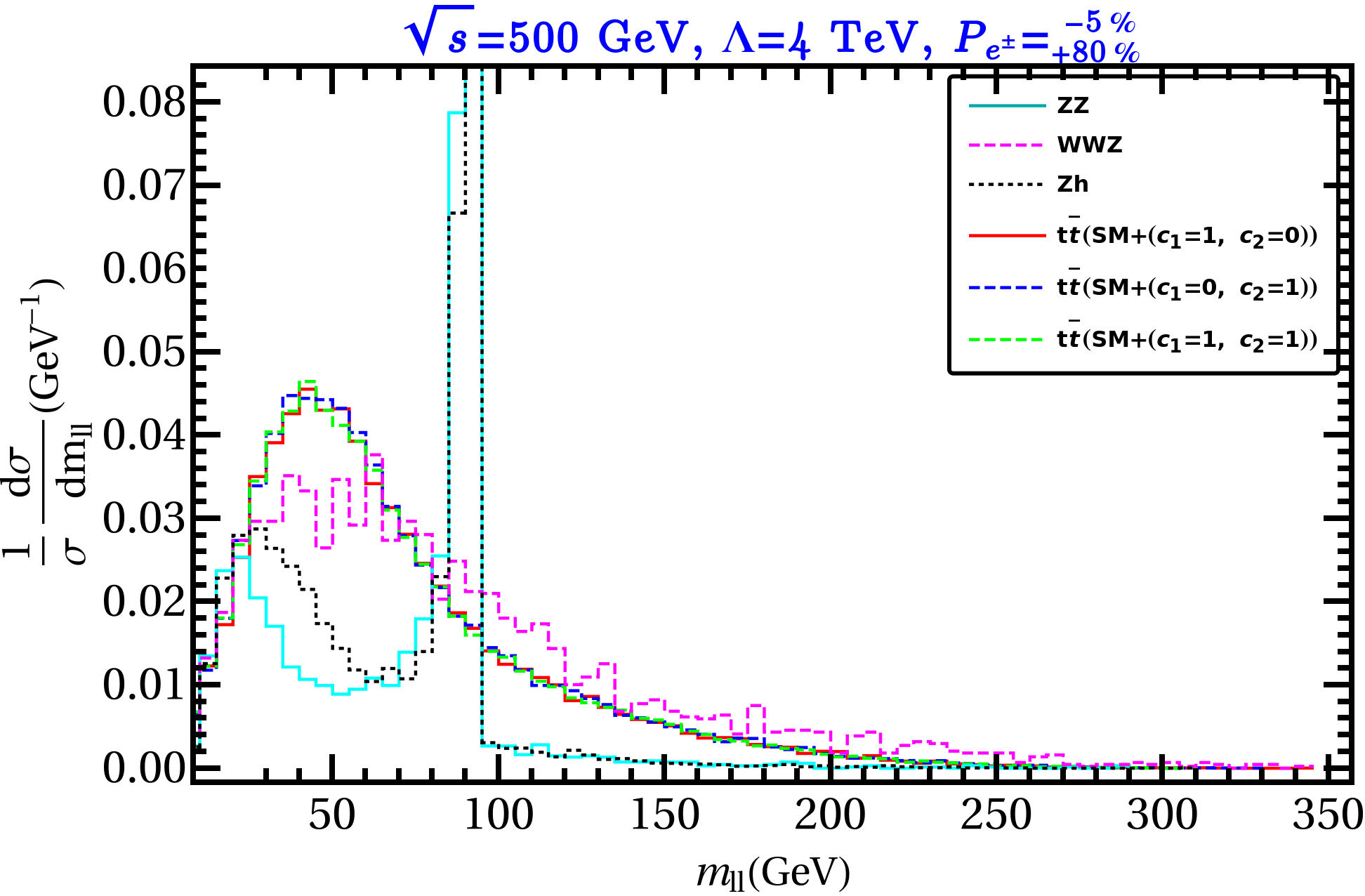}
	\includegraphics[width=6cm,height=4.5cm]{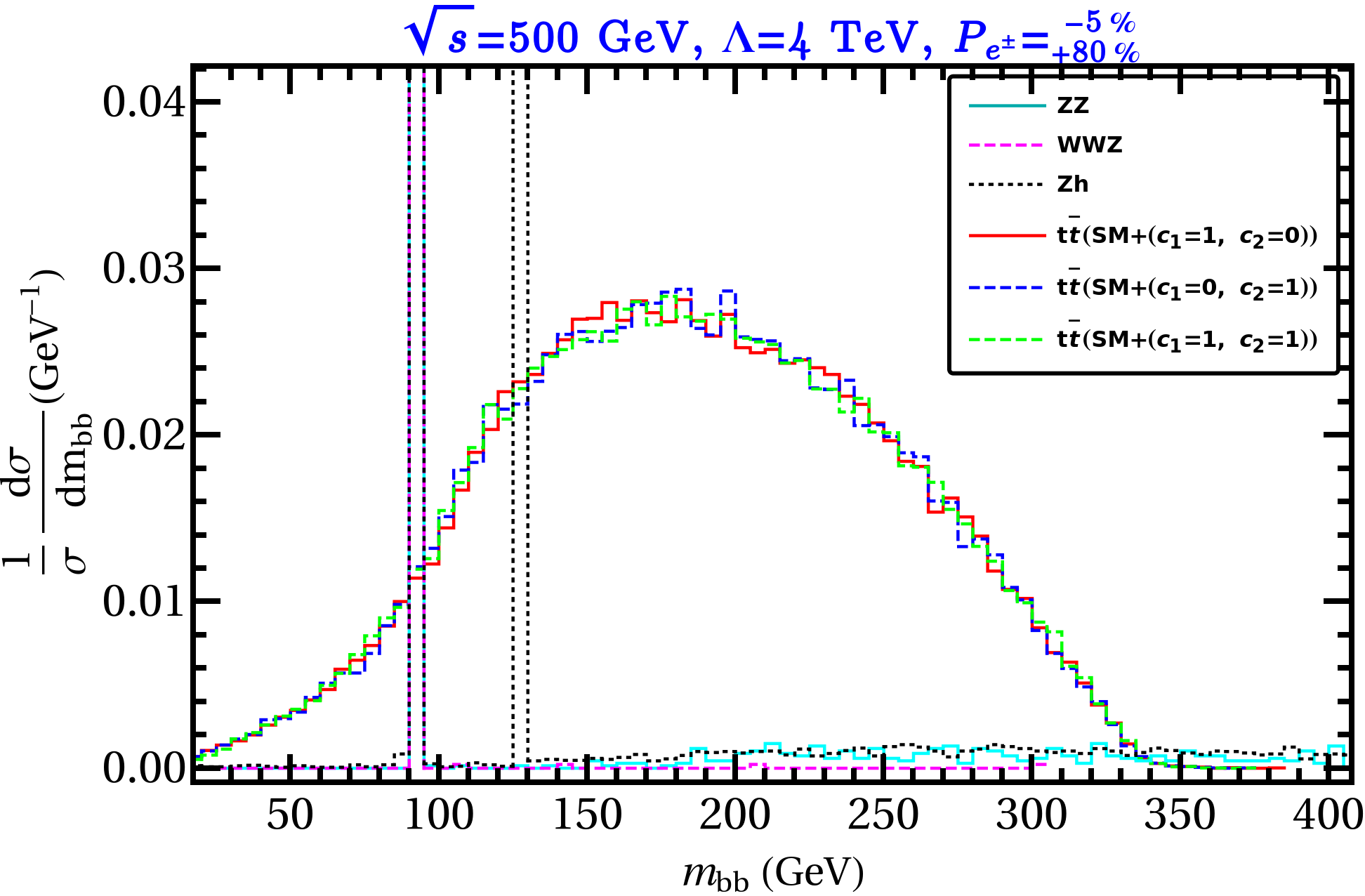}
	$$
	\caption{Invariant di lepton mass ($m_{ll}$) (left), Invariant di b-jet mass ($m_{b b}$) distributions (right) for $2l ~+2b$ + missing energy final state coming from $t\bar{t}$ signal with EFT 
		($\Lambda=4$ TeV) as well as dominant SM backgrounds at the $e^+\,e^-$ collider with $\sqrt s$ = 500 GeV. Top panel: Unpolarized beams; bottom panel: polarized beams 
		$P_{e^\pm}=^{-5\%}_{+80\%}$.}
	\label{fig:evntdis}
\end{figure}

We follow a standard approach, generating parton-level signal events using {\tt CalcHEP} \cite{Belyaev:2012qa}; the events are then showered and analyzed using 
{\tt Pythia} \cite{Alwall:2014hca}. For event reconstruction, and lepton and jet identification we use the following criteria:
\bit
	\item Events must have two opposite-sign leptons, and two $b$ jets; the $b$-tagging efficiency is chosen to be $0.6$ in accordance with the ILC TDR \cite{Behnke:2013xla}. 
	\item Lepton transverse momentum: $p^l_T > 10\,\gev$.
	\item Light jet transverse momentum: $p^j_T > 20\,\gev$.
	\item Leptons and light jets must be isolated: $ \Delta R_{ll'} > 0.2 $, $ \Delta R_{jl} > 0.4 $ and $ \Delta R_{jj'} > 0.4 $.
	\item Exclude events where the dilepton invariant mass is in the range $ 75\,\gev< m_{ll} < 105\,\gev$  to reduce the $Z \to \ell^+\ell^-$ background.
	\item Exclude events where the $b-\bar b$ invariant mass is in the range $ 115 \, \gev<  m_{bb} < 135\,\gev$, to reduce $h \to b\bar{b}$ contamination.
	\eit
where we defined  $\Delta R = \sqrt{\Delta \eta^2 + \Delta \phi^2} $ as the usual distance in the rapidity ($\eta$) - azimuthal angle ($\phi$) plane. The invariant mass cuts are designed to exclude the SM background; the values selected are based on the event distributions  plotted in \cref{fig:evntdis}.

Using these selection criteria we find the efficiency $ \epsilon $ as the ratio  of observed cross section  $ \sigma^{\tt FS} $  to the production cross section  $ \sigma^{\tt prod} $:
\beq
\epsilon=\frac{\sigma^{\tt FS}}{\sigma^{\tt prod}}.
\label{eq:eps}
\eeq
For the benchmark points in \cref{tab:BPs} we find $ \epsilon \sim 0.008 $, roughly independent of polarization. In the following we 
take a conservative approach and use $ \epsilon=0.001$ or $ \epsilon=0.005$.

\subsection{1$\sigma$ surfaces of EFT parameter uncertainties}
\label{sec:eft.err}
\begin{figure}[ht!]
	\centering
	$$
	\includegraphics[scale=0.23]{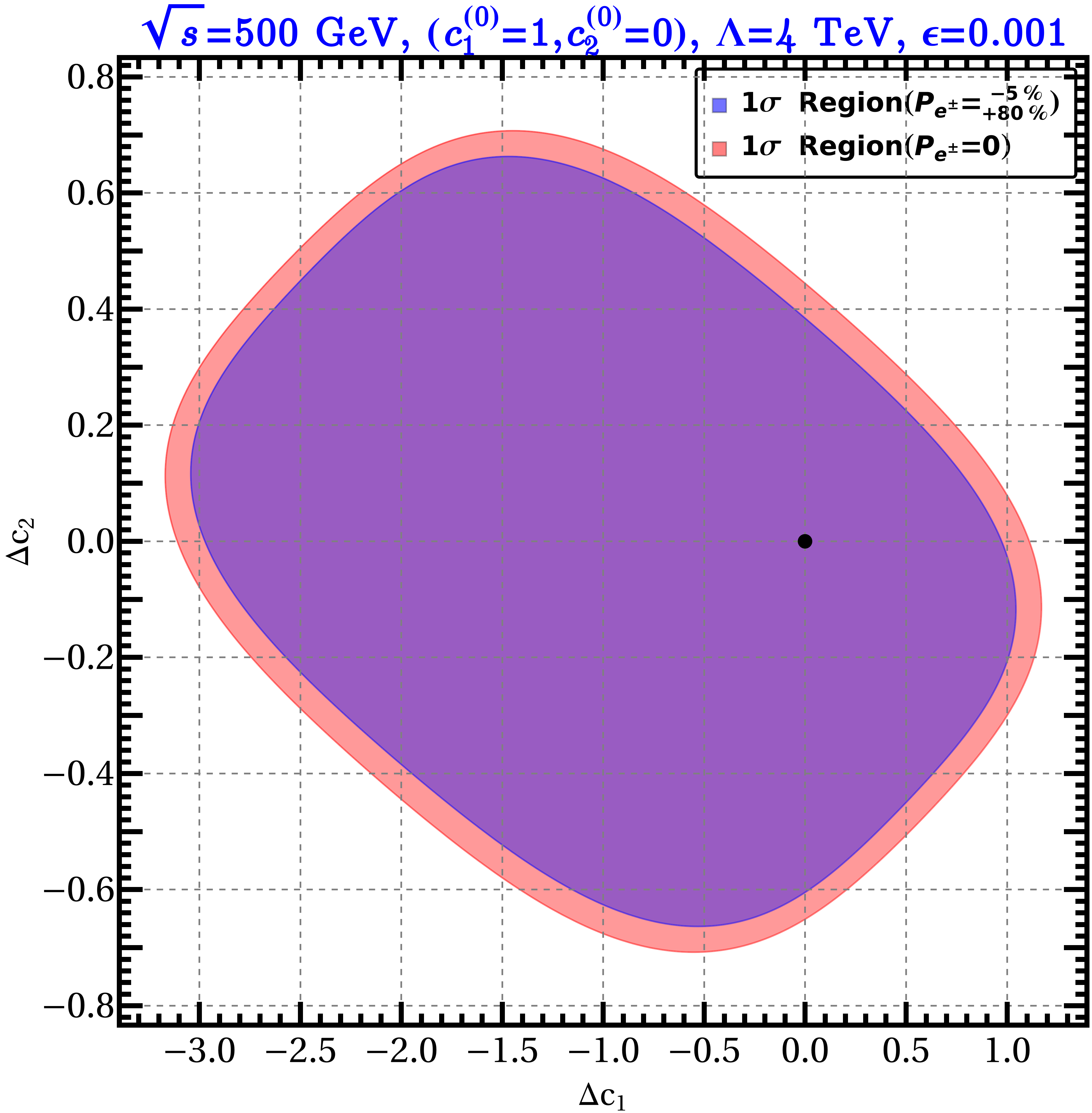}
	\includegraphics[scale=0.23]{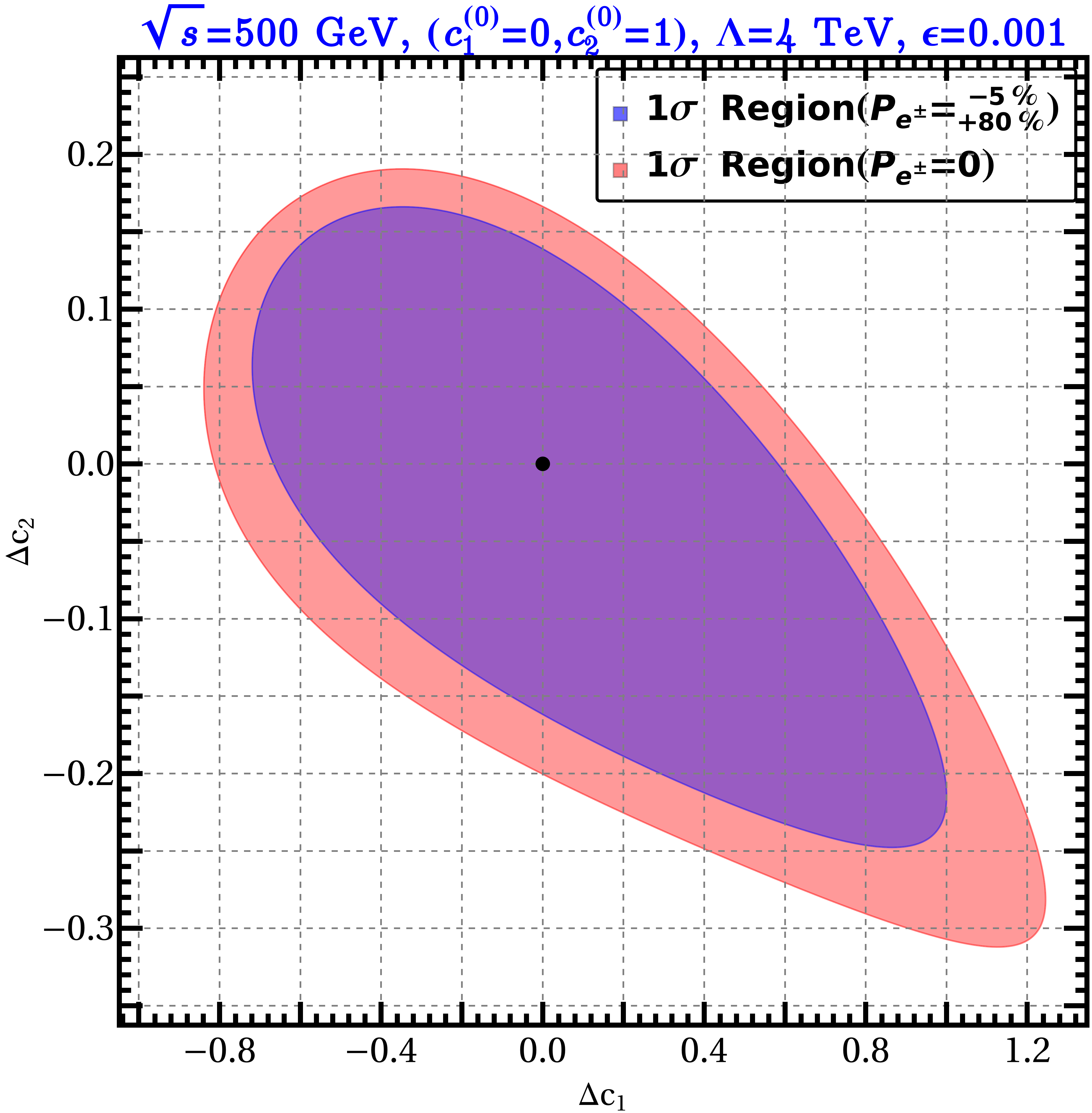}
	$$
	$$
	\includegraphics[scale=0.23]{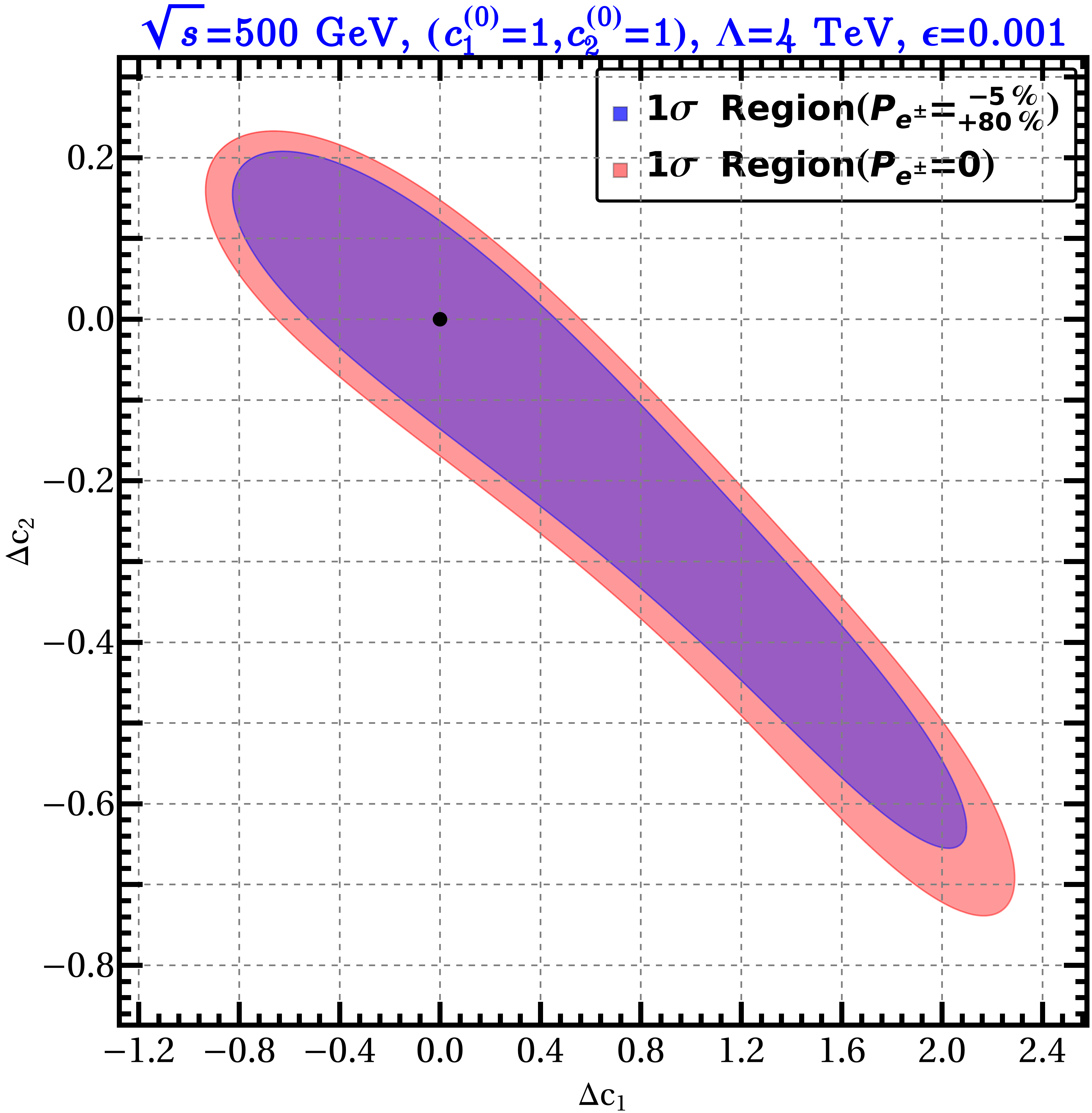}
	\includegraphics[scale=0.23]{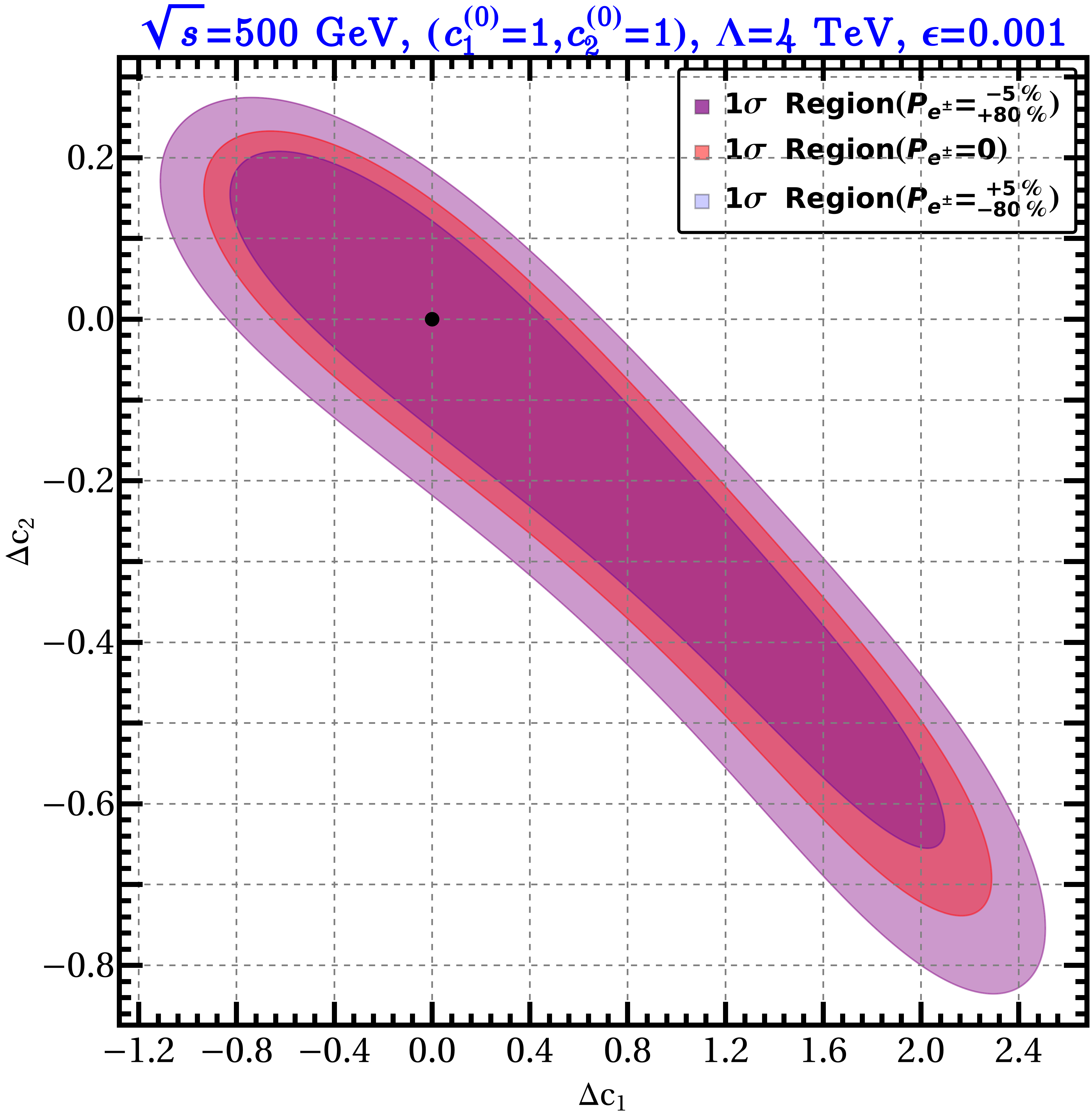}
	$$
	\caption{\small Optimal 1-$\sigma$ regions in $\Delta c_1-\Delta c_2$ plane for various choices of EFT parameters and choices of beam polarization. 
	See figure inset and heading for details.}
	\label{fig:1sigci}
\end{figure}

Using  \cref{eq:chi2}, we define the optimal 1$\sigma$ region of the statistical uncertainties of the EFT parameters $ c_{1,2} $ for the above described reaction. 
We choose three different combinations of the NP coupling seed values: $i)$ $c_1^0=1, c_2^0=0$, $ ii)$ $c_1^0=0, c_2^0=1$, 
and $iii)$ $c_1^0=1, c_2^0=1$. The 1$\sigma$ regions are plotted in \cref{fig:1sigci} for various choices of input values $ c_i\up0 $ and beam 
polarizations in $\Delta c_i - \Delta c_j$ plane where $\Delta c_i=c_i-c^0_i$. For this calculation we choose $\sqrt{s}$= 500 GeV and $\lcal_{int}$= 1000 $\rm fb^{-1}$.  
Note that, even though the SM dominates, the NP effects are significant even for $ \Lambda= 4\,\tev$. 

\begin{figure}[htb!]
	$$
	\includegraphics[scale=0.23]{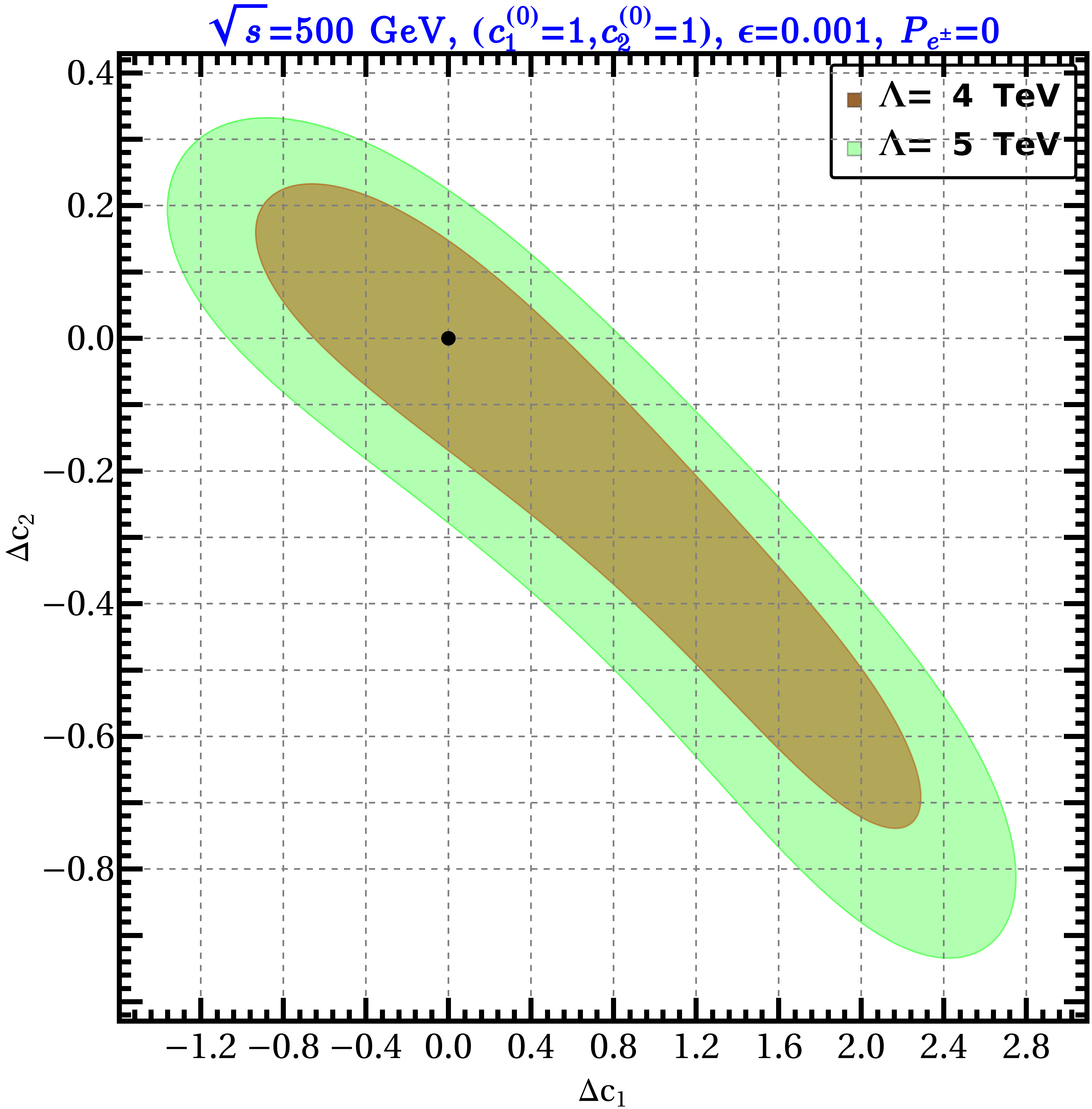}
	\includegraphics[scale=0.23]{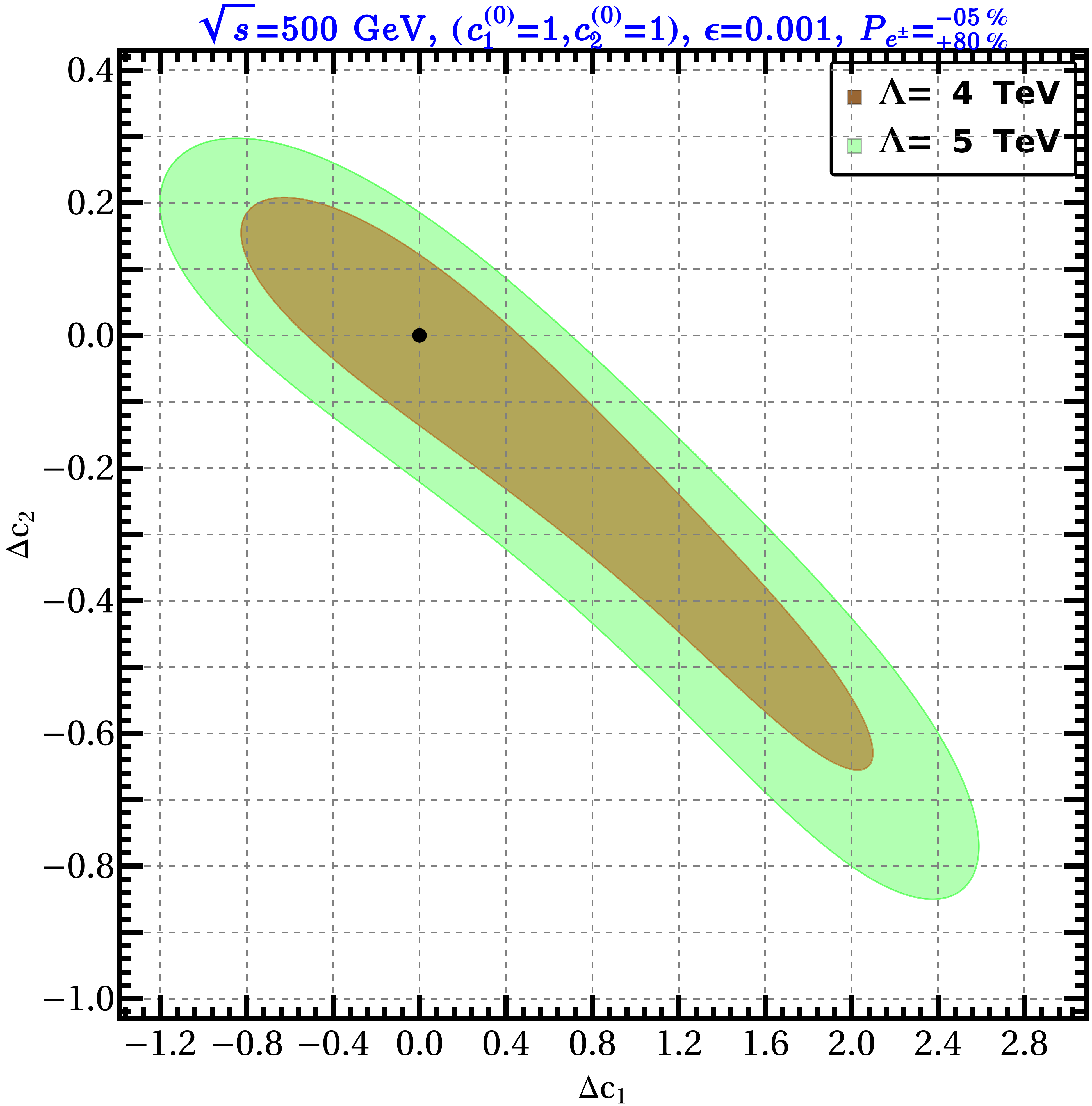}
	$$
	\caption{\small 1$\sigma$ surfaces in $\Delta c_1-\Delta c_2$ plane for two choices of $\Lambda$ 
	and unpolarized (left) and polarized $P_{e^\pm}=^{-5\%}_{+80\%}$ (right) beams. 
	See figure inset and heading for details.}
	\label{fig:lamvar}
\end{figure}

The 1$\sigma$ uncertainties for different combination of input values are listed in \cref{tab:1sigtt}, while 1$\sigma$ suraces are shown in \cref{fig:1sigci}. 
 The polarization of the initial beams in the lepton collider play a crucial role in determining the uncertainties of NP couplings. Precise extraction of NP relies on reducing the SM contribution and/or enhancing BSM contribution to the specific process. The uncertainties of NP couplings for different choices of beam polarization (within the possible ranges in accordance to collider TDR) are shown in the bottom right of fig.~\ref{fig:1sigci}. For $\left(P_{e^\pm}=^{+5\%}_{-80\%}\right)$, both SM and BSM contribution to the top-quark pair production increase, while if we flip the polarization sign, both contributions will decrease. But we observe that $\left(P_{e^\pm}=^{-5\%}_{+80\%}\right)$ provides more precise $\Delta c_i - \Delta c_j$, which implies that the reduction of SM contribution is larger than the reduction in BSM contribution. We note that for $\left(P_{e^\pm}=^{+5\%}_{-80\%}\right)$ choice of polarization combination, NP uncertainties are $\sim 10-20\%$ smaller compared to unpolarized beam. 
We also note that in all cases the uncertainties for the tensor operator coefficient $ c_2 $ are smaller than for the scalar one, as the tensor operator provides larger BSM contribution to $t\bar{t}$ production than the scalar operator. As $\Lambda$ 
increases, BSM contribution to the production cross-section is reduced, as a result, the uncertainties of the NP couplings are also increased. This 
behaviour is illustrated in \cref{fig:lamvar}. 

The OOT done here is primarily signal-based, where we assume that the non-interfering background effects are negligible, which is true to a great extent. However, uncertainty of NP couplings will increase when we include these remaining background effects. We take into account these effects by choosing
a lower efficiency ($\epsilon= 0.001$) than 
the one derived when ignoring such backgrounds
($\epsilon=0.008$) 
.

\begin{center}
	\begin{table}[h]
			{\renewcommand{\arraystretch}{1.4}%
		\begin{tabular}{| c | c  c | c  c | c  c | c   c | c | } 
			\hline
			\multicolumn{1}{|c}{} &
			\multicolumn{4}{|c|}{$P_{e^\pm}=0$} &
			\multicolumn{4}{c|}{$P_{e^\pm}=^{-5\%}_{+80\%}$} \\
			\cline{2-9}
			\multicolumn{1}{|c|}{Seed parameters} &
			\multicolumn{2}{c|}{$\epsilon=0.005$} &
			\multicolumn{2}{c|}{$\epsilon=0.001$}&
			\multicolumn{2}{c|}{$\epsilon=0.005$} &
			\multicolumn{2}{c|}{$\epsilon=0.001$}\\
			\cline{2-9}
			& $\pm\Delta c_1$ & $\pm\Delta c_2$ & $\pm\Delta c_1$ & $\pm\Delta c_2$ & $\pm\Delta c_1$ & $\pm\Delta c_2$ & $\pm \Delta c_1$ & $\pm\Delta c_2$\\ 
			\hline
			\multirow{2}{*}{$c\up0_1=1$, $c\up0_2=0$} & $+0.52$ & $+0.32$ &$+1.16$ & $+0.71$ & $+0.46$ & $+0.29$ & $+1.03$ & $+0.66$ \\
			&$-1.41$ & $-0.32$ & $-3.18$ & $-0.71$ &$-1.37$ & $-0.15$ &$-3.06$ & $-0.66$\\
			\hline
			\multirow{2}{*}{$c\up0_1=0$, $c\up0_2=1$} & $+0.55$ & $+0.08$ &$+1.24$ & $+0.19$ &$+0.44$&$+0.07$&$+1.00$&$+0.16$ \\
			& $-0.37$ & $-0.14$ & $-0.84$ & $-0.31$ &$-0.32$&$-0.11$&$-0.72$& $-0.25$\\
			\hline
			\multirow{2}{*}{$c\up0_1=1$, $c\up0_2=1$} & $+1.02$ & $+0.10$ &$+2.29$ & $+0.23$ &$+0.55$&$+0.09$&$+1.24$&$+0.19$ \\
			& $-0.42$ & $-0.33$ & $-0.94$ & $-0.74$ &$-0.37$&$-0.14$&$-0.84$& $-0.31$\\
			\hline
		\end{tabular}}
		\caption{\small Optimal 1$\sigma$ statistical uncertainties of the $c_1, c_2$ couplings for unpolarized and polarized $P_{e^\pm}=^{-5\%}_{+80\%}$ beams, and two values of $ \epsilon$; 
			we used the parameters in \cref{eq:params}.}
		\label{tab:1sigtt}
	\end{table}
\end{center}

\subsection{Differentiation of hypotheses}
\label{sec:hyp.dif}
\begin{center}
				{\renewcommand{\arraystretch}{1.4}%
	\begin{table}
		$$		 
		\begin{array}{| c | c | c | c | c |} 
			\hline
			\multirow{2}*{model}&
			\multirow{2}*{$\epsilon$}&
			\multirow{2}*{$\lcal_{\tt int}~(\text{fb}^{-1})$}&
			\multicolumn{2}{c|}{\text{significance($\Delta\sigma$)}} \\
			\cline{4-5}
			&&& P_{e^\pm}=0  &P_{e^\pm}=^{-5\%}_{+80\%}\\ 
			\hline
			\multirow{3}*{\minitab[l]{$\bar c_1=1$ \\ $\bar c_2=0$}}&0.001&1000 &0.29&0.35\\
			&0.001&2000&0.41&0.49\\
			&0.005&1000 &0.69&0.78\\
			\hline
			\multirow{3}*{\minitab[l]{$\bar c_1=0$ \\ $\bar c_2=1$}}&0.001&1000 &2.77&3.36\\
			&0.001&2000&3.92&4.75\\
			&0.005&1000 &6.19&7.51\\
			\hline
			\multirow{3}*{\minitab[l]{$\bar c_1=1$ \\ $\bar c_2=1$}}&0.001&1000 &3.93&4.83 \\
			&0.001&2000&5.56&6.82\\
			&0.005&1000 &8.79&10.80\\
			\hline
		\end{array} $$
		\caption{ Statistical significance $\Delta \sigma $ (see \cref{eq:signif-def}) of several hypotheses $\bar c_1,\,\bar c_2$ with respect to the  SM.}
		\label{table:discovery1}
	\end{table}}
\end{center}
An important feature of the OOT is that it provides a quantitative estimate of the degree to which a `base hypothesis' (characterized by NP coefficients $ g_i\up0$) can be distinguished from an alternate one (characterized by $ \bar g_i$). To this end we define the statistical significance by 
\beq
\Delta\sigma^2 \left(g\up0;\,\bar g \right) = \epsilon\sum_{i,j} \left(g^0_i - \bar g_i \right) \left( g^0_j - \bar g_j \right) \left( V^{-1}_0 \right)_{ij}\,, \quad V_0 = V(g=g^0)\,;
\label{eq:signif-def}
\eeq
with $V_0$ defined in \cref{eq:chi2}, and $ g^0_i= g_i(c_1^0,\,c_2^0)$ and $\bar g_i= g_i(\bar c_1,\,\bar c_2)$ as in \cref{fi}.

\begin{figure}[htb!]
	$$
	\includegraphics[height=5.5cm,width=5.8cm]{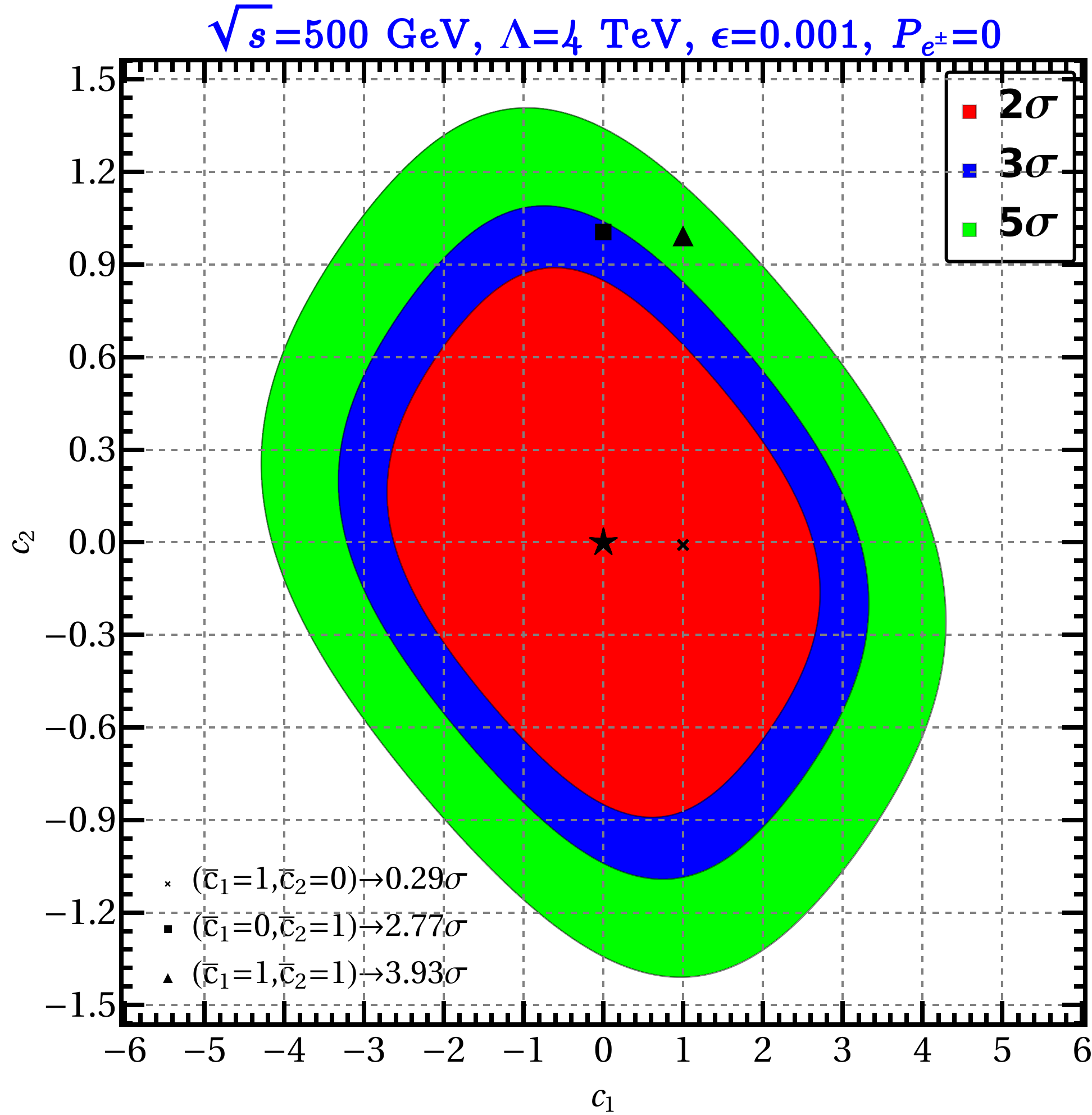}
	\includegraphics[height=5.5cm,width=5.8cm]{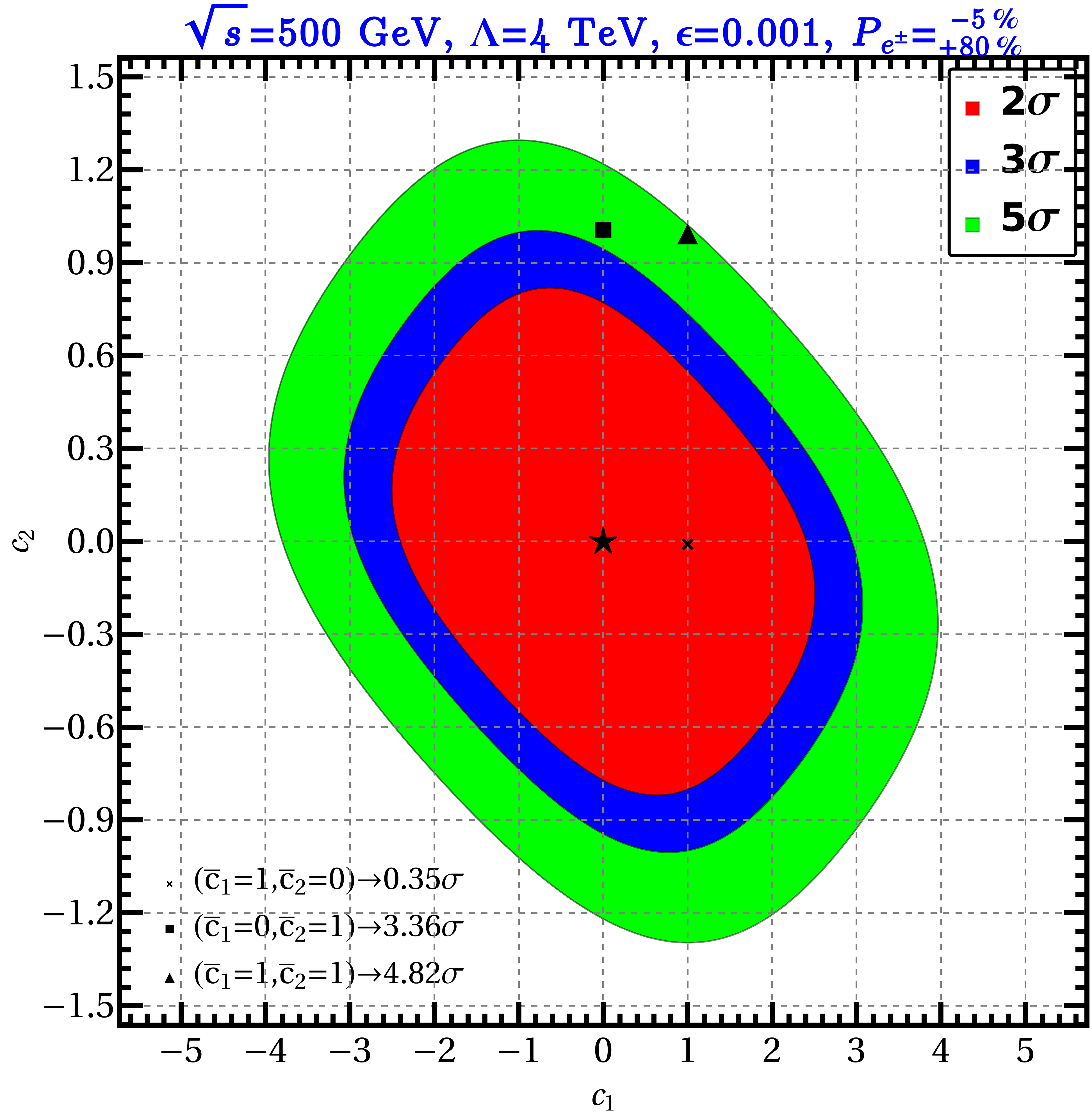}
	$$
	\caption{\small Statistical significance  ({\it cf.} \cref{eq:signif-def})  $\Delta \sigma \le 2,\,3,\,5 $  (respectively, red, blue and green areas) of alternate models with respect to the SM for unpolarized (left) and polarized $P_{e^\pm}=^{-5\%}_{+80\%}$ beams (right); also noted the statistical significance of 3 specific models. See figure inset and heading for details of the parameter choices.}
	\label{fig:discovery}
\end{figure}

As an example, we consider $c\up0_1 = c\up0_2=0$ as the base hypothesis (that corresponds to the SM), and various other choices of the seed values of NP couplings 
(given in the previous section) as alternative hypotheses. The degree of statistical differentiation of the alternate hypotheses from the base hypothesis are shown in \cref{fig:discovery} and listed in \cref{table:discovery1}, for both unpolarized and polarized beams. 
For unpolarized beams, hypothesis I ($\bar{c}_1=1,\bar{c}_2=0$) is indistinguishable form the SM even if we choose $\epsilon=0.005$ and $\mathfrak{L}_{int}=2000~\rm{fb^{-1}}$. With $\epsilon=0.001 (0.005)$ and $\mathfrak{L}_{int}=2000~\rm{fb^{-1}}$, hypothesis II ($\bar{c}_1=0,\bar{c}_2=1$) lies beyond 3$\sigma$ (5$\sigma$) the exclusion (discovery) limit. Due to larger BSM contribution, hypothesis III ($\bar{c}_1=1,\bar{c}_2=1$) is easier to distinguish than the other two hypotheses. Appropriate beam polarization, larger $\epsilon$ and high luminosity help the differentiation become more significant.

\subsection{Optimal versus standard $\chi^2$}
\label{sec:cut}
In this section we provide a comparison of the optimal coefficient uncertainties with those obtained using a basic analysis of collider data. 
To this end we imagine a collider experiment where the data obtained is organized in a number of bins $ j=1,\,2,\ldots,J $. We consider the differential cross-section with respect to the opening angle of the outgoing particles in the CM frame as our observable for this binned analysis (as was done in the OOT approach). We denote by $N_j^{\tt obs}$ the number of events in the $j$-th bin, and $N_j^{\tt theo}(g_i)$ 
the theoretical prediction for this number; using this we define \cite{Cowan:1998ji}
\begin{equation}
	\chi^2 =\sum^{\rm{bins}}_{j} \left(\frac{N_j^{\tt obs}-N_j^{\tt theo}(g_i)}{\sqrt{N^{\tt obs}_j}}\right)^2.
	\label{eq:chi2.cut}
\end{equation}

\begin{figure}[htb!]
	$$
	\includegraphics[height=5.5cm,width=5.2cm]{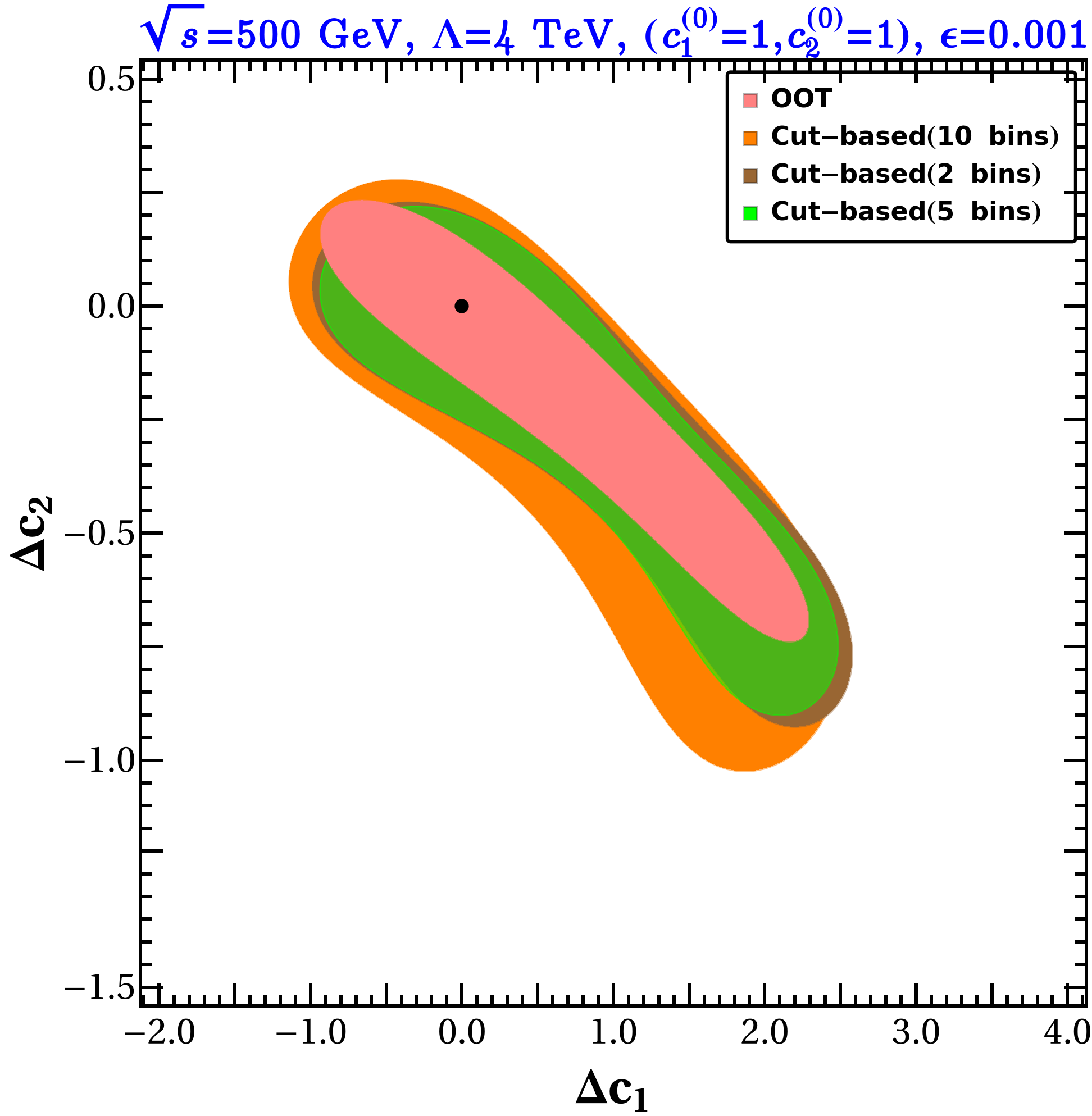}
	\includegraphics[height=5.5cm,width=5.2cm]{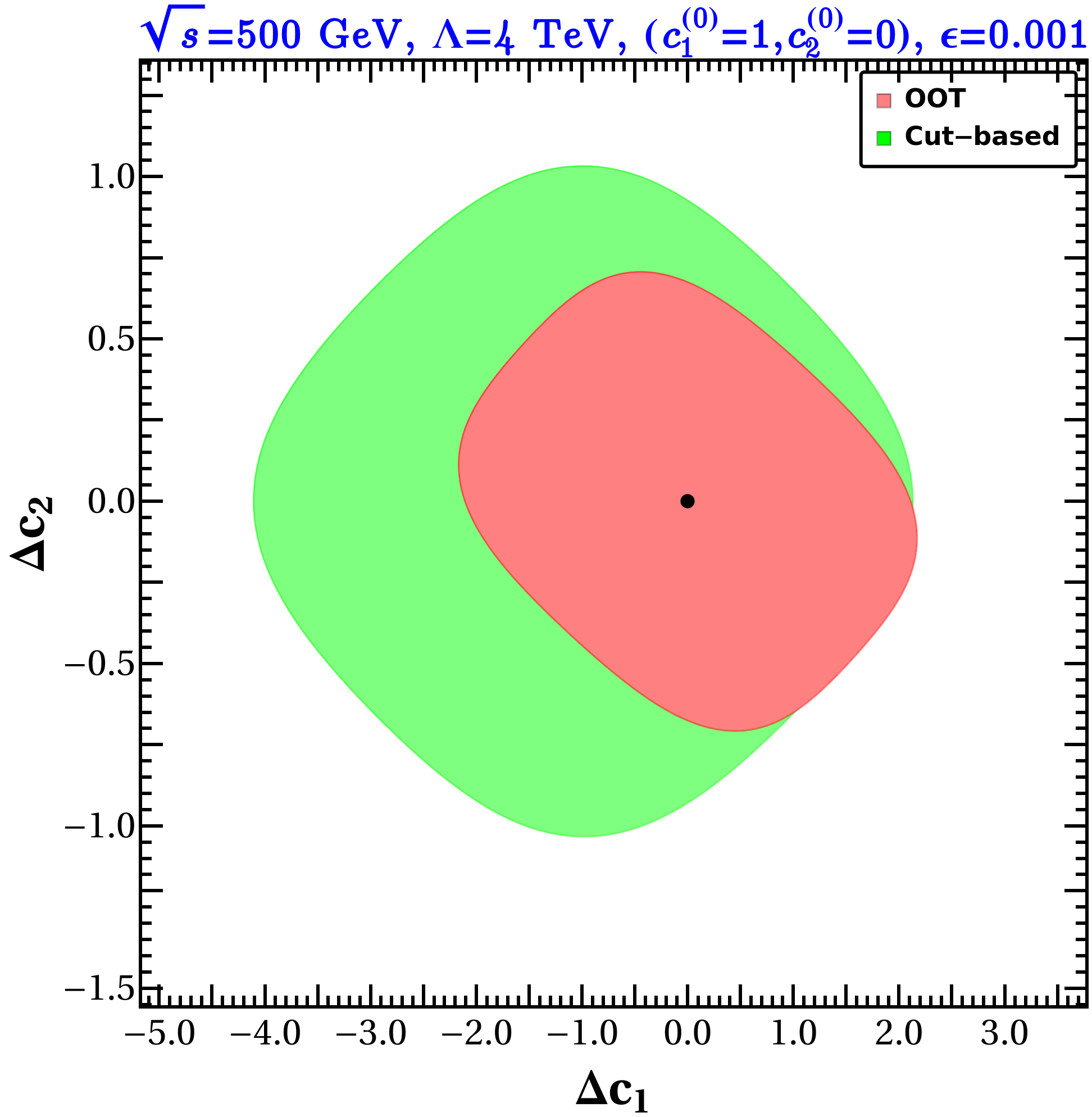}
	\includegraphics[height=5.5cm,width=5.2cm]{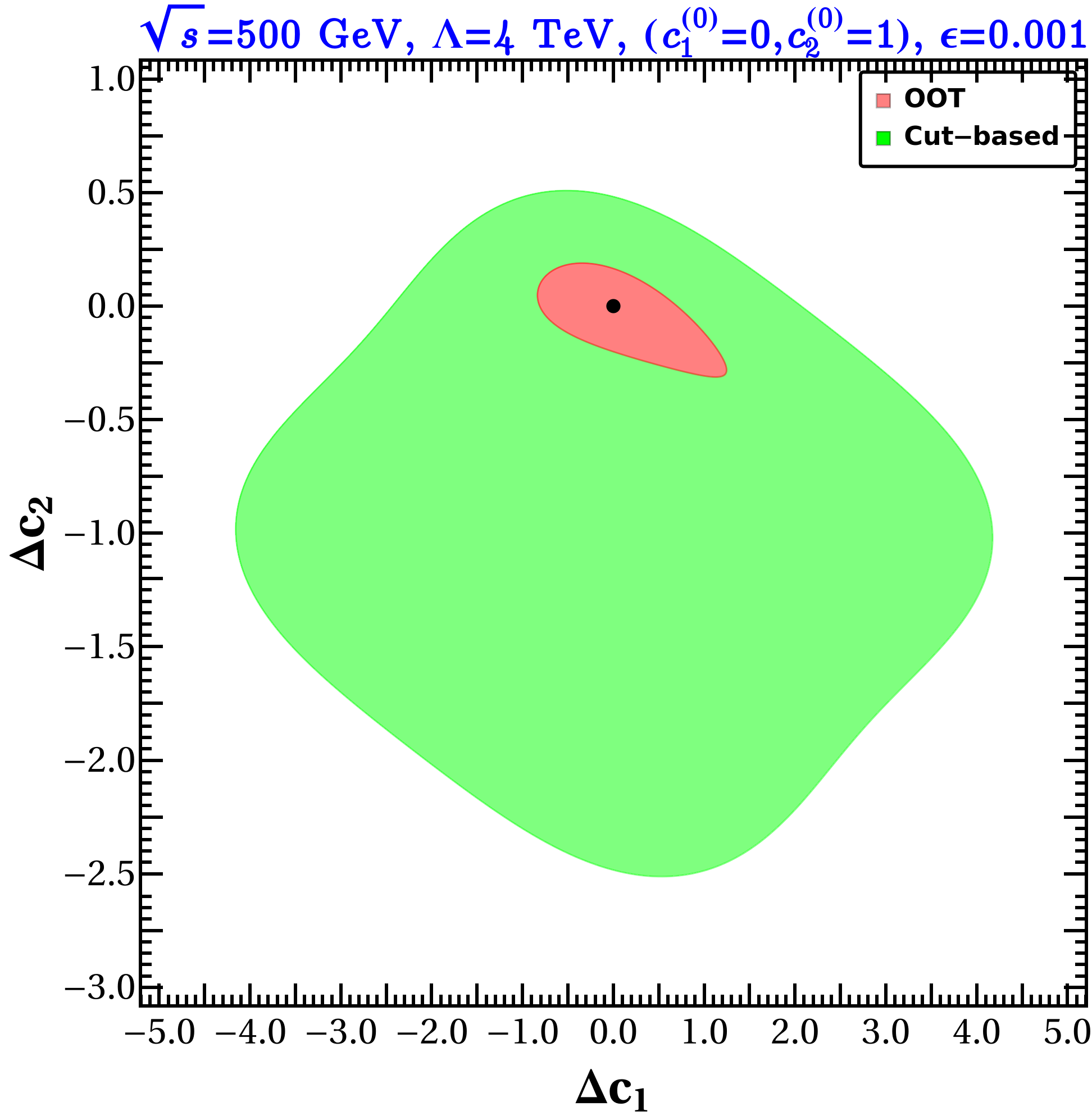}
	$$
	\caption{Comparison of 1$\sigma$ surfaces in $\Delta c_1-\Delta c_2$ plane between OOT (light red) and cut-based analysis as in \cref{eq:chi2.cut} (in green) 
	for unpolarized beams.}
	\label{fig:oot.cut}
\end{figure}

From this we define the 1$\sigma$ regions in parameter space by the condition  $\chi^2\le1$. In our case the $N_j^{\tt obs}$ are obtained using a numerical simulation that 
implements all the cuts described in \cref{sec:eft.col}. The results for three choices of model parameters, and the comparison with the corresponding OOT results are 
presented in \cref{fig:oot.cut}. The $\chi^2$ statistic defined in \cref{eq:chi2.cut} depends on the number of bins, a dependence that is also examined in \cref{fig:oot.cut}; 
we find that choosing 5 bins provides the most constrainted parameter space \footnote{This choice provides a balance between the number of events with the statistical fluctuations in each bin.} 
and thus we use this choice in the following examples. These results give a measure of the degree to which the analysis based on \cref{eq:chi2.cut} can be improved. In this case, as BSM contribution is less than SM to the signal final state, the statistical fluctuation in each bin is large, as a consequence, uncertainties in NP couplings estimated from binned analysis is worse than the OOT uncertainties.

\section{Example of NP dominance}
\label{sec:model}
We now consider the determination of NP parameters in cases where the NP dominates over the SM; 
specifically, we will consider the production of new particles at an $e^+e^-$ collider. For this a specific model must be selected, and we will use the well-understood
extension of the SM where an additional doublet is added to the scalar sector, the  two-Higgs doublet model\footnote{For a recent review see \cite{Bhattacharyya:2015nca,Krawczyk:2015xhl,Wang:2022yhm}.} (2HDM). 
As noted in \cref{sec:ttb}, the so-called `flipped' 2HDM \cite{Barger:1989fj,Grossman:1994jb,Akeroyd:1994ga,Akeroyd:1996he,Akeroyd:1998ui,Aoki:2009ha,Wang:2022yhm} 
can be used to describe the type of NP effects considered in that section.

The new particles in the 2HDM are a neutral CP-even scalar, a neutral CP-odd scalar, and a pair of charged scalars $ H^\pm$;  we will also assume the presence of 
right-handed neutrinos $N_R$ and assume that the $N_R$ and $H^\pm$ are light enough to be produced at the $e^+e^-$ collider. We then will study the degree to 
which this collider can be expected to determine some of the couplings associated with this extension of the SM.

\subsection{ $t\bar{t}$ production at $e^+e^-$ colliders within flipped-2HDM}
\label{sec:tt}
\begin{figure}[htb!]
	$$
	\includegraphics[height=4.6cm, width=6cm]{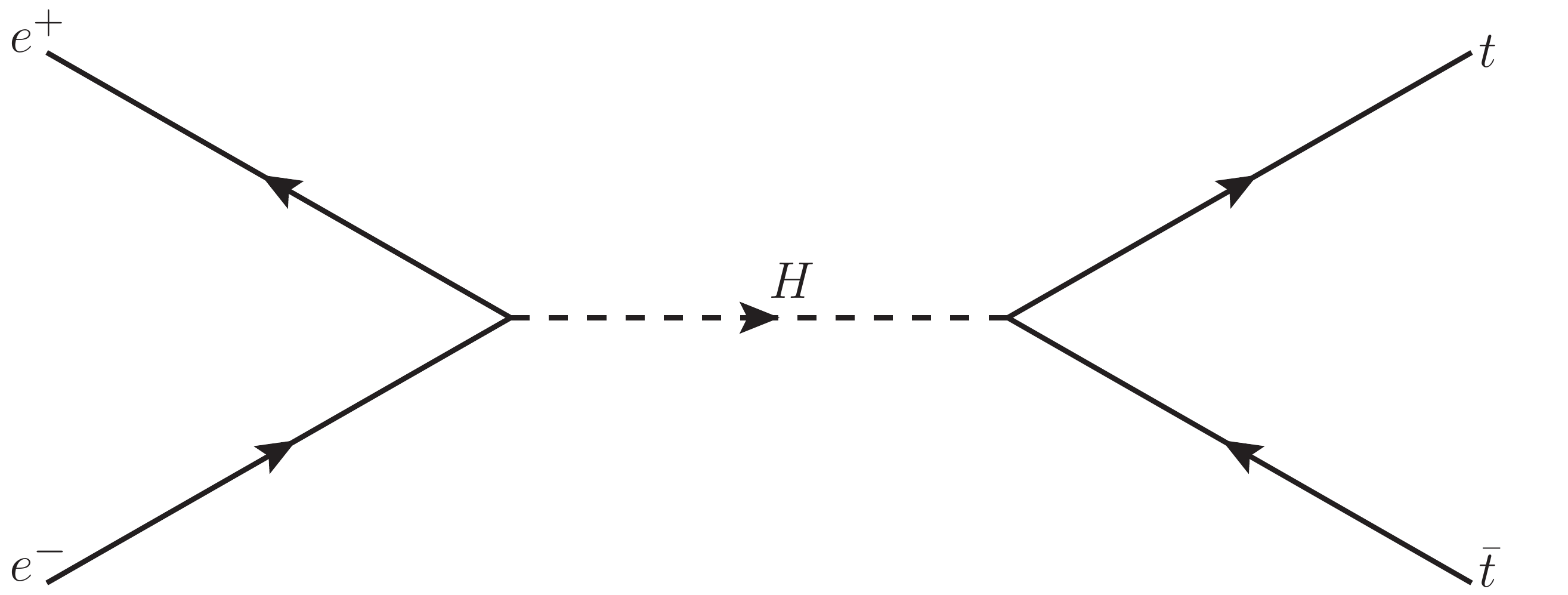} \qquad
	\includegraphics[height=4.7cm, width=6cm]{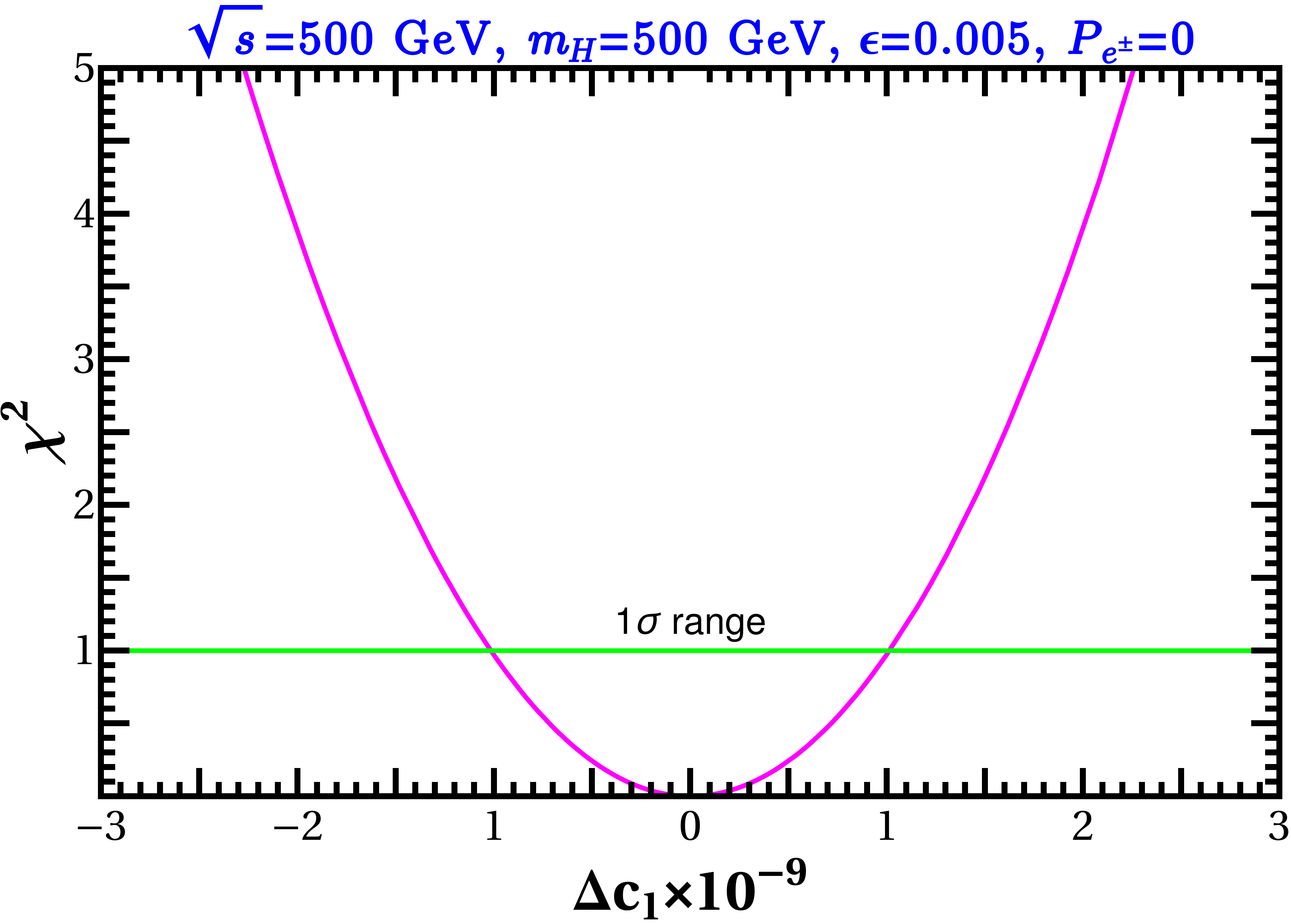}
	$$
	\caption{\small Left: NP contribution to $e^+ e^- \to t\bar t$ in a flipped 2HDM; where $H$ denotes a heavy scalar. Right: dependence of $ \chi^2$ (\cref{eq:chi2}) on the NP coefficient uncertainty $ \Delta c_1 $, when $ c_1\up0=5.71 \times 10^{-7} $, with the 1$\sigma$ bound indicated.}
		\label{fig:uv.th}
\end{figure}

We first briefly revisit the process considered in \cref{sec:ttb}, assuming now that the NP can be directed produced in colliders. 
In this case the contact interaction on the right-hand diagram on \cref{fig:ttdiag} is replaced by the heavy scalar exchange diagram in \cref{fig:uv.th}. We consider flipped 2HDM to elucidate this scenario. This model does not generate the effective interaction $Q_{lequ}\up3 $ in \cref{eq:leff}, so that $c_2=0$. The Yukawa couplings to the CP-even heavy Higgs ($H$) are given by \cite{Wang:2022yhm},
\beq
-\mathcal{L}_{\tt yuk}=\frac{m_f}{v}y_f \bar{f}fH\,,
\eeq
where $y_f=\cos(\beta-\alpha)-\sin(\beta-\alpha)\kappa_f$, with $\kappa_f=\text{cot}\beta = v_1/v_2$ and $ \alpha$ is the angle associated with the diagonalization of the neutral CP-even scalar mass matrix.

For this model $c_1=\frac{m_em_t}{v^2} k_f $. In order to estimate the the uncertainty of this NP coupling we will consider the resonant production of $H$ and its subsequent decay $\to t \bar{t}$. In the numerical analysis, we consider heavy Higgs mass $m_H = 500$ GeV,  
$\sin(\beta-\alpha)\sim 1$ in the  decoupling limit and small $\tan \beta~(\sim0.5)$. In this case we find $c_1=5.7\times10^{-7}$. Comparing uncertainty in $c_1$, $\Delta c_1$ in the flipped-2HDM is smaller than in the EFT scenario due to a greater NP contribution to $t \bar{t}$ production.

\subsection{$H^+H^-$ production at $e^+e^-$ colliders}
\label{sec:hh}
We consider next the pair production of $H^\pm$ which provides an additional channel to probe the NP in this model in which the NP contribution dominates.
We assume that the $H^\pm$ have the standard minimal coupling to the photon,
\beq
H^+(p_4)H^-(p_3)\gamma:- ~ie_0(p_4^{\mu}-p_3^{\mu})\,,
\eeq
and parameterize the leptonic-Yukawa\footnote{We assume that the masses of any other heavy neutrinos are large enough to have negligible effects at the colliders being considered.} and $Z$ couplings as follows:
\bal
H^+(p_4)H^-(p_3)Z:& - ~i a \left(p^{\mu}_4-p^{\mu}_3\right)~, \mcr
e^+ N_R H^-:& ~b;
\end{align}
where, $p_3$ and $p_4$ are the incoming momenta of $H^-$ and $H^+$, respectively.
For definiteness we will consider three different models : the flipped 2 HDM or inert doublet model (IDM) (doublet) \cite{Deshpande:1977rw}, the type-II seesaw model (triplet) \cite{Schechter:1981cv},  and the scotogenic model \cite{Ma:2006km}; these are characterized by\footnote{Here, $b$ is a free parameter.},
\beq
\begin{array}{l c c}
\qquad\text{\bf Model}& a & $\hspace{1cm}$b \cr
\text{Flipped 2 HDM/Inert-doublet (IDM):}$\hspace{0.5cm}$ & e_0\cot(2\theta_{\tt w})=0.21 & $\hspace{1cm}$ 0 \cr
\text{Scotogenic:}\quad& e_0\cot(2\theta_{\tt w})=0.21 & $\hspace{1cm}$ 0.1 \cr
\text{Type II seesaw:}\quad & e_0\tan(\theta_{\tt w})=0 .17 & $\hspace{1cm}$ 0 
\end{array}
\label{eq:2hdm.models}
\eeq

\begin{figure}[htb!]
$$
\includegraphics[height=4.5cm,width=7.5cm]{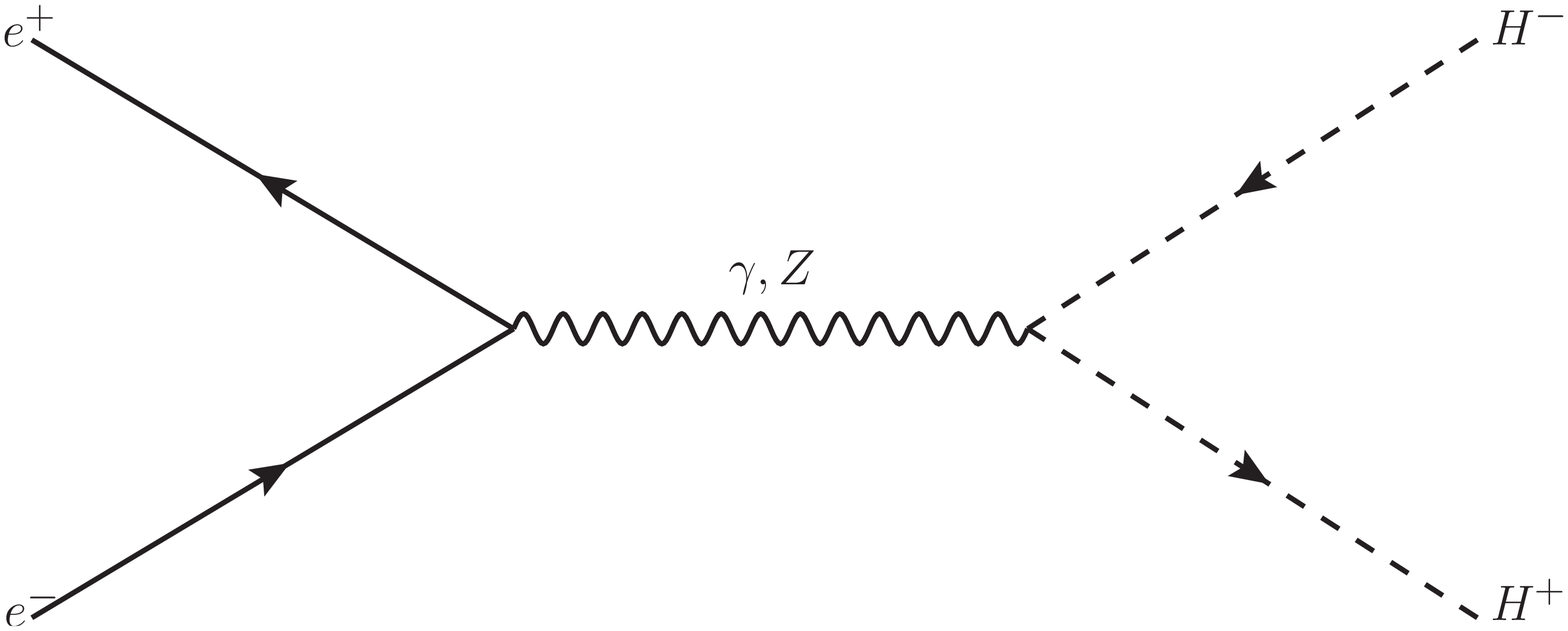} \quad
\includegraphics[height=4.5cm,width=5.5cm]{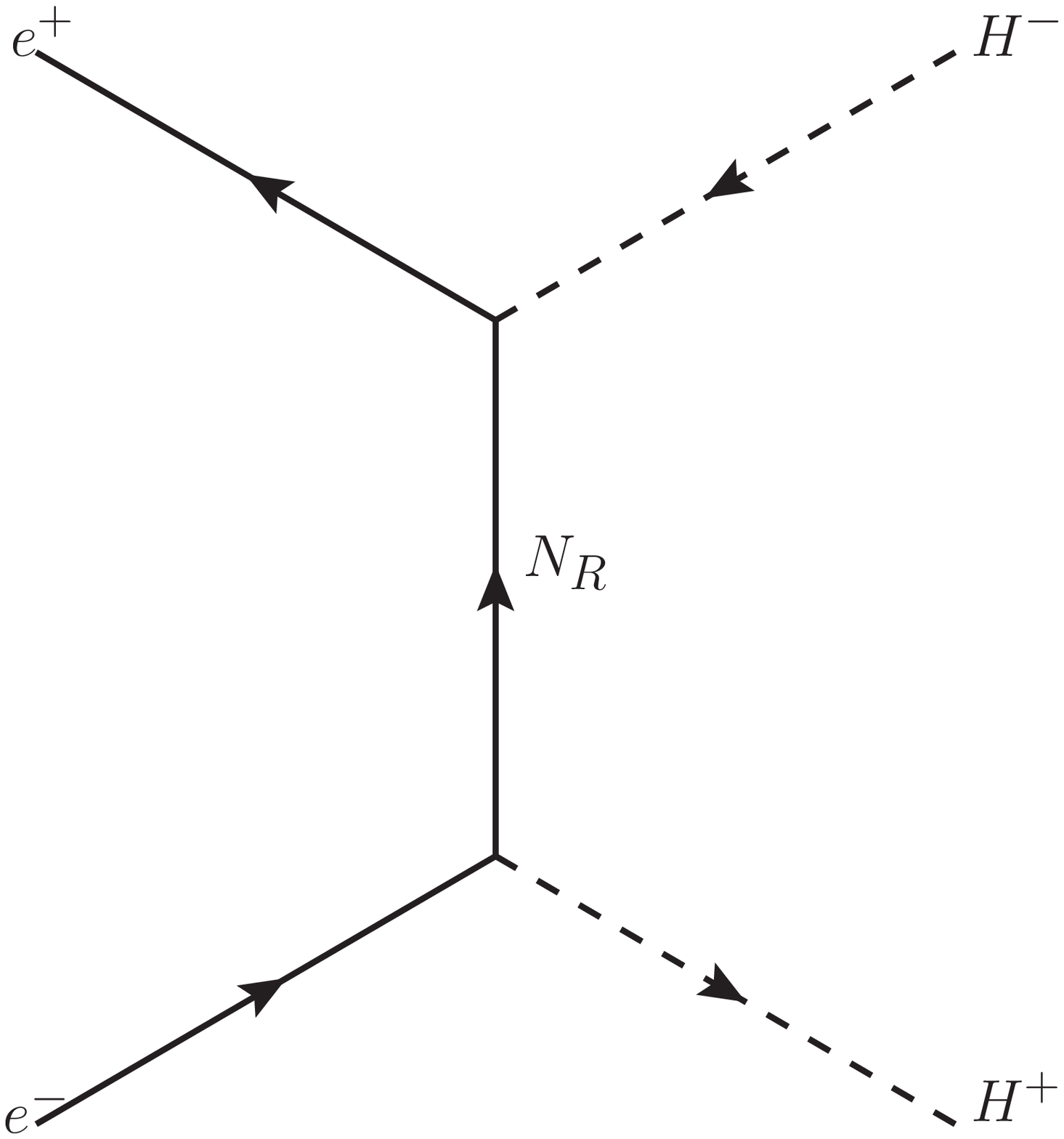}
$$
\caption{\small Leading pair-production mechanism for singly charged scalar pairs ($H^\pm$) at an $e^\pm$ collider.}
\label{fig:productn3}
\end{figure}

As a concrete application we will consider $H^+H^-$ pair-production at a linear $e^+e^-$ collider; the relevant diagrams are shown in fig.~\ref{fig:productn3}. The
corresponding helicity amplitudes, $ M'(\lambda_{e^-},\, \lambda_{e^+}) $, are easily obtained: $M'(\lambda,\,\lambda)=0$, and
\beq
M'(\lambda,\, -\lambda) = i \left[ e_0^2 + a e_0 \left( \frac{4\sw^2-1}{2 \stw}- \lambda \frac{1}{2 \stw} \right) \frac s{s-\mz^2} + \frac{(1+\lambda)b^2}4\frac s{t-m_{\tt N}^2}\right] \beta_{H^+} \sin\theta \,;
\eeq
where $\theta$ is the scattering angle of $H^\pm$ from the axis of collision, $ \beta_{H^+} = \sqrt{1 - 4 m_{H^+}^2/s} $, and 
\beq
t = \frac s2 \left( 1 + \beta_{H^+} \cos\theta  \right) + m_{H^+}^2\,.
\eeq

\begin{figure}[htb!]
$$
\includegraphics[height=5cm,width=5.2cm]{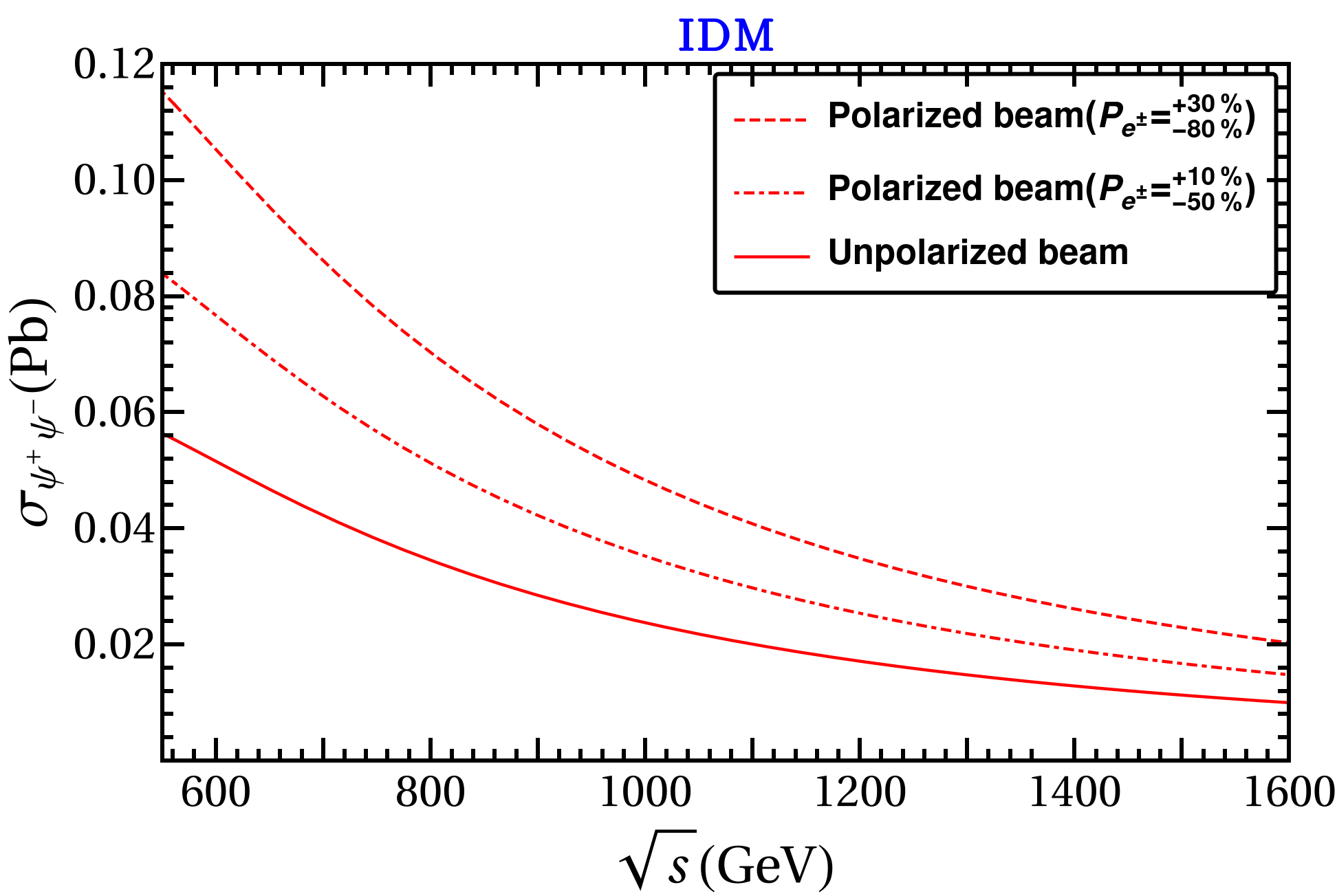}
\includegraphics[height=5cm,width=5.2cm]{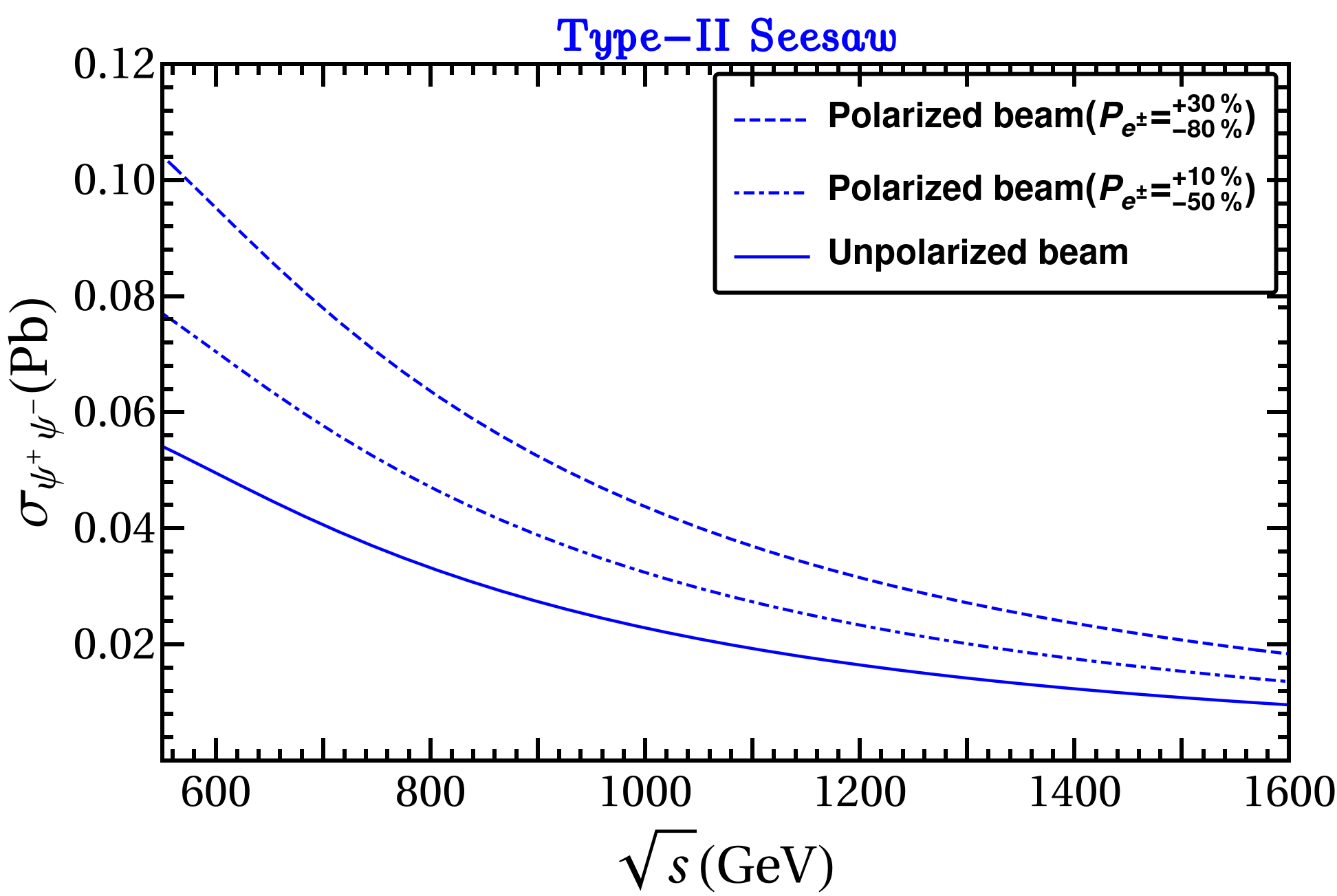}
\includegraphics[height=5cm,width=5.2cm]{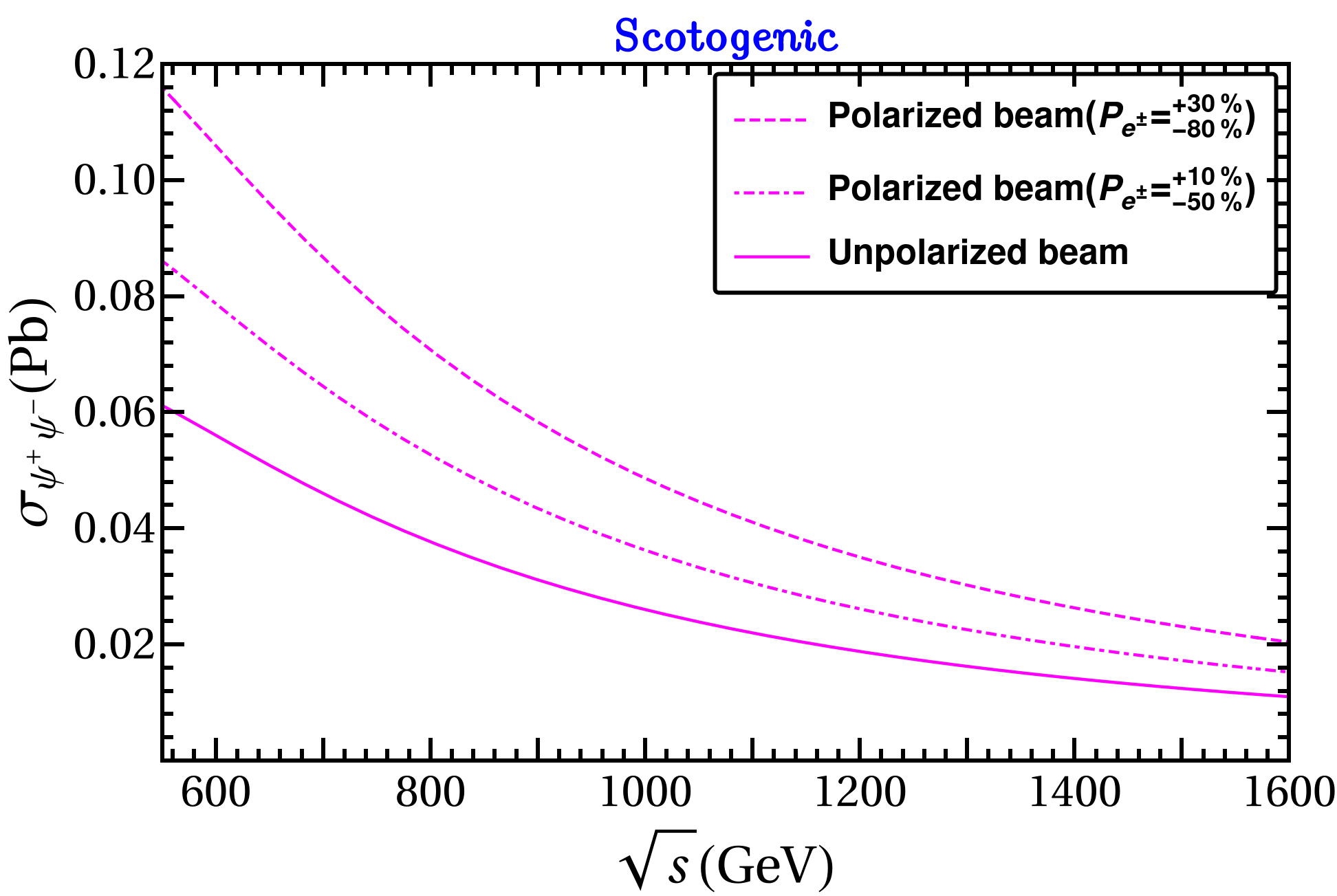}
$$
\caption{\small The total cross-section for charged scalar pair production as a function of the CM energy ($\sqrt{s}$) at an $e^+\,e^-$ collider for unpolarized beams and two different choices of beam polarization. Left: IDM;  middle: type-II seesaw;  right: scotogenic (see \cref{eq:2hdm.models}).}
\label{fig:hhprod}
\end{figure}

In the numerical examples we will choose a charged-Higgs mass of $m_{H^\pm}=200 \,\gev$ and the right-handed neutrino mass of $m_N = 200 \, \gev$. 
As the $Z$ boson couples strongly to left-handed (right-handed) electron (positron) compared to right-handed electron, left polarized electron, and right polarized positron beam enhances the $H^+ H^-$ cross-section. We study the effect of beam polarization by comparing the results for unpolarized beams to those with $ P_{e^\pm} = ^{+30\%}_{-80\%} $ and $ P_{e^\pm} = ^{+10\%}_{-50\%} $. The plots of the total cross section as a function of the CM energy are presented in \cref{fig:hhprod}.

\subsection{Collider analysis of the inert doublet model}
\label{sec:collider-model}

In this section we estimate the efficiency factor $\epsilon$ for $ H^+H^-$ production within the inert doublet model (IDM). This model is one of the simplest extension 
of the SM where the scalar sector is assumed to include an additional doublet $H_2$, and an unbroken discrete $\zBB_2$  symmetry under which $H_2$ is odd 
whereas all other fields are even. This discrete symmetry forbids Yukawa interactions between the inert doublet $H_2$ and SM fermions and ensures that the 
lightest physical component of $H_2$ can serve as a Dark Matter (DM) candidate. The Lagrangian consisting the scalar dark sector can be written as 
\beq
\lcal= \lcal_{\tt SM} + |D_{\mu}H_2|^2 - V(H_1,H_2),
\eeq
where,
\bea
V(H_1, H_2) &=& -\mu_H^2(H_1^{\dagger}H_1)+\lambda_{H}(H_1^{\dagger}H_1)^2+\mu^2 (H_2^{\dagger}H_2) + \lambda (H_2^{\dagger}H_2)^2 
+ \lambda_1 (H_1^{\dagger}H_1) (H_2^{\dagger}H_2) \nonumber \\ 
&+& \lambda_2 (H_1^{\dagger}H_2)(H_2^{\dagger}H_1) + \frac{\lambda_3}{2} [(H_1^{\dagger}H_2)^2 + \text{H.c.}] .
\eea

We assume $\mu^2 > 0$  that ensures that the \vev\ of the inert doublet vanishes and guarantees that $\zBB_2$ remains unbroken. In contrast the SM scalar doublet $H_1$ does acquire a \vev\ $v$. The physical modes consist of a  singly-charged scalar $ H^\pm$, a CP-odd neutral scalar $A^0$, and CP-even neutrals $h,\,H^0$, where $h$ is the physical SM scalar field. The corresponding tree-level masses are given by
\bal
m_{H^0}^2 &= \mu^2 + \lambda_L v^2, \cr
m_{H^\pm}^2 &= \mu^2 + \half\lambda_1 v^2, \cr
m_{A^0}^2 &= \mu^2 + \half (\lambda_1 + \lambda_2 - 2 \lambda_3) v^2,
\end{align}
where $\lambda_L=\frac{1}{2}(\lambda_1+\lambda_2+\lambda_3)$.

\begin{figure}[htb!]
	$$
	\includegraphics[scale=0.35]{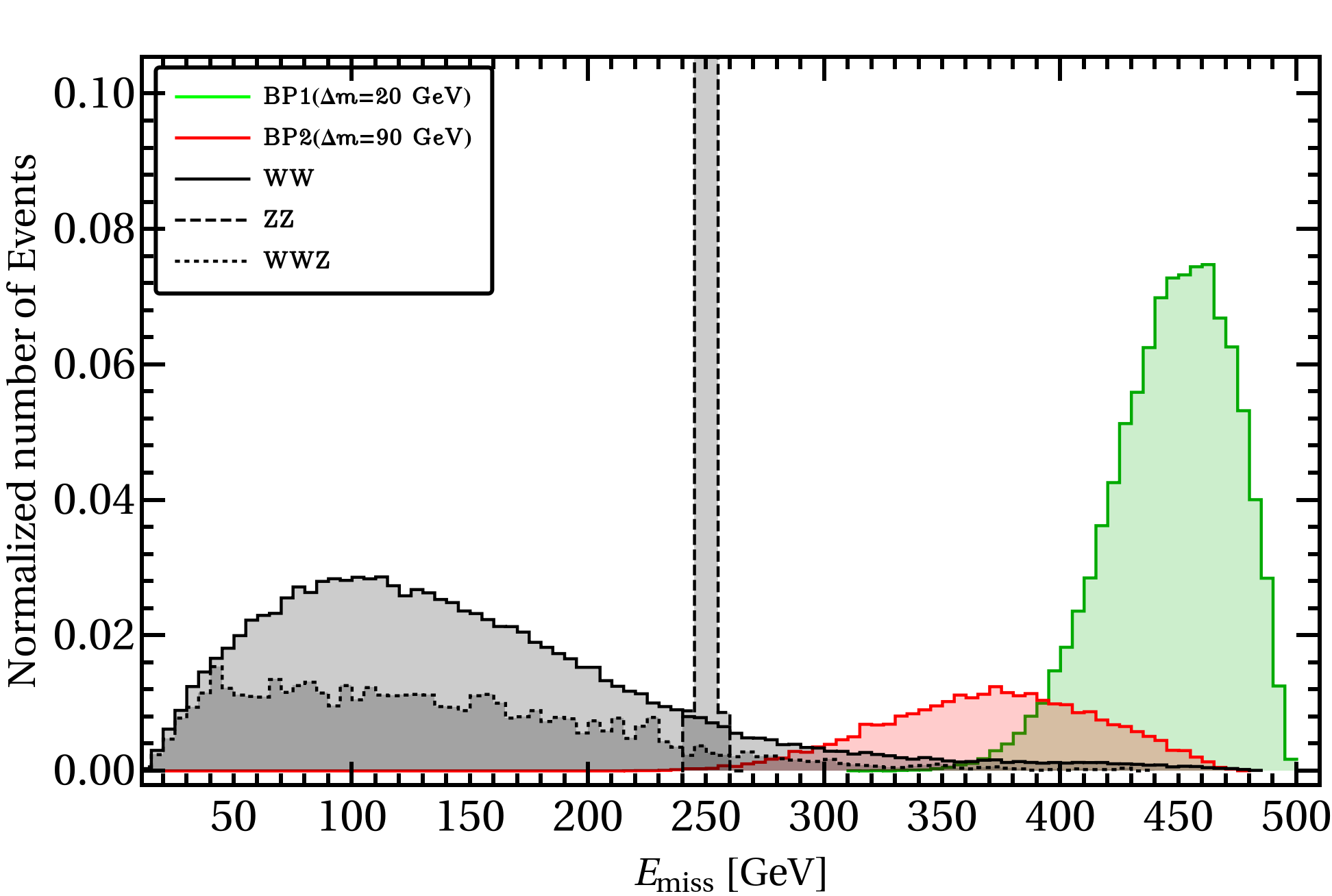}
	$$
	\caption{\small Normalized missing energy distribution of OSL + missing energy signal for IDM at the $e^+\,e^-$ collider with $\sqrt{s}=500$ GeV and unpolarized beams.}
	\label{fig:histidm}
\end{figure}

The signal we will use to identify the production of an $H^+ H^-$ pair will be two opposite-sign $e$ or $\mu $ leptons (OSL) plus missing energy ($\slashed E$). 
The decay chain can be written as
\beq
e^+ e^- \to H^+ H^-, \, \quad H^{\pm} \to H^0 W^\pm, \, \quad W^- \to l^- \bar\nu_l\,,~ W^+ \to l^+ \nu_l\,;
\eeq
where we focus on the leptonic decay modes of $W$. The dominant (non-interfering) SM background arises from $W^+W^-$, $ZZ$ and $W^+W^-Z$ production. The analysis of $H^+H^-$ production in an $ e^+e^-$ collider for the IDM has been studied in detail in several publications, see, {\it e.g.}, \cite{Aoki:2013lhm,Hashemi:2015swh,Kalinowski:2018ylg,Kalinowski:2018kdn}.

\begin{table}[htb!]
	\centering
		{\renewcommand{\arraystretch}{1.4}%
	\begin{tabular}{|l|c|c | c|c|c|c|c|c|c|c|}
		\hline
		\multicolumn{1}{|c}{\multirow{2}*{BPs}} &
		\multicolumn{1}{|c}{\multirow{2}*{$\Delta m$}} &
		\multicolumn{1}{|c}{\multirow{2}*{$m_{H^\pm}$}}&
		\multicolumn{2}{|c|}{Cross-section (fb)} \\
		\cline{4-5}
		\multicolumn{1}{|c }{} &
		\multicolumn{1}{|c|}{} &
		\multicolumn{1}{ c|}{} &
		\multicolumn{1}{ c|}{$P_{e^\pm}=0$} &
		\multicolumn{1}{ c|}{$P_{e^\pm}=_{-80\%}^{+30\%}$} \\
		\hline
		BP1 & 20 & 90 & 101.2  & 201.2    \\
		BP2 & 90 & 160 & 56.5 & 112.4   \\
		\hline
	\end{tabular}}
	\caption{\small Total cross-section of charged scalar pair-production ($H^+\,H^-$) for  different benchmark points for unpolarized and polarized ($P_{e^\pm}=_{-80\%}^{+30\%}$) beams. }
	\label{tab:BP}
\end{table}

We use the criteria for events reconstruction as in section \ref{sec:eft.col} in order to reduce the SM background we impose the following cuts:
\begin{itemize}
\item $\ccal_1$: events must contain two opposite sign leptons in the final state.
\item $\ccal_2$: missing energy $\slashed{E} \le 370 \, (300)$ for BP1 (BP2).
\end{itemize}   
It follows form \cref{fig:histidm} that $ \ccal_{1,2}$ strongly reduce the (non-interfering) SM backgrounds. We also chose charged scalar masses equal to $m_{H^\pm}$ = $90 \, \rm{GeV} \, \rm{and} \, 160 \, \rm{GeV}$ in view of the latest bounds \cite{Pierce:2007ut}.  Using the the definition of $\epsilon$ in \cref{eq:eps}, the efficiency factor for the two benchmark points is tabulated in \cref{tab:idm}. These results justify the conservative choice $ \epsilon=0.005$ we used above for all the different NP scenarios.
\begin{table}[htb!]
\centering
	{\renewcommand{\arraystretch}{1.4}%
\begin{tabular}{|l|c|c|c|c|c|c|c|c|c|c|c|c|r|}
	\hline
	\multicolumn{1}{|c}{} &
	\multicolumn{5}{|c|}{Cross-section} &
	\multicolumn{2}{c|}{\multirow{2}*{Efficiency factor ($\epsilon$)}} \\
	\cline{2-6}
	\multicolumn{1}{|c}{} &
	\multicolumn{2}{|c}{Signal} &
	\multicolumn{3}{|c|}{SM background} &
	\multicolumn{2}{c|}{}\\
	\cline{2-8}
	Cuts &BP1 & BP2 &  $W^+\,W^-$ & $ZZ$ & $W^+ \, W^- \, Z$ & \multirow{2}*{$\epsilon^{\tt BP1}$} & \multirow{2}*{$\epsilon^{\tt BP2}$}  \\
	& (fb) & (fb) & (fb) & (fb) & (fb) & &   \\
	\hline\hline
	$\mathcal{C}_1$ & 2.02 & 1.13 & 211.4 & 11.2 & 0.8 &0.02 & 0.02  \\
	$\mathcal{C}_2$ & 1.96 & 0.85 & 4.01 & 0.0 & 0.002 &0.019 & 0.015 \\
	\hline
\end{tabular}}
\caption{\small Event cross-section and efficiency factor ($\epsilon$) for two different benchmark points of IDM and corresponding SM background estimation background estimation for unpolarized beams with CM energy ($\sqrt{s}$) = 500 GeV.}
\label{tab:idm}
\end{table}
\subsection{$1\sigma$ surfaces in the $a-b$ plane}
\label{sec:onesig}

\begin{figure}[htb!]
	\begin{align*}
		\includegraphics[scale=0.22]{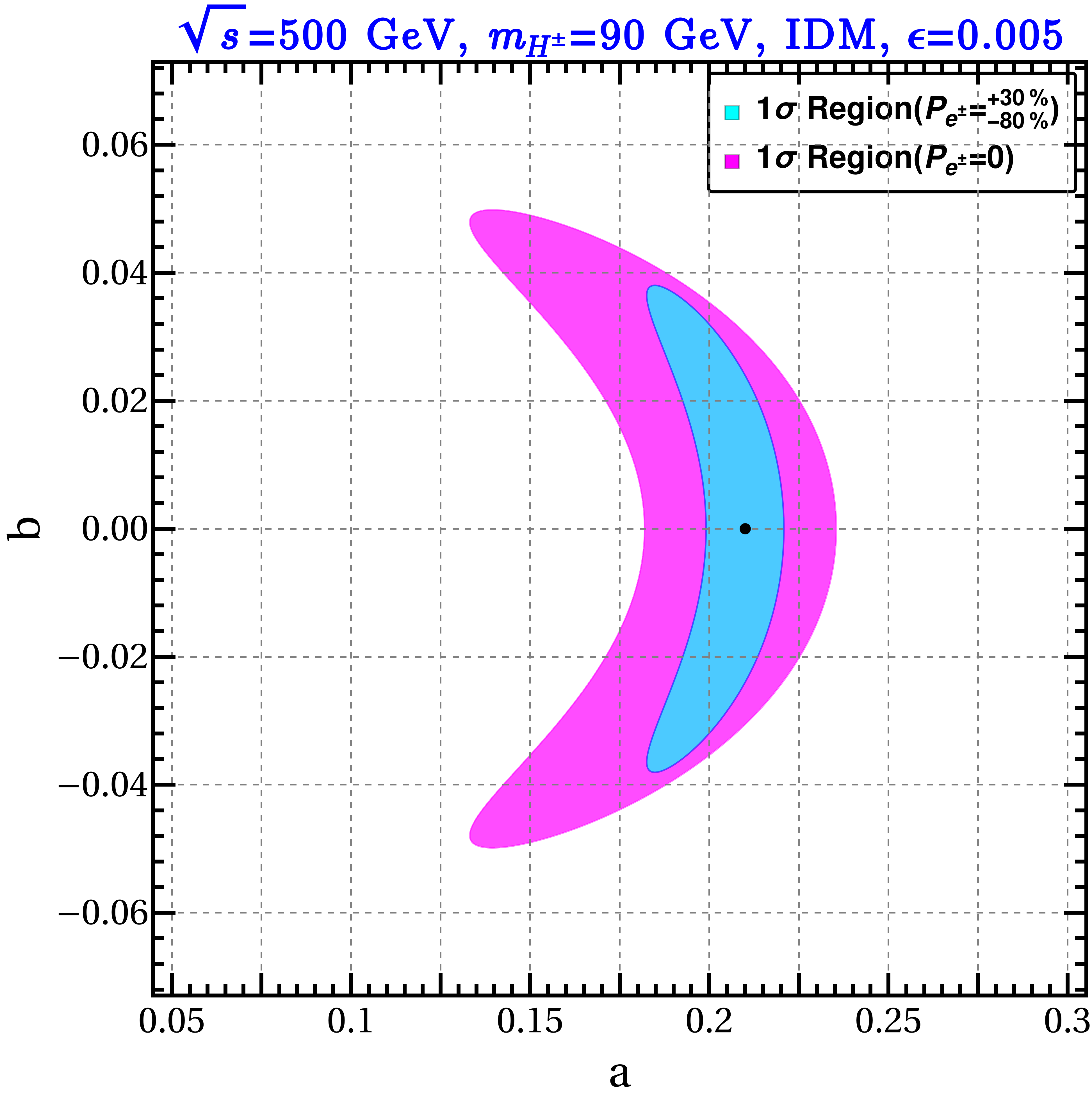} \quad&\quad \includegraphics[scale=0.22]{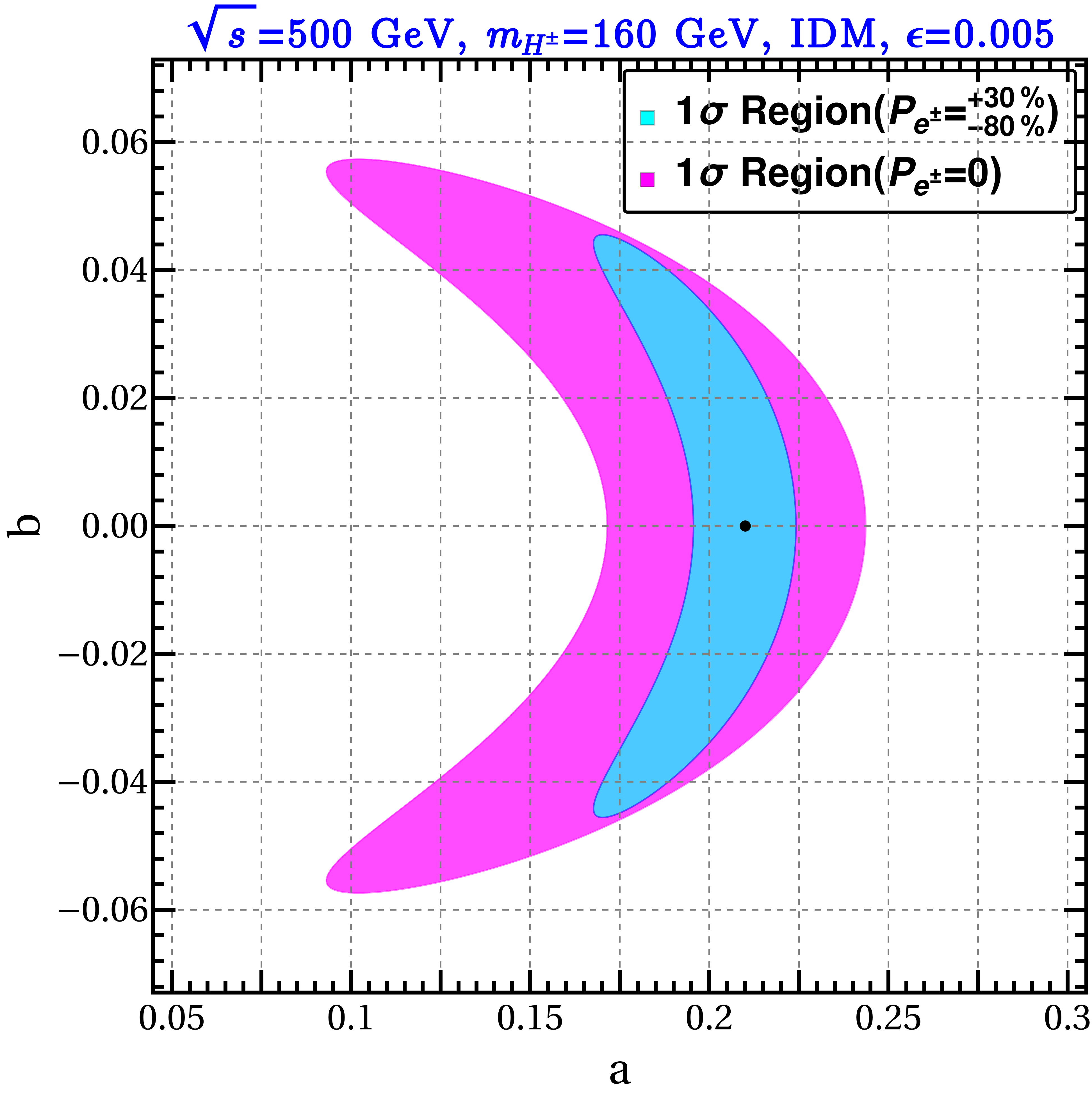} \cr
		\includegraphics[scale=0.22]{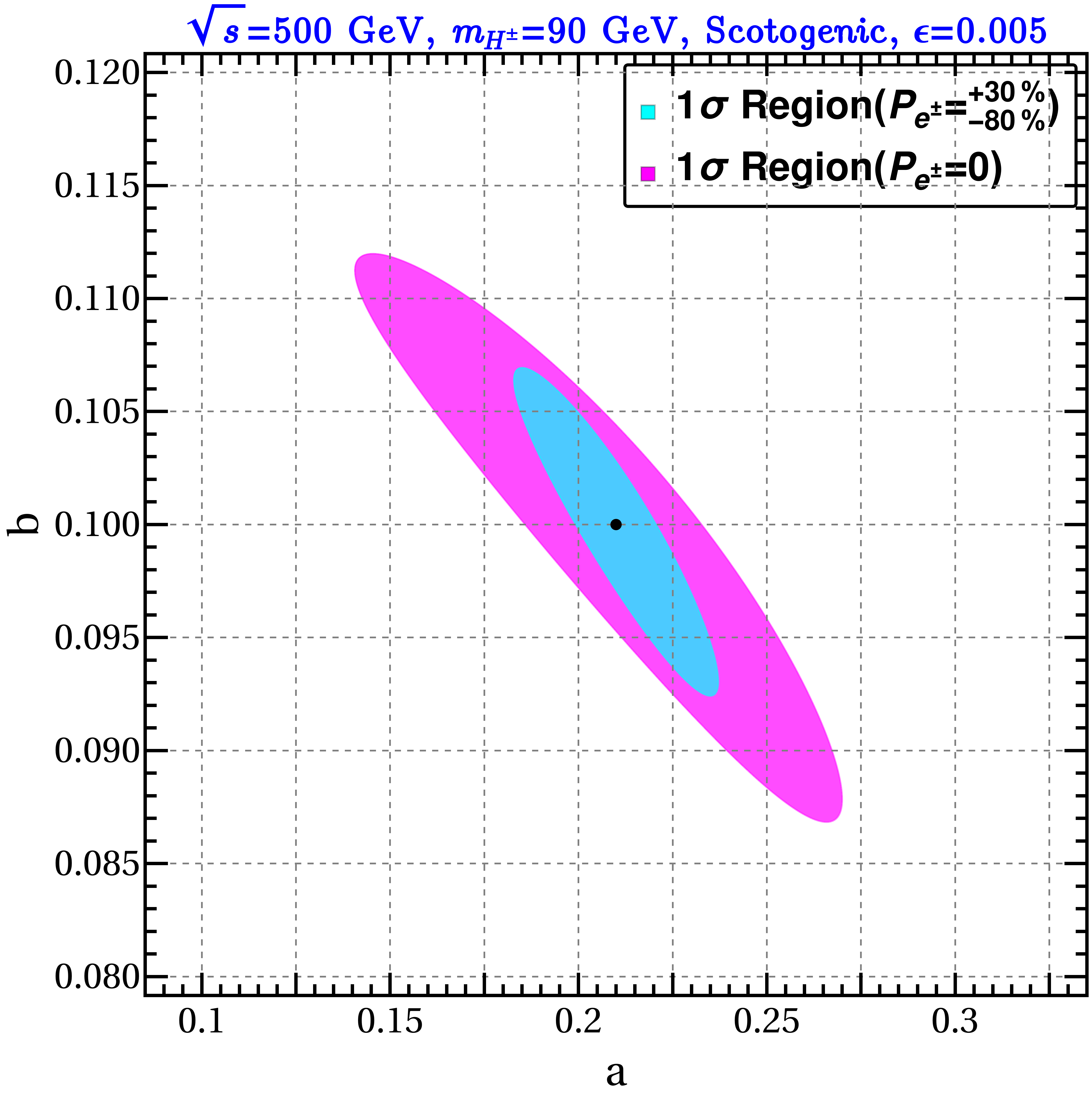} \quad&\quad \includegraphics[scale=0.22]{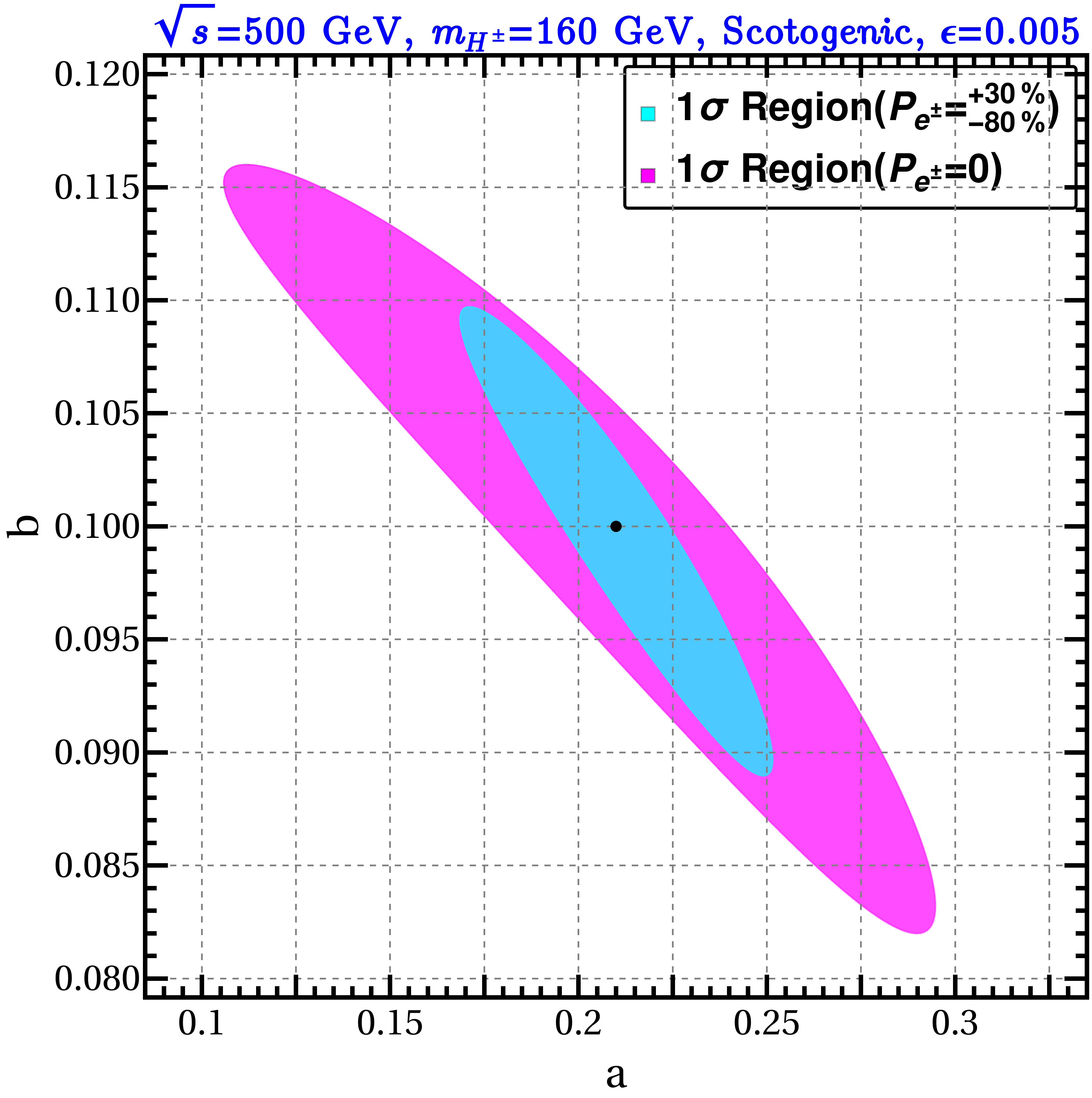} \cr
		\includegraphics[scale=0.22]{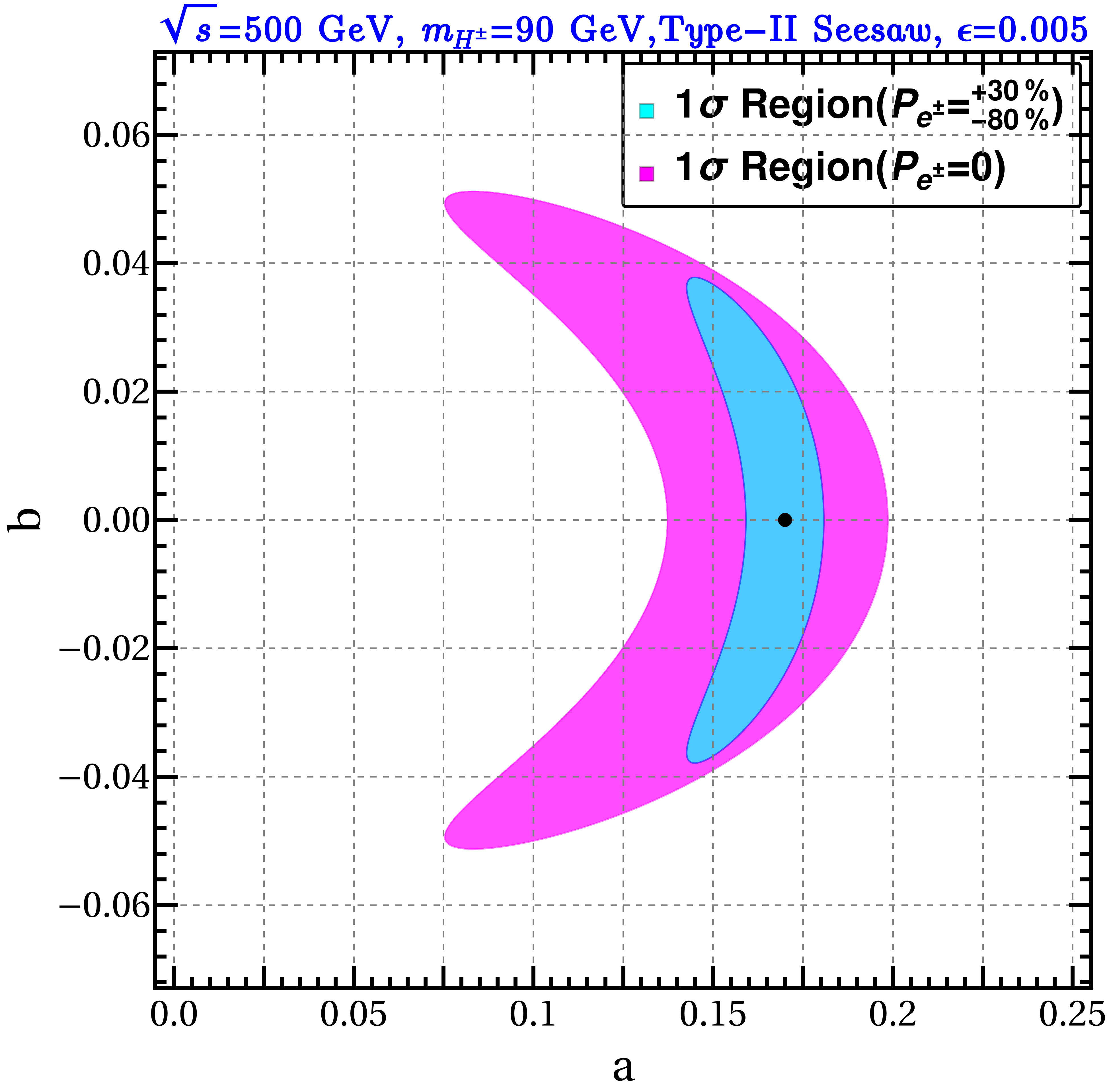} \quad&\quad  \includegraphics[scale=0.22]{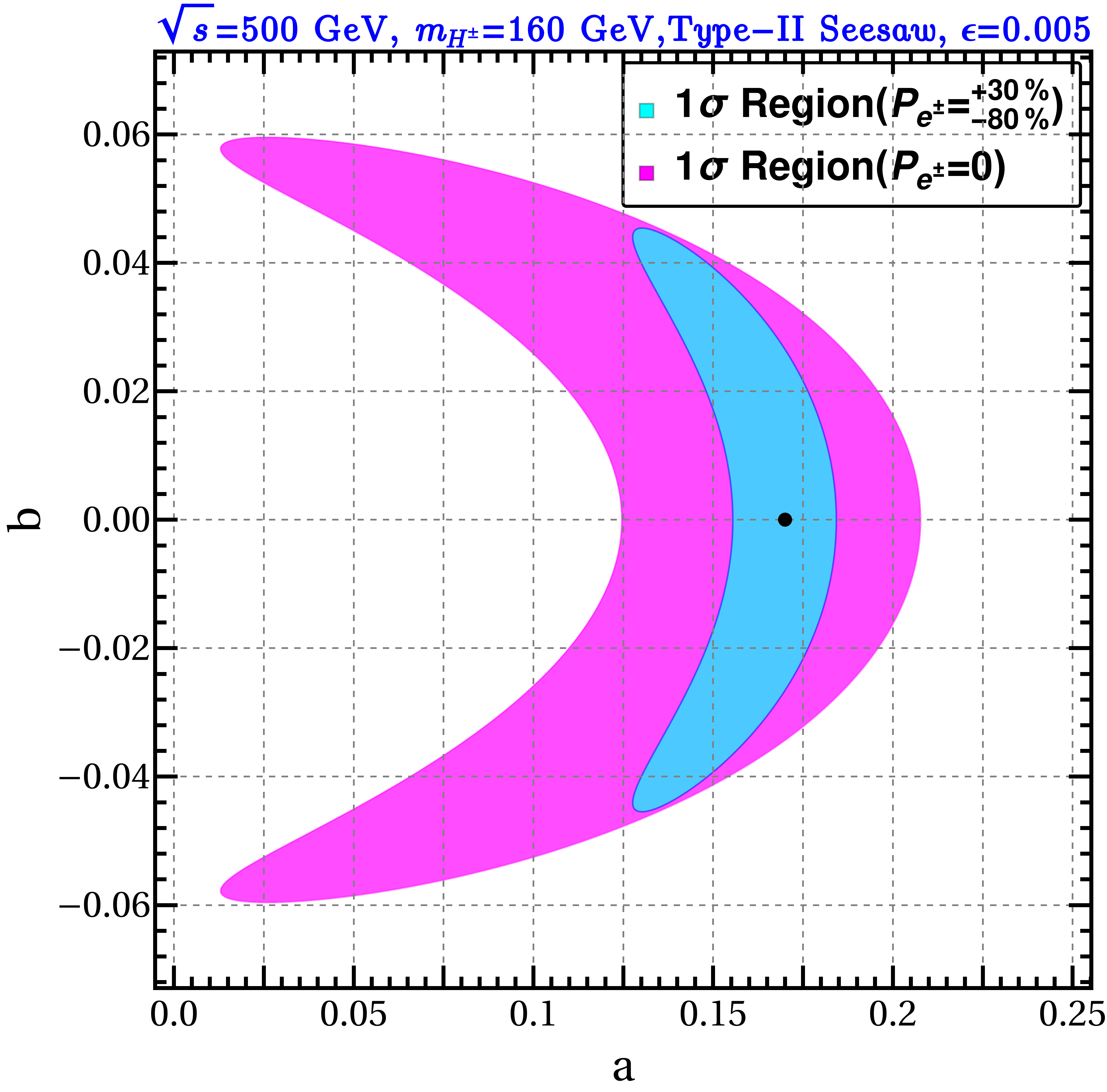} 
	\end{align*}
	\caption{ Optimal 1$\sigma$ surfaces for different models with unpolarized and polarized  beams $P_{e^\pm}=^{+30\%}_{-80\%} $,  $ \epsilon = 0.005$, and  $ m_{H^\pm} = 90 \, \gev$ (left column) or $ m_{H^\pm}= 160 \, \gev$ (right column).}	
	\label{fig:1sighh}
\end{figure}

We now apply the OOT to obtain the the optimal statistical uncertainty regions for the NP couplings $(a,\,b)$; the results are presented in \cref{tab:1sighh} and the  corresponding 1$\sigma$ 
regions are shown in \cref{fig:1sighh} with CM energy ($\sqrt{s}$) = 500 GeV, luminosity ($\mathcal{L}_{int}$) = 1000 $\rm{fb}^{-1}$ and signal efficiency ($\epsilon$) = 0.005. For the IDM and 
type II models the uncertainties in NP couplings are similar since the cross sections are almost equal. For the scotogenic model where $b\not=0$, the t-channel diagram contributes and 
enhances the cross section, resulting in an increased sensitivity to $b$. In table~\ref{tab:1sighh}, 1$\sigma$ uncertainties of NP couplings for two different charged Higgs masses 
($m_{H^\pm}=90,160$ GeV) are listed. For $m_{H^\pm}$ =160 GeV, due to smaller production cross-section, the uncertainties of NP couplings are increased by 
$\sim $30\% compared to the case of $m_{H^\pm}=90$ GeV. We can also see that a judicious choice of 
polarization also enhances the cross section, leading to a reduced statistical uncertainty in the determination of the NP coefficients. For $ P_{e^\pm}=^{+30\%}_{-80\%}$ the
uncertainty in the parameter $a~(b)$ is  reduced by $\sim$ 50-55\% ($\sim$25-40\%).
Finally, as a function of the luminosity $ \lcal$ and the efficiency $ \epsilon $, the uncertainties scale as $ 1/(\sqrt{\lcal_{int} \epsilon} )$ with the expected result that a larger luminosity 
and/or efficiency also lead to reduction of the uncertainties.

\begin{table}[htb!]
\centering
{\renewcommand{\arraystretch}{1.4}%
\begin{tabular}{| c | c | c  c | c  c | c  c |  c c | } 
	\cline{1-6}
	\multicolumn{2}{|c|}{\multirow{2}*{}} &
	\multicolumn{4}{c|}{Uncertainties} \\
	\cline{3-6}
	\multicolumn{2}{|c|}{} &
	\multicolumn{2}{c}{$P_{e^\pm}=0$} &
	\multicolumn{2}{|c|}{$P_{e^\pm}=^{+30\%}_{-80\%}$} \\
	\cline{1-6}
	\bf Model &$m_{H^\pm}$(GeV) & $\pm\Delta a$ & $\pm\Delta b$ & $\pm\Delta a$ & $\pm\Delta b$ \\ 
	\cline{1-6}
	\multirow{4}*{\minitab[l]{Inert doublet}}
	& \multirow{2}*{90}$ $ & $+0.025$ & $+0.050$ &$+0.011$&$+0.038$& \\ 
	& &$-0.077$ & $-0.050$ & $-0.027$ & $-0.038$  \\ 
	\cline{2-6}
	&\multirow{2}*{160} & $+0.033$ & $+0.058$ & $+0.014$ & $+0.046$ \\
	&  &$-0.117$ & $-0.058$  &$-0.042$ & $-0.046$\\
	\cline{1-6}
	\multirow{4}*{\minitab[l]{Scotogenic}}
	&  \multirow{2}*{90}  &$+0.060$ & $+0.012$ & $+0.028$ & $+0.007$\\
	& &$-0.070$ & $-0.012$ & $-0.028$ & $-0.007$\\
	\cline{2-6}
	& \multirow{2}*{160} &$+0.085$ & $+0.016$ & $+0.041$ & $+0.010$ \\
	& &$-0.105$ & $-0.016$ & $-0.042$ & $-0.010$\\
	\cline{1-6}
	\multirow{4}*{\minitab[l]{Type-II \\ Seesaw}}
	&  \multirow{2}*{90}   &$+0.028$ & $+0.050$ & $+0.017$ & $+0.038$ \\
	&  &$-0.095$ & $-0.050$ & $-0.028$ & $-0.038$\\
	\cline{2-6}
	&\multirow{2}*{160}  &$+0.037$ & $+0.058$ & $+0.015$ & $+0.046$\\
	& & $-0.157$ & $-0.058$  &$-0.043$ & $-0.046$\\   
	\cline{1-6}
\end{tabular}}
\caption{\small Optimal 1$\sigma$ statistical uncertainty in the $a,b$ couplings for both unpolarized and polarized $P_{e^\pm}=^{+30\%}_{-80\%}$ beams with $ \epsilon=0.005$.}
\label{tab:1sighh}
\end{table}

\subsection{Differentiation of models}
\label{sec:modeldiff}

We can now follow the same approach as in \cref{sec:hyp.dif} and use the OOT to estimate the extent to which different hypotheses can be distinguished. Specifically, 
given a ``base'' and alternate hypotheses, $ \{a^0,\,b^0\}$ and $\{\bar a,\,\bar b\}$, respectively, we define the significance as in \cref{eq:signif-def} (where now
$ g^0_i= g_i(a^0,\,b^0),\,\bar g_i= g_i(\bar a,\,\bar b)$) and again assume that these hypotheses can be distinguished at the $ \ge \ell \sigma $ level if  $ \Delta \sigma \ge \ell $.

\begin{figure}[htb!]
	$$
	\includegraphics[scale=0.23]{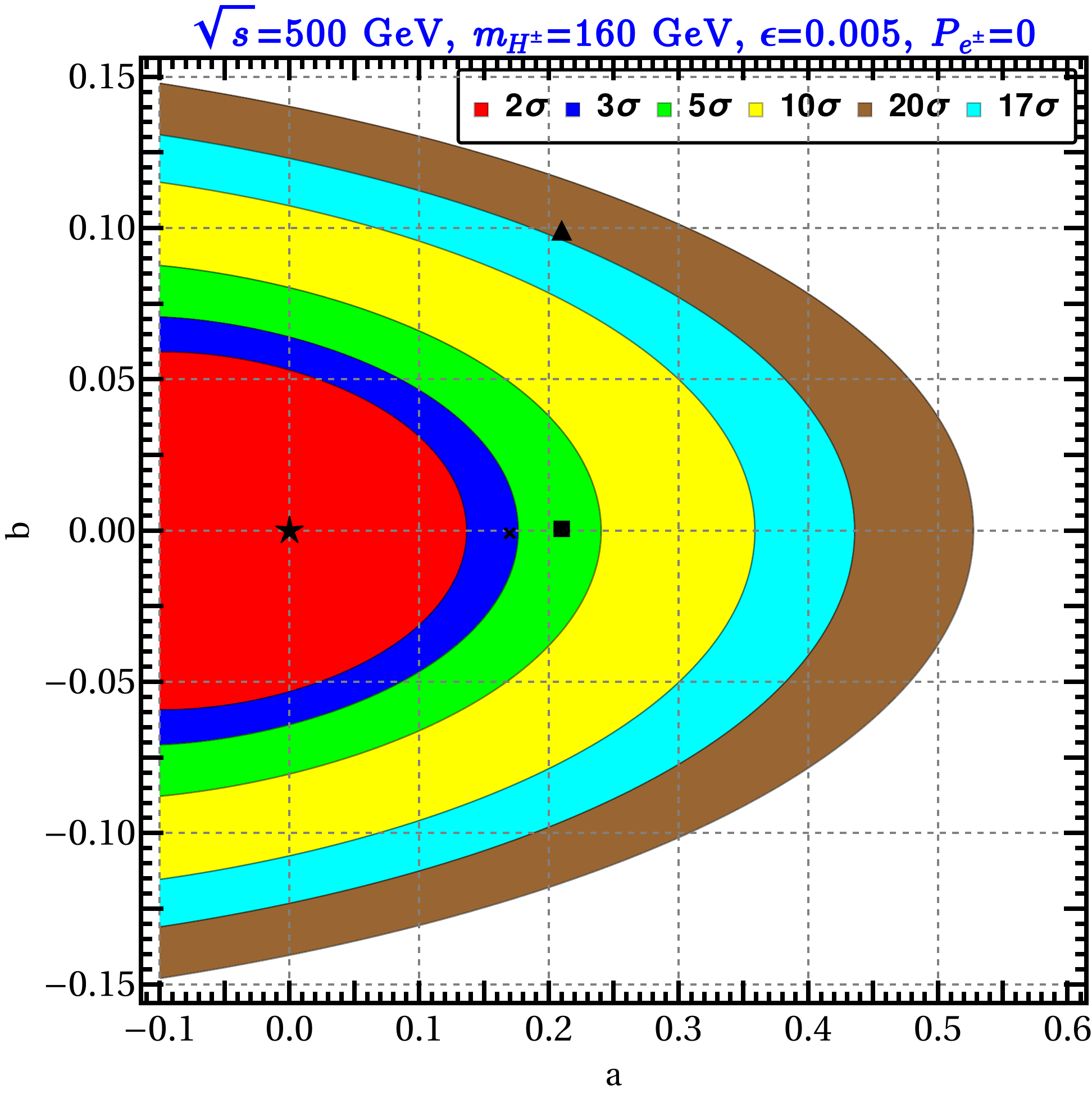} 
	\includegraphics[scale=0.23]{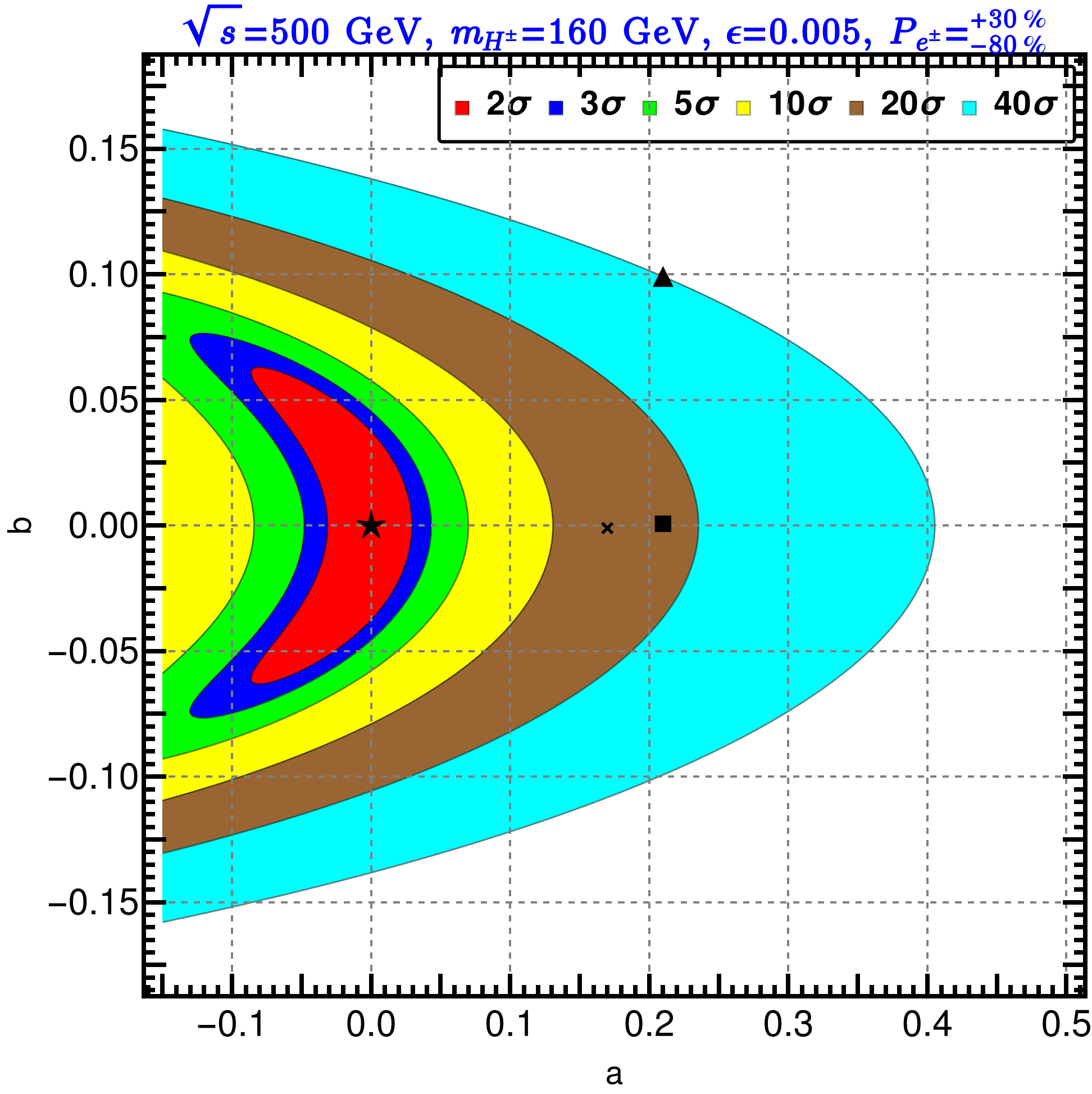} 
	$$
	\caption{\small Statistical significance of alternate models with respect to the $a^0=b^0=0$ model ({\it cf.} \cref{eq:signif-def}), $\Delta\sigma \le \ell$, for various choices of $ \ell$. Cross, square and triangle correspond to the IDM, type-II seesaw and scotogenic models, respectively. Left: unpolarized beams, right: polarized beams $ P_{e^\pm} =^{+30\%}_{-80\%} $.}
	\label{fig:Dscvry160GeV}
\end{figure}

Taking $ a^0=b^0=0$ (similar to the SM) as a base hypothesis, we determine the statistical separation of the models listed above. For CM energy $\sqrt{s} = 500 $ GeV,  $\lcal_{int} = 1000  \  \text{fb}^{-1} $, $m_{H^{\pm}}=160$ GeV,
and $ \epsilon=0.005$, the separation significance are listed in  \cref{tab:sigh} and corresponding plots are shown in \cref{fig:Dscvry160GeV}. 
For unpolarized beams, we can see that for both inert doublet and type-II seesaw models are under discovery limit (that is, $\Delta\sigma < 5 $), but the scotogenic model is above 
this limit  due to the enhancement of the cross-section by the  t-channel $N_R$ contribution. Polarized beams ($P_{e^\pm}=^{+30\%}_{-80\%}$) enhances the production cross-section,  
which in turn provides a clear distinction (above 5$\sigma$) of the three different models from the base hypothesis. Also note that with larger luminosity, total number of events increases 
to provide higher significance.

\begin{table}[htb!]
	\centering
	{\renewcommand{\arraystretch}{1.4}%
	\begin{tabular}{|l|c|c | c|c|c|c|c|}
		\hline
		\multicolumn{1}{|c}{\multirow{2}*{Models}} &
		\multicolumn{2}{|c|}{significance ($\Delta \sigma$)} \\
		\cline{2-3}
		\multicolumn{1}{|c}{\multirow{2}*{}}&
		\multicolumn{1}{| c}{$P_{e^\pm}=0$} &
		\multicolumn{1}{| c|}{$P_{e^\pm}=_{-80\%}^{+30\%}$} \\
		\hline
		Inert doublet  & 4$\sigma$  & 17$\sigma$ \\
		Type-II Seesaw & 2.82$\sigma$ & 13$\sigma$   \\
		Scotogenic & 17$\sigma$ & 40$\sigma$   \\
		\hline
	\end{tabular}}
	\caption{\small Statistical significance of three different models with respect to the $a^0=b^0=0$ model.}
	\label{tab:sigh}
\end{table}

\subsection{Comparison between optimal and standard $\chi^2$}
\label{sec:comp.lim}
The statistical uncertainties obtained using the OOT can be compared to those obtained using a basic collider analysis based on \cref{eq:chi2.cut}, as was done in \cref{sec:cut}.
Here also we consider the differential cross-section for $H^+H^-$ pair production  after subsequent decays (with the same cuts as in 
	section \ref{sec:collider-model}) to OSL + missing energy for the binned analysis.
\begin{figure}[htb!]
	$$
	\includegraphics[height=5.4cm, width=5.3cm]{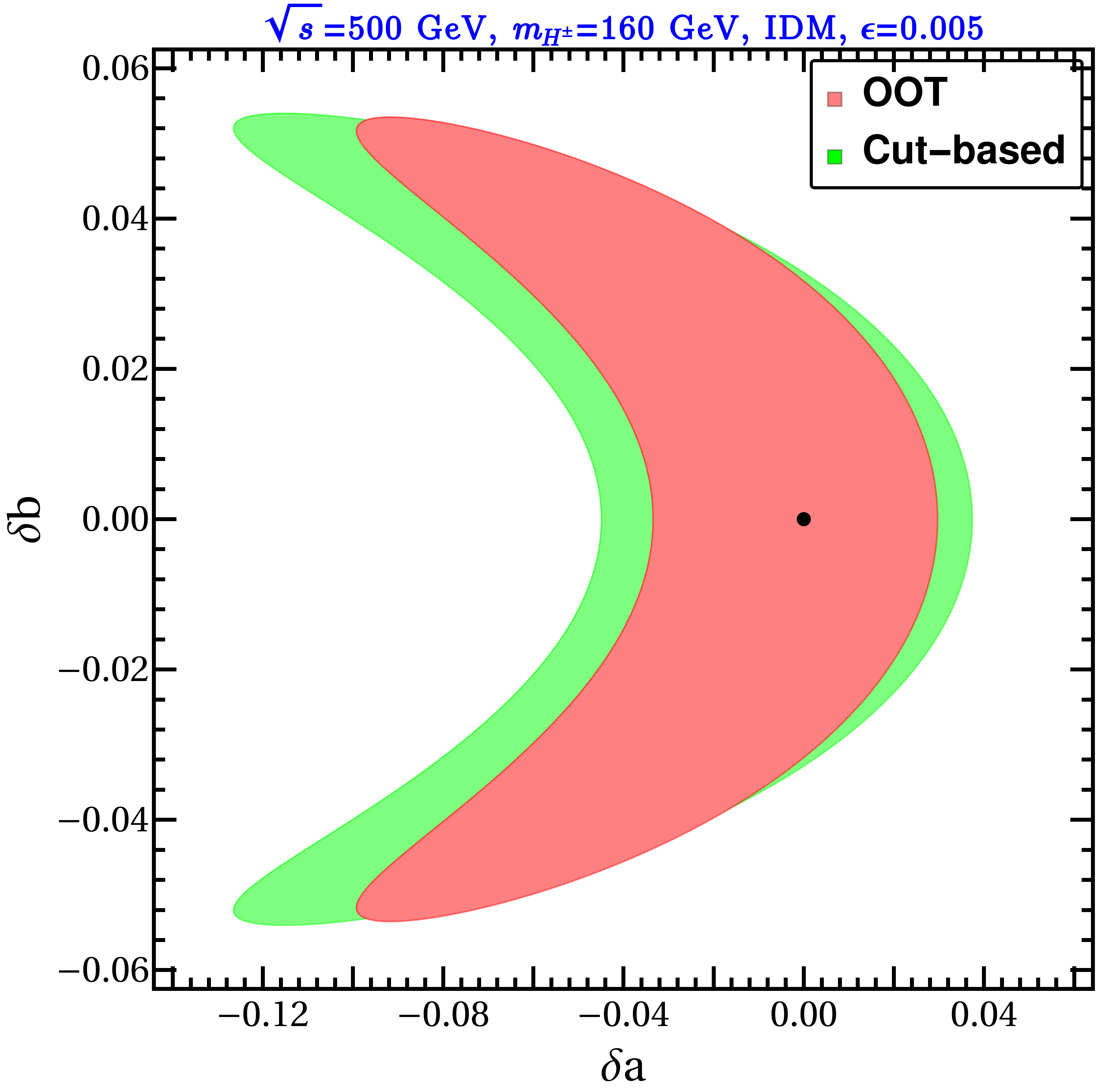}
	\includegraphics[height=5.4cm, width=5.3cm]{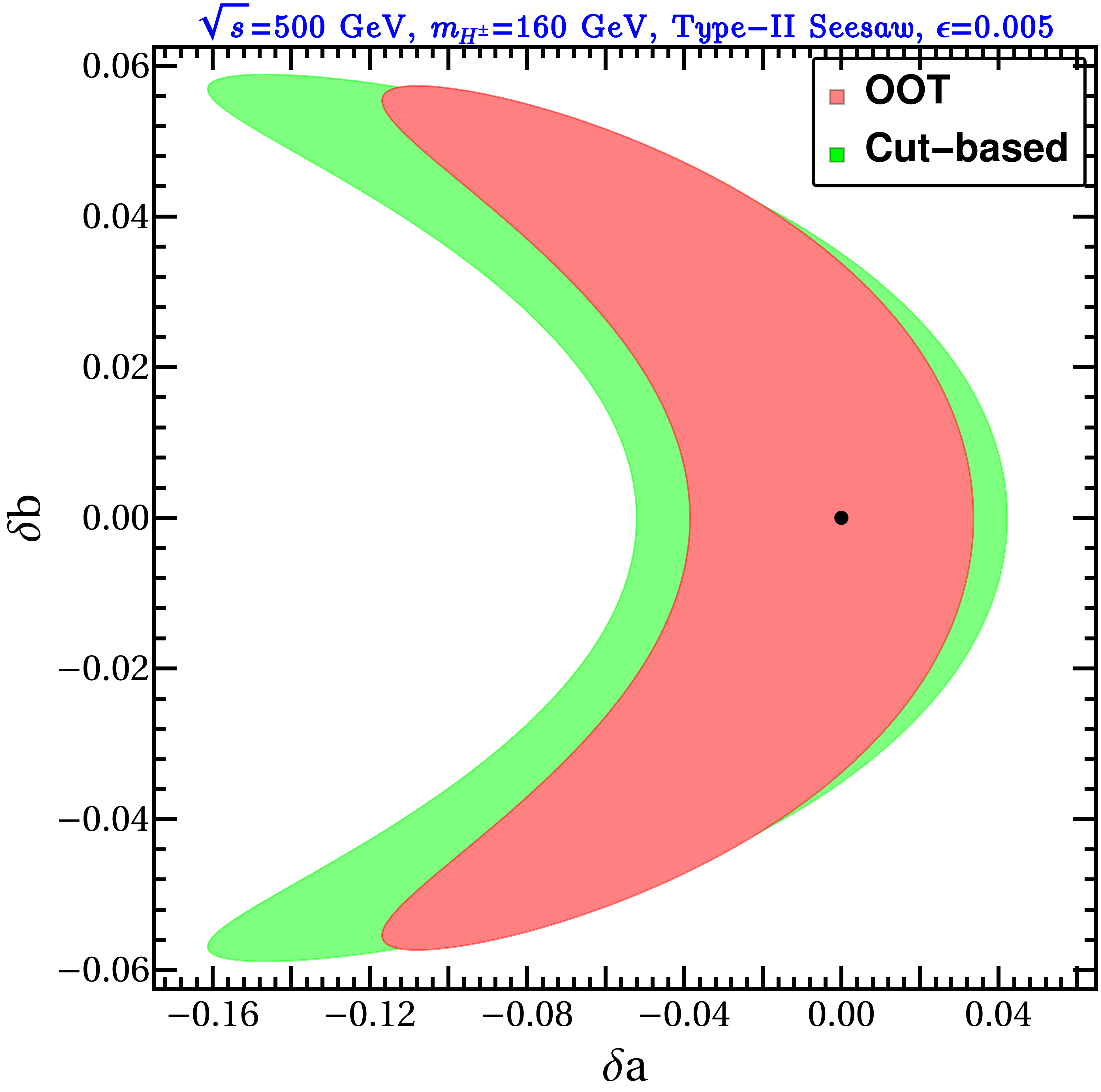}
	\includegraphics[height=5.4cm, width=5.3cm]{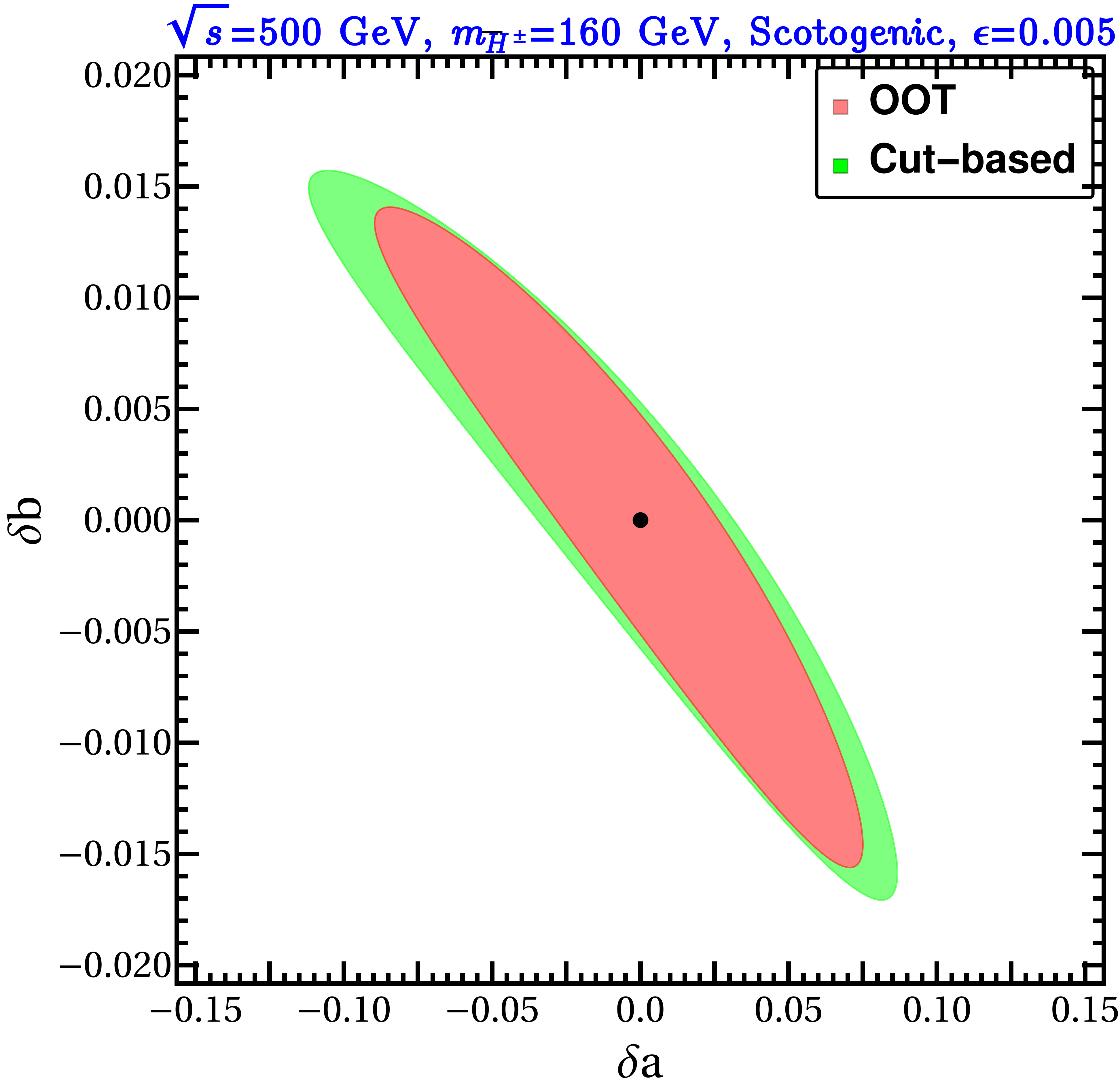}
	$$
	\caption{\small Comparison of 1$\sigma$ statistical limit for OOT (pink) and collider (green) $\chi^2$ in $a-b$ plane for charged scalar pair production at 
		$e^+e^-$ colliders for unpolarized beam. All the relevant parameters are written in the caption. Left: IDM;  middle: type-II seesaw;  right: scotogenic model 
		(see \cref{eq:2hdm.models}).}
	\label{fig:comp.lim}
\end{figure}
The resulting 1$\sigma$ regions in the $a-b$ plane for three different models are shown 
in fig.~\ref{fig:comp.lim} for both the binned and OOT analyses (the collider parameters as in section \ref{sec:collider-model}). The 1$\sigma$ contours for binned analysis (OOT) are  shown by green (pink) color contours. In this scenario, with purely NP productions, the cross-section is larger, statistical fluctuation in each bin is less, making the NP uncertainties determined through binned analysis close to OOT.


\section{Summary and Conclusions}
\label{sec:summary}
The OOT determines the smallest statistical uncertainty with which a NP coupling can be extracted from experiment. In this work we studied
 two limits of this technique, one in SMEFT ($ t \bar t $ production in an $ e^+e^-$ collider) where the SM dominates over the NP contribution, 
 and one in UV complete NP models ($ H^+H^- $ production in an $ e^+e^-$ collider) where SM contribution is subdominant.

For the first application we used an effective Lagrangian parameterization of the NP effects, including for simplicity only those operators that do not interfere with the SM 
contributions in this process. We find that for realistic collider parameters the $1\sigma$ statistical uncertainty of the NP parameters lie in the 20\%\ to 100\%\ range 
(depending on the values of the coefficients, the beam polarizations and the efficiency of signal background estimation). The possibility of distinguishing different NP models 
(defined by their values of the NP coefficients) is equally modest, with a significance below $5\sigma $ in all cases studied. 

In the second application the NP particles are assumed to be light enough to be directly produced. Here we find that the sensitivity is much higher 
so that the NP parameters could be measured to a precision of 1\%\ to 10\%\ (depending on the values of the coefficients, the beam polarizations and the 
efficiency). Moreover, different models can be easily distinguished, provided the beams are strongly polarized, which provides a useful comparison of how 
efficiently the NP couplings can be extracted at the proposed future $e^+\,e^-$ colliders. The calculations for both applications require the estimation of an 
efficiency factor $ \epsilon $ which we obtain by performing standard collider analyses of the corresponding reactions by studying a cut based signal background 
analysis.


The statistical uncertainties in the NP coefficients obtained using the OOT are $ O(0.5)$ in the first case we considered (\cref{sec:ttb}), white significantly smaller, $O(0.05)$, for the second example (\cref{sec:model}). This is imply due to the different values of the NP contributions to the cross sections in each case: in the first the NP effects are small, this leads to a relatively small inverse-correlation matrix $M$ in \cref{eq:V.M}, while in the second example NP effects dominate leading ot a larger $M$ and correspondingly smaller uncertainties.

We also compared the OOT results with those obtained a simple collider estimate of the parameter uncertainties based on the $ \chi^2$ statistic of \cref{eq:chi2.cut}. We found that the optimal uncertainties are significantly smaller in the case where the NP effects are subdominant, but that in the case where the NP dominates the results are comparable. The collider analysis, however, can be improved by optimizing the data binning, and possibly by selecting a better suited statistic; such investigations, however, lie outside the scope of this paper.

\acknowledgments

SB would like to acknowledge the grant CRG/2019/004078 from DST-SERB, Govt. of India. SJ thanks Jayita Lahiri for useful discussions. 

\appendix 

\section{OOT for large NP}
\label{sec:app}

In this appendix we provide a derivation of the correlation matrix that determines the optimal statistical uncertainties of the model parameters. We interpret the differential cross section $d\sigma/d\phi $ (where $ \phi$ denote the relevant phase-space coordinates) as an unnormalized probability density function; its normalized counterpart is then
\beq
\ft  = \inv\Nt \frac{d\sigma }{d\phi}\,; \quad \Nt= \vevof{\frac{d\sigma}{d\phi}}\,,
\eeq
where, for any $X(\phi)$,
\beq
\vevof X = \int d\phi \, X(\phi)\,.
\eeq

Denoting by $ \lum $  the integrated luminosity over a period, the  average event number in that period is
\beq
\er = \lum \Nt\,.
\eeq
Assuming the events occur at a constant rate and are independent of one another~\footnote{Situations where this may not be true usually occur when the number of events is very large, so that effects such as Bose enhancement of Pauli blocking become important.}, the probability of observing $A$ events is given by a Poisson distribution
\beq
\frac{\er^A}{A!} e^{-\er}\,.
\eeq
Therefore the probability of observing $A$ events in regions $ \Delta \phi_a$ around phase-space points $ \phi_a,\,a=1,\ldots, A $ is

\bal
& \FFt(A;\xibf_A) \, \Delta\xibf_A = \frac{\er^A}{A!} e^{-\er} \prod_{a=1}^A \ft (\phi_a) \Delta\phi_a\,; \quad \xibf_A=(\phi_1,\ldots,\phi_A)\,, ~~ \Delta\xibf_A = \prod_{a=1}^A \Delta\phi_a \,,\mcr
&\then \FFt(A;\xibf_A)   = \frac{e^{-\er} }{A!} \prod_{a=1}^A \fft(\phi_a)\,; \quad \fft = \ert \ft = \lum \frac{d\sigma_{\tt theo}}{d\phi} \,. 
\end{align}

Now, for all cases of interest the cross-section can be written in form
\beq
\frac{d\sigma_{\tt theo}}{d\phi} =  \Gg\cdot\Ff(\phi)\,,
\eeq
where $\Gg=(g_1,\,\ldots,\, g_n)$ are a set of parameters to be determined through observations, and $\Ff$ a set of convenient linearly-independent phase-space functions~\footnote{The choice of $ f_i $ is not unique, but all observable results are unambiguous \cite{Diehl:1993br}.} The coefficients $ g_i $ will be  (not necessarily linear) functions of the Lagrangian parameters and, for processes that receive SM contributions, some or all of them will depend on the SM parameters. We make no additional assumptions on the magnitude of the $ g_i $.

The goal is to obtain a set observables that allow the determination of the $\Gg$ with the smallest statistical uncertainty possible. To do this we first choose a generic set of observables $ \op_i(\phi),\, (i=1,\,\ldots,\,n) $ and let
\beq
\Op_i(A;\xibf_A) = \sum_{a=1}^A \op_i(\phi_a)\,.
\eeq
Then
\beq
\vvevof{\FFt \Op_i} = \sum_{A=1}^\infty \frac{e^{-\er} }{A!} \er^{A-1} A \vevof{\fft \op_i} =  \vevof{\fft \op_i} =\lum \vevof{\op_i f_j} g_j \,,
\eeq
where, given a function $\ucal(\xibf_A) $,
\beq
\vvevof \ucal = \sum_{A=1}^\infty \frac{e^{-\er} }{A!} \int d\xibf_A \, \ucal(\xibf_A)\,.
\eeq
Let then
\beq
\Gamma_i = C^{-1}_{ij} \Op_j\,, \qquad C_{ij} = \lum \vevof{\op_i f_j} \then g_i = \vvevof{\FFt \, \Gamma_i} \,,
\eeq
so the $ \Gamma_i $ can be used to determine the $g_i$. The correlation matrix of these observables is given by
\beq
V_{ij} =  \vvevof{\FFt \, \Gamma_i \Gamma_j} - g_i g_j \,,
\eeq
which we extremize as a function of the $\op_i $:
\beq
\frac{\delta   V_{ij}}{\delta\op_k}  =0 \,.
\eeq
A straightforward calculation shows that the solution is
\beq
\op_k = \frac{f_k}\fft \,,
\eeq
for which the condition $ \vevof{f_k} = C_{kl} g_l  $ is satisfied and
\beq
V = \inv\lum C^{-1} \,; \qquad C_{ij} = \lum \vevof{ \frac{f_i f_j}\fft} = \vevof{ \frac{f_i f_j}{(d\sigma/d\phi)} }\,.
\label{eq:v.oot.gen}
\eeq

If the cross section receives a contribution from both the SM and the new physics, and when the relevant SM parameters are well measured we write
\beq
\frac{d\sigma}{d\phi} = \frac{d\sigma_{\tt MS}}{d\phi} + \Gg\cdot\Ff(\phi)
\eeq
where the second term contains pure NP contributions, and, in general, SM-NP interference terms as well. In this case a similar derivation to the one above gives again \cref{eq:v.oot.gen}.

\bibliographystyle{JHEP}
\bibliography{ref_scalar}
\end{document}